\documentclass[12pt,a4paper]{article}
\usepackage{amsmath}
\usepackage{graphicx,color,subfigure}
\usepackage{amssymb,amsthm}
\usepackage{epsfig}
\usepackage{hyperref}
\def\half{\frac{1}{2}}
\newcommand{\nc}{\newcommand}
\nc{\renc}{\renewcommand} \nc{\nt}{\newtheorem}
\newlength{\defbaselineskip}
\nc{\setlinespacing}[1]%
           {\setlength{\baselineskip}{#1 \defbaselineskip}}
\nc{\doublespacing}{\setlength{\baselineskip}%
                           {2.0 \defbaselineskip}}
\nc{\singlespacing}{\setlength{\baselineskip}{\defbaselineskip}}
\renc{\baselinestretch}{1.5}    
\nc{\state}[1]{\left<#1\right>} \nc{\set}[1]{\left\{#1\right\}}
\nc{\bset}[1]{\left(#1\right)} \nc{\sbset}[1]{\left[#1\right]}
\nc{\tol}{\rightarrow} \nc{\To}{\Rightarrow}
\nc{\lr}{\Leftrightarrow} \nc{\nn}{\nonumber} \nc{\bo}{\textbf} \nc{\p}{\prime}

\nc{\delbar}{\delta\mkern-9mu\mathchar'26} \nc{\ho}{\hbar\omega}
\nc{\mz}{\mathcal{Z}} \nc{\md}{\mathcal{D}} \nc{\ml}{\mathcal{L}} \nc{\mS}{\mathcal{S}} \nc{\mU}{\mathcal{U}}
\nc{\mx}{\mathbf{x}} \nc{\mh}{\mathcal{H}} \nc{\mxi}{\mathbf{\xi}}

\nc{\mk}{\mathbf{k}} \nc{\order}[1]{\mathcal{O}(#1)}
\nc{\oi}{\omega_i} \nc{\frtwo}{\frac{1}{\rtwo}} \nc{\Eo}{E_\omega}
\nc{\So}{S_\omega} \nc{\eo}{\epsilon_\omega} \nc{\mo}{m_\omega}
\nc{\Uo}{U_\omega} \nc{\bsig}{\bar{\sigma}}
\nc{\wt}[1]{\tilde{#1}} \nc{\bx}{\bar{x}}
\nc{\wtPhi}{\wt{\Phi}} \nc{\wtMG}{\wt{M_G}} \nc{\dDx}{\int d^{D+1}x
 \sqrt{-g}} \nc{\sqg}{\sqrt{\abs{g}}}
\nc{\bq}{\bar{q}}
 \nc{\notts}{School of Physics
and Astronomy, University of Nottingham,\\ University Park, Nottingham
NG7 2RD, United Kingdom}

\nc{\etal}{\mbox{\it et al. }}
\nc{\ie}{{\it i.e. }} \nc{\eg}{{\it e.g. }}
\renc{\thefootnote}{\arabic{footnote}} \nc{\capt}[1]{{\bf Figure.}
{\small\sl #1}}

\nc{\eqss}[3]{\mbox{Eqs.~(\ref{#1}, \ref{#2}, \ref{#3})}}
\nc{\eqs}[2]{\mbox{Eqs.~(\ref{#1}, \ref{#2})}}
\nc{\eqm}[2]{\mbox{Eqs.~(\ref{#1}-\ref{#2})}}
\nc{\eq}[1]{\mbox{Eq.~(\ref{#1})}}
\nc{\figs}[2]{\mbox{Figs.~\ref{#1} and \ref{#2}}}
\nc{\fig}[1]{\mbox{Fig.~\ref{#1}}}
\nc{\tbl}[1]{\mbox{Table:~\ref{#1}}}
\nc{\tbls}[2]{\mbox{Tables:~\ref{#1} and \ref{#2}}}

\renc{\baselinestretch}{1.5}
\newlength{\overeqskip}
\newlength{\undereqskip}
\nc{\be}[1]{\begin{equation} \mbox{$\label{#1}$}}
\nc{\bea}[1]{\begin{eqnarray} \mbox{$\label{#1}$}}

\nc{\Section}[2]{\section{#2}\label{#1}}
\nc{\Bibitem}[1]{\bibitem{#1}} \nc{\Label}[1]{\label{#1}}
\nc{\ee}{\vspace{\undereqskip}\end{equation}}
\nc{\eea}{\vspace{\undereqskip}\end{eqnarray}}

\nc{\dpsty}{\displaystyle} \nc{\bc}{\begin{center}}
\nc{\ec}{\end{center}} \nc{\ba}{\begin{array}} \nc{\ea}{\end{array}}
\nc{\bab}{\begin{abstract}} \nc{\eab}{\end{abstract}}
\nc{\btab}{\begin{tabular}} \nc{\etab}{\end{tabular}}
\nc{\bit}{\begin{itemize}} \nc{\eit}{\end{itemize}}
\nc{\ben}{\begin{enumerate}} \nc{\een}{\end{enumerate}}
\nc{\bfig}{\begin{figure}} \nc{\efig}{\end{figure}}
\nc{\arreq}{&\!=\!&} \nc{\arrmi}{&\!-\!&} \nc{\arrpl}{&\!+\!&}
\nc{\arrap}{&\!\!\!\approx\!\!\!&} \nc{\non}{\nonumber\\*}
\def\dtil{\; \raise0.3ex\hbox{$d$\kern-0.75em
       \raise-1.1ex\hbox{$\tilde$}}\; }
\def\lsim{\; \raise0.3ex\hbox{$<$\kern-0.75em
       \raise-1.1ex\hbox{$\sim$}}\; }
\def\gsim{\; \raise0.3ex\hbox{$>$\kern-0.75em
       \raise-1.1ex\hbox{$\sim$}}\; }
\def\nar{\; \raise0.1ex\hbox{$\nabla$\kern-1.05em
       \raise1.5ex\hbox{$\rightarrow$}}\;}
\def\nal{\; \raise0.1ex\hbox{$\nabla$\kern-1.05em
       \raise1.5ex\hbox{$\leftarrow$}}\; }
\nc{\DOT}{\hspace{-0.08in}{\bf .}\hspace{0.1in}} \nc{\Laada}{\hbox
{$\sqcap$ \kern -1em $\sqcup$}}
\nc\loota{{\scriptstyle\sqcap\kern-0.55em\hbox{$\scriptstyle\sqcup$}}}
\nc\Loota{{\sqcap\kern-0.65em\hbox{$\sqcup$}}} \nc\laada{\Loota}

\nc{\real}{{\rm I \! R}} \nc{\Z}{{\sf Z \!\!\! Z}}
\nc{\complex}{{\rm C\!\!\! {\sf I}\,\,}}
\def\bigid{\leavevmode\hbox{\small1\kern-3.8pt\normalsize1}}
\def\id{\leavevmode\hbox{\small1\kern-3.3pt\normalsize1}}
\nc{\slask}{\!\!\!/} \nc{\bis}{{\prime\prime}} \nc{\pa}{\partial}
\nc{\na}{\nabla} \nc{\ra}{\rangle} \nc{\la}{\langle}
\nc{\goto}{\rightarrow} \nc{\swap}{\leftrightarrow}

\nc{\EE}[1]{ \mbox{$\cdot10^{#1}$} } \nc{\abs}[1]{\left|#1\right|}
\nc{\at}[2]{\left.#1\right|_{#2}} \nc{\norm}[1]{\|#1\|}
\nc{\abscut}[2]{\Abs{#1}_{\scriptscriptstyle#2}}
\nc{\vek}[1]{{\rm\bf #1}} \nc{\integral}[2]{\int\limits_{#1}^{#2}}
\nc{\inv}[1]{\frac{1}{#1}} \nc{\dd}[2]{{\frac{\partial #1}{\partial
#2}}} \nc{\ddd}[2]{{\frac{{\partial}^2 #1}{\partial {#2}^2}}}

\nc{\al}{\alpha} \nc{\g}{\gamma} \nc{\Del}{\Delta}
\nc{\eps}{\epsilon} \nc{\lam}{\lambda} \nc{\om}{\omega}
\nc{\Om}{\Omega} \nc{\ve}{\varepsilon} \nc{\mn}{{\mu\nu}}
\nc{\vp}{\varphi} 
\nc{\nb}{\nonumber}
\renewcommand{\baselinestretch}{1.1}
\makeatletter
\newcommand{\figcaption}[1]{\def\@captype{figure}\caption{\scriptsize{#1}}}
\newcommand{\tblcaption}[1]{\def\@captype{table}\caption{\scriptsize{#1}}}
\makeatother
\begin{document}

\begin{center}
{\Large {\bf Some stationary properties of a $Q$-ball\\ in arbitrary space dimensions} }

\vspace{10mm}
Mitsuo I. Tsumagari${}^\dag$\footnote{ppxmt@nottingham.ac.uk}, Edmund~J.~Copeland${}^\dag$\footnote{ed.copeland@nottingham.ac.uk} and Paul M. Saffin${}^\dag$\footnote{paul.saffin@nottingham.ac.uk}
\\[3mm]
\vspace{.5cm}

{\em  ${}^\dag$\it\notts}
\end{center}
\begin{abstract}

Introducing new physically motivated ans\"{a}tze, we explore both analytically and numerically the classical and absolute stabilities of
a single $Q$-ball in an arbitrary number of spatial dimensions $D$, working in both the thin and thick wall limits.

\end{abstract}
\section{Introduction}
\hspace{12pt}
In a pioneering paper published in 1985 \cite{Coleman:1985ki}, Sidney Coleman showed that it was possible for a new  class of non-topological solitons \cite{Friedberg:1976me} to exist within a self-interacting system by introducing the notion of a $Q$-ball . His model had a continuous unbroken global U(1) charge $Q$ (for reviews see \cite{Lee:1991ax, raj, Dine:2003ax, Enqvist:2003gh}), which corresponds to an angular motion with angular velocity $\omega$ in the U(1) internal space. The conserved charge stabilises the $Q$-ball, unlike the case of topological solitons whose stability is ensured by the presence of conserved topological charges. Once formed, a $Q$-ball is absolutely stable if four conditions are satisfied: (1) \textit{existence condition} \cite{Coleman:1985ki} - its potential should grow less quickly than the quadratic mass term, and this can be realised through a number of routes such as the inclusion of  radiative or finite temperature corrections to a bare mass, or  non-linear terms in a polynomial potential \cite{Coleman:1985ki, Battye:2000qj},
(2) \textit{absolute stability condition} - the energy $E_Q$ (or mass) of a $Q$-ball must be lower than the corresponding energy that the collection of the lightest possible scalar particle quanta (rest mass $m$) could have, (3) \textit{classical stability condition} \cite{Friedberg:1976me} - the $Q$-ball should be stable to linear fluctuations; with the threshold of the stability being located at the saddle point of the $D$-dimensional Euclidean action, the bounce action $\So$, \cite{Callan:1977pt}, (4) \textit{fission condition} \cite{Lee:1991ax} - the energy of a single $Q$-ball must be less than the total energy of the smaller $Q$-balls that it could in principle fragment into. It turns out that for each of the four conditions to be satisfied we require:
\be{four}
\omega_-\leq \abs{\omega}<\omega_+,\ \ E_Q< mQ,\ \ \frac{\omega}{Q}\frac{dQ}{d\omega}\leq 0 \lr \frac{d^2\So}{d\om^2}\ge 0,\ \ \frac{d^2E_Q}{dQ^2}<0 \lr \frac{d\omega}{dQ}<0
\ee
where $\omega_\mp$ are the lower and upper limits of $\om$ that the $Q$-ball can have. The lower limit, $\om \simeq \om_-$, can define thin wall $Q$-balls, either without \cite{Coleman:1985ki} or with
\cite{Spector:1987ag, Shiromizu:1998rt}
the wall thickness being taken into account, while the upper limit, $\om\simeq \om_+$, can define thick wall $Q$-balls in \cite{Kusenko:1997ad} which may be approximated by a simple Gaussian ansatz \cite{Gleiser:2005iq}.

There is a vast literature on non-topological solitons, including $Q$-balls. They have been seen to be solutions in Abelian gauge theories \cite{Shiromizu:1998rt, Lee:1988ag, Levi:2001aw, Anagnostopoulos:2001dh, Li:2001he, Shiromizu:1998eh}, in non-Abelian theories \cite{Safian:1987pr, Safian:1988cz, Axenides:1998fc}, in self-dual (Maxwell-) Chern-Simons theory \cite{Jackiw:1990pr, Jackiw:1990aw, Hong:1990yh, DeshaiesJacques:2006ae}, in noncommutative
 complex scalar field theory \cite{Kiem:2001ny}, in models which include fermionic interactions \cite{Lee:1988ag, Levi:2001aw, Anagnostopoulos:2001dh, Shiromizu:1998eh, Friedberg:1976eg, Friedberg:1977xf}, as well as in the presence of gravity \cite{Lee:1986tr, Lynn:1988rb}. $Q$-balls themselves have been quantized either by canonical \cite{Friedberg:1976me} or by path integral schemes \cite{PaccettiCorreia:2001uh, Benson:1989id, Rajaraman:1975qr}. With thermal effects, it has been shown that $Q$-balls coupled to massless fermions are able to evaporate away \cite{Cohen:1986ct}; however at sufficiently low temperatures they become stable, and indeed they then tend to grow \cite{Benson:1991nj, Laine:1998rg}. The authors in \cite{Friedberg:1976me, Shiromizu:1998rt, Volkov:2002aj, Brihaye:2007tn} have discussed and analysed the spatially excited states of $Q$-balls, including radial modes as well as spatially dependent phase excitations. A more general mathematical argument concerning the stability of solitary waves can be found in \cite{Shatah:1985vp, Blanchard:87}. A related class of objects to $Q$-balls are known as oscillons \cite{Bogolyubsky:1976nx, Bogolyubsky:1976sc, Bogolyubsky:1976yu, oscillon1, Gleiser:1993pt,Copeland:1995fq} or as I-balls \cite{Kasuya:2002zs}, and recent attention has turned to the dynamics of these time dependent non-linear meta-stable configurations \cite{Saffin:2006yk, Gleiser:2007ts, Hindmarsh:2007jb}.

Standard $Q$-balls exist in an arbitrary number of space dimensions $D$ and are able to avoid the restriction arising from  Derrick's theorem  \cite{Derrick:1964ww} because they are time-dependent solutions. A few examples include polynomial models both for $D=3$ \cite{Axenides:1999hs, Multamaki:1999an} and for arbitrary $D$ \cite{Gleiser:2005iq}; models with supersymmetry broken by gravity mediation \cite{Multamaki:1999an}; and models with supersymmetry broken by gauge interaction \cite{Laine:1998rg, Multamaki:2000ey, Asko:2002phd}. Returning to the case of $D=3$, phenomenologically, it turns out that the $Q$-balls present in models with gravity mediated supersymmetry breaking are quasi-stable but long-lived, allowing in principle for these $Q$-balls to be the source of both the baryons as well as the lightest supersymmetric particle (LSP) dark matter particle \cite{Enqvist:1997si}. On the other hand, $Q$-balls in models of gauge mediated supersymmetry breaking can be a dark matter candidate as they can be absolutely stable \cite{Enqvist:2003gh}. Both types of $Q$-balls have been shown to be able to provide the observed baryon-to-photon ratio \cite{Laine:1998rg}.

The dynamics and formation of $Q$-balls involve solving complicated non-linear systems, which generally require numerical simulations. The dynamics of two $Q$-balls in flat Minkowski space-time depends on parameters, such as the relative phases between them, and the relative initial velocities \cite{Battye:2000qj, Multamaki:2000ey, Multamaki:2000qb}. In addition, the main formation process through the Affleck-Dine mechanism \cite{Affleck:1984fy} has been extensively examined in both gauge mediated  \cite{Kasuya:1999wu}, gravity mediated \cite{Kasuya:2000wx, Enqvist:2000cq, Multamaki:2002hv}, and
running inflaton mass models \cite{Enqvist:2002si}. As analysing individual $Q$-balls is difficult in its own right, it is extremely challenging to deal with multiple $Q$-balls. A number of analytical approaches to address that issue have been made over the past few years, \eg \cite{Lee:1994qb, Koutvitsky:2006mp, Griest:1989cb}. Multiple thermal $Q$-balls have been described in a statistical sense in \cite{Enqvist:2000cq, Christ:1975wt}.

In this paper we aim to analytically and numerically address stationary properties of a single $Q$-ball in an arbitrary number of spatial dimensions. The work will draw on earlier work of Correia and Schmidt \cite{PaccettiCorreia:2001uh} who derived analytic properties for the thin and thick wall limits of $Q$-balls in $D=3$. Recently, Gleiser and Thorarinson \cite{Gleiser:2005iq} proved the absolute stability for thin wall $Q$-balls using the virial theorem. We will generalise the main results of \cite{Gleiser:2005iq, PaccettiCorreia:2001uh} to the case of arbitrary spatial dimensions, and in doing so will both analytically predict and numerically confirm the unique values of the angular velocity $\om_a$ for the absolute stability of the $Q$-balls via both the virial relation and thin wall $Q$-ball approximations.

Moreover, we will obtain the classical stability conditions for the thin and thick wall approximations, and  discover the connections between the virial relation and thin or thick wall approximation for the characteristic slopes $E_Q/\om Q$. In a companion paper \cite{next:paper} we will investigate dynamical properties of multiple $Q$-balls, including thermal effects, and their formation (see \cite{mitmovie} for movies showing the dynamics of single $Q$-balls).

This paper is divided into the following sections. In section \ref{qball} we review the properties of a $Q$-ball in an arbitrary number of spatial dimensions $D$, including the existence, and stability conditions. By introducing a number of different ans\"{a}tze, in section \ref{thin-thick} we present a detailed analysis of the solutions in the thin and thick wall limits. We then demonstrate the advantages of using two particular modified ans\"{a}tze in section \ref{numerical} where we present detailed numerical results for the case of both degenerate and non-degenerate underlying potentials. Finally we conclude in section \ref{conc}.
\section{\textit{Q}-ball in $D$ dimensions}
\label{qball}
We shall begin with a standard $Q$-ball ansatz \cite{Coleman:1985ki} which satisfies a Laplace equation called the $Q$-ball equation, and we will introduce the Legendre relations \cite{Friedberg:1976me} which will make some computations easier. The existence of $Q$-balls places constraints on the allowed form of the potential, and introduces limiting values of $\omega$, \ie $\omega_\pm$, near which we may describe the $Q$-balls analytically using either a thick or thin wall approximation. We will then introduce three conditions for $Q$-balls to be stable \cite{Lee:1991ax}. Finally we will obtain the characteristic slope $E_Q/\om Q$ and minimum charge $Q_{min}$, and propose approximate values for $\om_a$, the limiting frequency for absolute stability, using a virial theorem and showing that it does not rely on detailed analytic profiles and potential forms.
\subsection{\textit{Q}-ball ansatz}
We consider a complex scalar field $\phi$ in Minkowski spacetime of arbitrary spatial dimensions $D$ with a U(1) potential bounded by $U(\abs{\phi})\ge 0$ for any values of $\phi$:
\bea{act}
S&=& \int d^{D+1}x \sqrt{-g}\; \ml,\\
\textrm{where} \hspace{10pt} \ml&=& -\half g^{\mn}\pa_\mu \phi^{\dag} \pa_\nu \phi - U(|\phi|).
\eea
The metric is $ds^2=g_{\mn}dx^\mu dx^\nu=-dt^2+h_{ij}dx^idx^j$ and $g$ is the determinant of $g_{\mn}$ where $\mu,\, \nu$ run from $0$ to $D$, and $i,\, j$ denote spatial indices running from $1$ to $D$. Now using the standard decomposition of $\phi$ in terms of two real fields $\phi=\sigma e^{i \theta}$, the energy momentum tensor $T_{\mn}\equiv -\frac{2}{\sqrt{-g}}\frac{\delta S}{\delta g_{\mn}} + (symmetrising\; factors)$ and the conserved U(1) global current $j_{\mu,U(1)}$ via the N\"{o}ether theorem, we obtain
\bea{}
T_{\mn}&=&(\pa_\mu \sigma \pa_\nu \sigma + \sigma^2 \pa_\mu \theta \pa_\nu \theta)+g_{\mn} \ml,\\
j_{\mu,U(1)}&=&\sigma^2 \pa_\mu \theta.
\eea
Using a basis of vectors $\{n^\mu_{(a)}\}$ where $n^\mu_{(t)}$ is time-like and
  $n^\mu_{(i)}$ are space-like unit vectors oriented along the spatial $i$-direction, the above currents give the definitions of energy density $\rho_E$, charge density $\rho_Q$, momentum flux $\hat{P}_i$ and pressure $p$:
\be{defeq}
\rho_E \equiv T_{\mn} n^\mu_{(t)} n^\nu_{(t)};\hspace{10pt} \rho_Q\equiv j_\mu n^\mu_{(t)};\hspace{10pt} \hat{P}_i \equiv T_{\mn} n^\mu_{(t)} n^\mu_{(i)};\hspace{10pt} p\equiv T_{\mn} n^\mu_{(i)} n^\nu_{(i)}.
\ee
Defining the $D$ dimensional volume $V_D$ bounded by a $(D-1)$-sphere, the Noether charges (energy, momenta, and U(1) charge) become
\be{epq}
E=\int_{V_D}\rho_E,\hspace{15pt} P_i=\int_{V_D} \hat{P}_i,\hspace{15pt} Q=\int_{V_D}\rho_Q,
\ee
where $\int_{V_D}\equiv \int d^Dx \sqrt{h}$. Minimising an energy with a fixed charge $Q$ for any degrees of freedom, we find the $Q$-ball (lowest) energy $E_Q$ by introducing a Lagrange multiplier $\omega$ and setting $n^\mu_t=(-1,0,0,\dots,0)$:
\bea{EQ}
E_Q&=&E+\omega\bset{Q-\int_{V_D} \rho_Q},\\
\label{EQ2} &=& \omega Q + \int_{V_D} \bset{\half \set{ \dot{\sigma}^2+\sigma^2(\dot{\theta}-\omega)^2 +(\nabla \sigma)^2 + \sigma^2 (\nabla \theta)^2} + \Uo }, \\
\label{legtrns}&=& \omega Q + \So
\eea
where $\Uo=U-\half \omega^2 \sigma^2$, $\dot{\sigma} \equiv \frac{d\sigma}{dt}$ etc... and $\omega$ will turn out to be the rotation frequency in the U(1) internal space. The presence of the positive definite terms in \eq{EQ2} suggests that the lowest energy solution is obtained by setting
$\dot{\sigma}=0=\dot{\theta}-\omega=\nabla \theta$. The Euclidean action $\So$ and the effective potential $U_\omega$ in \eqs{EQ2}{legtrns} are finally given by
\be{Uo}
\So=\int_{V_D} \half (\nabla \sigma)^2 + U_\omega,\hspace{10pt} U_\omega \equiv U -\half \omega^2 \sigma^2.
\ee
The second term in $U_\omega$ comes from the internal spin of the complex field. Following Friedberg et. al \cite{Friedberg:1976me}, it is useful to define the functional
\be{legtrns2}
G_I\equiv \int_{V_D} \half (\nabla \sigma)^2 + U
=  E_Q - \bset{\half \omega^2} I = \So + \bset{\half \omega^2} I
\ee
where $\half \omega^2$ is the corresponding Lagrange multiplier and $I\equiv \int_{V_D} \sigma^2$.

Given that the spherically symmetric profile is the minimum energy configuration  \cite{spherical}, we are lead to the standard stationary $Q$-ball ansatz
\be{qansatz}
\phi=\sigma(r)e^{i\omega t}.
\ee
Substituting \eq{qansatz} into \eq{defeq}, we find
\bea{rho_Q}
\label{cheg}    \rho_E&=& \half \sigma^{\p 2} + U+\half \sigma^2 \om^2 ,\hspace*{10pt} \rho_Q=\omega \sigma^2,\\
\label{radialp}  p_r&=&\half \sigma^{\p 2}-\Uo, \hspace*{10pt} P_i=0
\eea
where $\sigma^{\p} \equiv \frac{d\sigma}{dr}$ and $p_r$ is a radial pressure given in terms of the radially oriented unit vector $n^\mu_s=(0,1,0,\dots,0)$.
Without loss of generality, we set both $\om$ and $Q$ as positive.
\subsection{Legendre relations}
It is sometimes difficult to compute $E_Q$ directly, but using Legendre relations often helps \cite{Friedberg:1976me, PaccettiCorreia:2001uh}. In our case, from \eq{legtrns} and \eq{legtrns2} we find
\be{legendre}
\left.\frac{d E_Q}{d Q}\right|_{\So} =\omega,\hspace{10pt} \left.\frac{d \So}{d\omega}\right|_{E_Q}=-Q,\hspace{10pt} \left.\frac{d G_I}{dI}\right|_{\So}=\half \omega^2
\ee
because $Q$-ball solutions give the extrema of $E_Q,\; \So$, and $G_I$ as regards  $Q,\; \om,$ and $I$, respectively.
These variables match the corresponding "thermodynamic" ones:
$E_Q,\; \omega,\; Q,\; \So$, and $G_I$ correspond to the internal energy, chemical potential, particle number,
and "thermodynamic" potentials \cite{Laine:1998rg}. After computing $\So$ or $G_I$,
one can calculate $Q$ or $\half \om^2$ using the second or third relation in \eq{legendre},
and can compute $E_Q$ using \eq{legtrns} or \eq{legtrns2}, \ie
\be{easycalc}
\So \to Q=-\frac{d \So}{d \omega} \to E_Q=\omega Q + \So,
\ee
or similarly $G_I \to \half \om^2=\frac{d G_I}{dI}\to E_Q=G_I+\bset{\half \om^2} I,\; \So =G_I -\bset{\half  \om^2} I.$ We shall make use of this powerful technique later.
\subsection{\textit{Q}-ball equation and existence condition}

\begin{figure}[ht]
  \begin{center}
   \subfigure{\label{fig:deg}\includegraphics[angle=-90,scale=0.32]{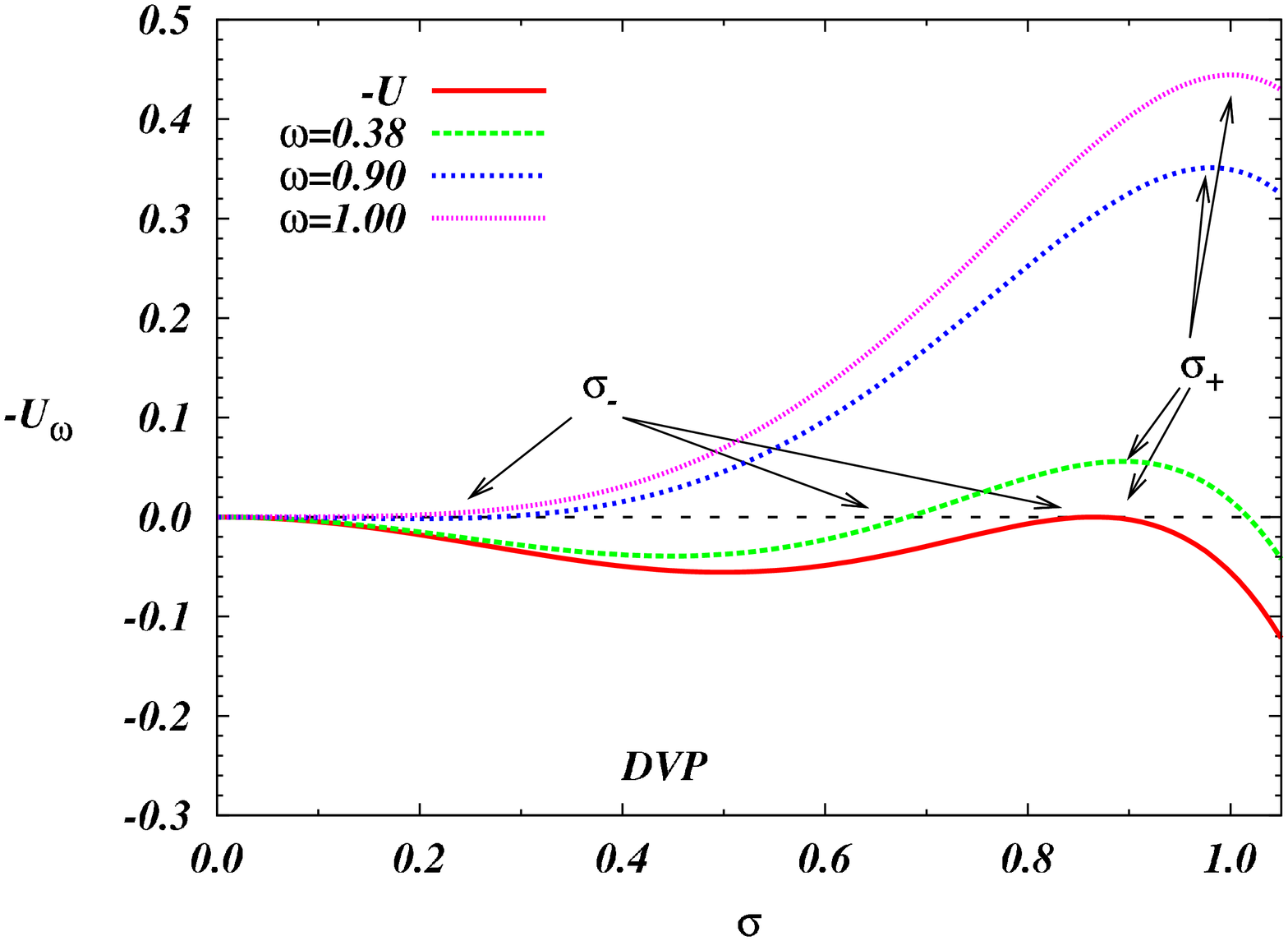}}
   \subfigure{\label{fig:nondeg} \includegraphics[angle=-90, scale=0.32]{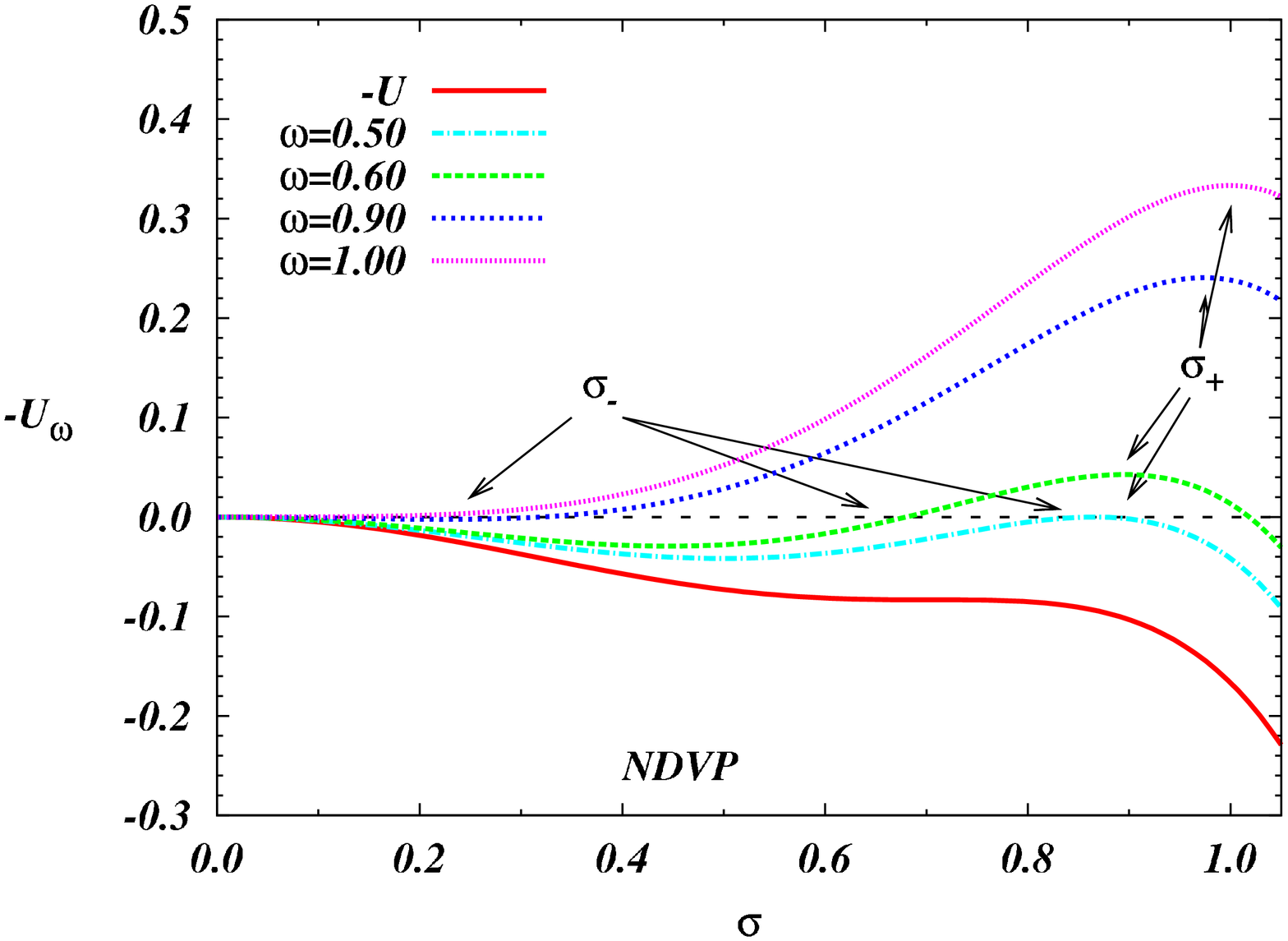}}
  \end{center}
  \figcaption{Parameters $\sigma_\pm$ in two typical
  potentials $U(\sigma)=\half \sigma^2-A \sigma^4+ B \sigma^6$ where $m=\om_+=1$ and the effective potentials $-\Uo$ are plotted for various values of $\om$: degenerate vacua potential (DVP) with $A=\frac43,\; B=\frac89$ on the left and non-degenerate vacua potential (NDVP) with $A=1,\; B=\frac23$ on the right. The DVP has degenerate vacua in the original potential $-U$ (red-solid line) where we set $\om_-=0$. The NDVP does not have degenerate vacua, but with $\om=\om_-=0.5$ (sky-blue dot-dashed line) the effective potential $-\Uo$ does have degenerate vacua. The two lines in the lower limit $\om=\om_-$ show that $\sigma_-\to \sigma_+$ where we have defined the maximum of the effective potential to be at $\sigma_+$ and $\Uo(\sigma_-)=0$ for $\sigma_-\neq 0$. The purple lines show $\sigma_- \to 0$ with the thick wall limit $\om=\om_+$. With some values of $\om$ (green dotted lines) satisfying the existence condition \eq{QBeq}, both potentials show the values of $\sigma_\mp$ clearly.}
\label{fig:twomdl}
\end{figure}

Let us consider the action
$S=-\int dt  \So$ in \eq{act} with our ansatz \eq{qansatz} and the following boundary condition on a $(D-1)$-sphere which represents spatial infinity
\be{nontbdry}
\sigma^\p|=0\; \textrm{on the ($D-1$)-sphere}.
\ee
Varying $\So$ with respect to $\sigma$ we obtain the $Q$-ball equation:
\bea{QBeq}
\frac{d^2\sigma}{dr^2}+\frac{D-1}{r}\frac{d\sigma}{dr}-\frac{dU_\omega}{d\sigma}&=&0,\\
\label{dampsig}\lr \frac{d}{dr}\bset{\half \bset{\frac{d\sigma}{dr}}^2-U_\omega}&=&-\frac{D-1}{r}\bset{\frac{d\sigma}{dr}}^2 \le 0.
\eea
There is a well known mechanical analogy for describing the $Q$-ball solution of \eq{QBeq} \cite{Coleman:1985ki}, and that comes from viewing \eq{QBeq} in terms of the Newtonian dynamics of an unit-mass particle with position $\sigma$, moving in potential $-U_\omega$ with a friction $\frac{D-1}{r}$, where $r$ is interpreted as a time co-ordinate. Moreover $\rho_Q=\omega \sigma^2$ can be considered as the angular momentum \footnote{$I$ is realised as an inertia moment in this mechanical analogy \cite{Coleman:1985ki, Friedberg:1976me}.}. Note that the friction term is proportional to $\frac{D-1}{r}$, hence becomes significant for high $D$ and/or small $r$. According to \eq{dampsig}, the "total energy", $\half \bset{\frac{d\sigma}{dr}}^2-U_\omega$, is conserved for $D=1$ and/or $r \tol \infty$, implying that in that limit the $Q$-balls have no radial pressure. Of course these are really field theory objects, and consequently, more restrictions apply: (i) no symmetry breaking, in other words $\sigma(r \to large)=0;\ U^{\p\p}(\sigma=0)\equiv m^2>0$ with an effective mass $m$, (ii) regularity condition: $\sigma^\p(r=0)=0$, (iii) reflection symmetry under $\sigma \to -\sigma$.  Note that \eq{QBeq} coupled with the boundary condition \eq{nontbdry} implies  $\sigma(r)$ is a monotonically decreasing function, \ie $\sigma^\p<0$. In fact, according to \eqs{nontbdry}{QBeq} and the above conditions, our mechanical analogy implies that a static particle with a unit mass should be released somewhere on its potential, eventually reaching the origin at large (but finite) time and stopping there due to the presence of a position- and $D$- dependent friction. These requirements constrain the allowed forms of the U(1) potentials: for example if the local maximum of the effective potential $-\Uo$ is less than $0$, the "particle" can not reach the origin, a process known as $undershooting$. To avoid undershooting we require
\be{LEFT}
max(-U_\omega) \geq 0 \lr \min\bset{\frac{2U}{\sigma^2}}\leq \omega^2.
\ee
If $-U_\omega$ is convex at $\sigma=0$, the "particle" cannot stop at the origin, a situation  termed \\$overshooting$ such that
\be{RIGHT}
\left.\frac{d^2 U_\omega}{d\sigma^2}\right|_{\sigma=0}<0 \lr \omega^2 < \left.\frac{d^2U}{d\sigma^2}\right|_{\sigma=0}.
\ee
Combining \eqs{LEFT}{RIGHT}, we find the condition on $\omega$ for the existence of a single $Q$-ball:
\be{EXIST}
 \omega_-\leq \abs{\omega} < \omega_+
\ee
where we have defined the lower and upper bounds of $\omega$ as $\omega_\mp$, \ie $\omega^2_- \equiv \min\bset{\frac{2U}{\sigma^2}} \ge 0$ and $\omega^2_+\equiv \frac{d^2U}{d\sigma^2}|_{\sigma=0}= m^2$. The case, $\om_-=0$, corresponds to degenerate vacua potentials (DVPs), while the other case, $\om_-\neq 0$, does not have degenerate vacua (NDVPs). The \textit{existence condition} in \eq{EXIST} shows that U(1) potentials must have a non-linear interaction and $U_\omega$ is weakly attractive \cite{Lee:1991ax}. For convenience we define the maximum of the effective potential to be at $\sigma_+$ (i.e. $\frac{dU_\om}{d\sigma}|_{\sigma = \sigma_+}=0)$, thus $\om^2_-=\frac{2U_+}{\sigma^2_+}$ and $U_{\omega_-}(\sigma_+)=0$. Moreover  $\sigma_-$ satisfies $\Uo(\sigma_-)=0$ for $\sigma_-\neq 0$. Notice $\sigma_-\simeq \sigma_+$ when $\om \simeq \om_-$. In \fig{fig:twomdl}, we indicate the above introduced parameters, $\sigma_\pm,\; \om_-$ using typical original and effective potentials for both DVP (left) and NDVP (right).
To proceed with analytical arguments, we consider the two limiting values of $\om$ or $\sigma_0\equiv \sigma(0)$ which describe
\be{}
\begin{cases}
  \bullet \hspace*{5pt} \textrm{thin wall $Q$-balls when}\; \om\simeq \om_-\; or\; \sigma_0\simeq \sigma_+\\
  \bullet \hspace*{5pt} \textrm{thick wall $Q$-balls when}\; \om\simeq \om_+\; or\; \sigma_0 \simeq \sigma_-.
\end{cases}
\ee
We will not be considering $Q$-ball solutions that exist in a false vacua where $\om^2_-<0$ \cite{Spector:1987ag} or in flat potentials. When it comes to obtaining $Q$-ball profiles numerically, we will adopt a standard shooting method which fine-tunes the "initial positions" $\sigma_0$ subject to $\sigma_-(\om) \le \sigma_0<\sigma_+(\om)$, in order to avoid \emph{undershooting} and \emph{overshooting}.

\subsection{Three kinds of stability}
\paragraph*{Absolute stability}
When the volume $V_D$ approaches infinity \cite{Lee:1991ax} and/or $\om$ is outside the limits of  \eq{EXIST}, then plane wave solutions can exist around the vacua of $U(|\phi|)$. The equation of motion for $\phi$ becomes a free Klein-Gordon equation whose solution can be written as $\phi=N e^{i(\textbf{k}\cdot \textbf{x}-\omega_k t)}$ where $\omega_k=\sqrt{m^2+\mathbf{k}^2}$ and the normalisation factor $N=\sqrt{\frac{Q}{2\omega_k V_D}}$ has been calculated from $Q$. Then the energy of the plane wave solution is proportional to $\omega_k$ and $Q$ linearly: $E_{free}=\omega_k Q \tol E_{free} \simeq mQ$ where we have taken the infra-red limit, to obtain the second relation. The energy $E_{free}$ can be interpreted as the energy of
a number $Q$ of free particle quanta with the rest masses $m$. Furthermore, one might expect that the thick wall $Q$-ball energy with $\om \simeq \om_+=m$ approaches $E_{free}$ because the $Q$-ball profiles approach zero exponentially at infinity \cite{Lee:1991ax}:
\be{freeengy}
E_Q(\om=\om_+) \simeq E_{free}\simeq mQ.
\ee
Hence the $absolute\; stability$ condition for a $Q$-ball becomes
\be{ABSCOND}
E_Q(\om)<E_{free}.
\ee
We would expect \eq{ABSCOND} to be the strongest condition which a $Q$-ball solution has to satisfy.
\paragraph*{Classical stability}
The \textit{classical stability} \cite{Friedberg:1976me, Lee:1991ax} can be
defined in terms of the mass-squared of the fluctuations around a $Q$-ball
solution. For zero mass fluctuations this corresponds to a zero mode, \ie translation and phase transformation of the $Q$-ball solution. Using collective coordinates and \eq{QBeq} which extremises $\So$, such a mode can be removed. Since a detailed analysis can be found in the literature \cite{Friedberg:1976me, Lee:1991ax}, we simply state the final result which implies the \emph{classical stability} condition is
\be{CLS}
\frac{\omega}{Q}\frac{dQ}{d\omega}\leq 0 \lr \frac{d^2\So}{d\om^2}\ge 0
\ee
where we have used \eq{legendre} in the second relation of \eq{CLS}. Since $\omega$ and $Q$ have the same sign, the sign of $\frac{dQ}{d\om}$ signals whether the solution is classically stable. The first relation of \eq{CLS} indicates the presence of an extreme charge in the parameter space of $\om$, (we will later see that the extreme charge at some critical value $\om=\om_c$ turns out to be the minimum allowed and will be denoted by $Q_{min}$). Let us remark on the characteristic slope of $E_Q/Q$ as a function of $\om$:
\be{chslope}
\frac{d}{d\om} \bset{\frac{E_Q}{Q}}=-\frac{\So}{Q^2}\frac{dQ}{d\om} \ge 0
\ee
where we have used \eq{legtrns} and \eq{legendre}. Since, as we will see, $\So>0$ for all possible values of $\om$ within \eq{EXIST} (in a specific potential \cite{Alford:1987vs}), the classically stable $Q$-balls should satisfy $\frac{d}{d\om} \bset{\frac{E_Q}{Q}}\ge 0$. The conditions from both \eq{CLS} and \eq{chslope} must be same.
\paragraph*{Stability against fission}
Suppose the total energy of two $Q$-balls is less than the energy of a single $Q$-ball carrying the same total charge. The single $Q$-ball naturally decays into two or more with some release of energy. As shown in \cite{Lee:1991ax}, the stability condition against \emph{fission} for a $Q$-ball is given by
\be{SAF}
\frac{d^2E_Q}{dQ^2}<0 \lr \frac{d\omega}{dQ}<0
\ee
where we have used \eq{legendre} in the second relation of \eq{SAF}. Note that this is the same condition as we found above in \eqs{CLS}{chslope}, so the condition for classical stability is identical to that of stability against fission.

\vspace{10pt}
Trying to summarise the stability we can categorise three types of a $Q$-ball: \ie absolutely stable, meta-stable, or unstable $Q$-balls. Absolutely stable $Q$-balls are stable quantum mechanically as well as classically; meta-stable $Q$-balls decay into free particle quanta, but are stable under small fluctuations; whereas unstable $Q$-balls sometimes called $Q$-clouds \cite{Alford:1987vs} decay into lower energy $Q$-balls or free particle quanta.

\subsection{Virial theorem}
Derrick's theorem restricts the existence of static non-trivial scalar field solutions in terms of a number of spatial dimensions. For example in a real scalar field theory, non-trivial solutions exist only in one-dimension, \eg Klein-Gordon kink.
$Q$-balls (or any non-topological solitons), however, avoid this constraint because they are time-dependent (stationary) solutions \cite{Kusenko:1997ad, Gleiser:2005iq}. We can easily show this and in doing so obtain  useful information about the scaling properties of the $Q$-balls as a function of dimensionality as well as the ratio between the surface and potential energies. Following \cite{Gleiser:2005iq}, we begin by scaling the $Q$-ball ansatz \eq{qansatz} using a one-parameter family $r \to \alpha r$, whilst keeping $Q$ fixed. Defining a surface energy $\mS\equiv \int_{V_D} \half \sigma^{\p 2}$, a potential energy $\mU\equiv \int_{V_D} U$, and recalling that the charge satisfies $Q=I\omega$, we see that the energy of the $Q$-ball, \eq{legtrns}, becomes
\bea{}
E_Q &=& \mS+\mU + \frac{Q^2}{2I}.
\eea
Now, under the scaling $r \to \alpha r$, then $E_Q \to E_Q^\p$ where $\left.\dd{E_Q^\p}{\alpha}\right|_{\alpha=1}=0$ because the $Q$-ball solutions are the extrema (minima) of $E_Q$. Evaluating this, we obtain the virial relation relating $\mU$ and $\mS$
\be{virial}
 D\ \mU = -(D-2)\mS +D\frac{Q^2}{2I} \geq 0 \To Q^2 \geq Q^2_{min}
\ee
where we have used our earlier notation, $U\ge 0$, for any values of $\sigma$ and defined the minimum charge (corresponding to $\mU =0$) as $Q^2_{min}\equiv \frac{2I(D-2)}{D}\mS > 0 \lr  D > 2$. Since $Q$ is taken as real and positive, no conditions appear for $D=1,\; 2$. Notice this does not mean that one- and two- dimensional $Q$-balls do not exist, as can be seen from \eqs{QBeq}{EXIST}.
The case of $Q=0$ recovers Derrick's theorem, showing no time-independent
solutions for $D\geq2$ \cite{Gleiser:2005iq}.
Using $\mS=\frac{D Q^2}{2I}\bset{D-2+D\; \frac{\mU}{\mS}}^{-1}$ from \eq{virial}, the characteristic slope $E_Q/\om Q$ becomes
\be{linear}
\frac{E_Q}{\omega Q}=1+\bset{D-2+D\; \frac{\mU}{\mS}}^{-1}.
\ee
Let us consider three cases: (i)$\,\mU \gg \mS$, (ii)$\,\mU \simeq \mS$, and (iii)$\,\mU \ll \mS$. They lead  to predictions for the ratio of the $Q$-ball energy $E_Q$ to the energy contribution from the charge $\om Q$:
\be{virieq}
\frac{E_Q}{\omega Q}\simeq
    \left\{
    \begin{array}{ll}
    1 &\; \textrm{for (i)}, \\
    \frac{2D-1}{2(D-1)} &\; \textrm{for (ii)},\\
    \frac{D-1}{D-2} &\;	 \textrm{for (iii)}.
    \end{array}
    \right.
\ee
All of the $Q$-balls in the range of $\om$ are classically stable because the terms, $E_Q/Q$, monotonically increase as a function of $\om$, see \eqs{chslope}{virieq}.
The first case (i) corresponds to the extreme thin and thick wall limits $\omega \simeq \omega_\mp$ as will see. In the second case (ii), the potential energy is of the same order as the surface energy which means $\mS$ and $\mU$ have equally virialised. This case will turn out to be that of the thin wall limit for DVPs when the surface effects are included.

Suppose $\mS/\mU=const.$ over the large range of $\om$ within the existence condition \eq{EXIST} except $\om\simeq \om_+$ where $E_Q/\om_+ Q\simeq 1$.  We can find an approximate threshold value $\om_a$ for a $Q$-ball to be absolutely stable using \eqs{freeengy}{virieq}:
\be{viriwa}
\frac{\om_a}{m} \simeq
    \left\{
    \begin{array}{ll}
    1 &\; \textrm{for (i)}, \\
    \frac{2(D-1)}{2D-1} &\; \textrm{for (ii)},\\
    \frac{D-2}{D-1} &\;	 \textrm{for (iii)}.
    \end{array}
    \right.
\ee
Roughly speaking, $Q$-balls are classically and absolutely stable if $\om< \om_a$ because of \eqs{ABSCOND}{chslope} and \eq{virieq}. These approximations can and will be justified by our numerical results. We will find that the virial relation is a powerful tool enabling us to find appropriate values of $\om_a$ as opposed to the rather complicated computations we will have to adopt in the next section. We should point out a caveat in this argument, the assumption we are making here, that most of the $Q$-balls have an identical energy ratio $\mS/\mU$ over a range of $\om$, does of course rely on the specific form of the potential. We have to remind the readers that the virial relation \eq{virial} gives only the relation between $\mS$ and $\mU$, and states the presence of the minimum charge $Q_{min}$ if the system allows the time-dependent ($Q$-ball) solutions \eq{qansatz} to exist.
\section{Thin and thick wall approximations}
\label{thin-thick}
In this section we will obtain approximate solutions for $Q$-balls in $D$-dimensions based on the well known thin and thick wall approximations for the radial profiles $\sigma(r)$ of the fields. Moreover we will show how we can then use these results to verify the solutions we obtained in the previous section for $E_Q/\om Q$ in \eq{virieq}. Further we will then be able to test the solutions against detailed numerical solutions in section \ref{numerical}. We start with two simple ans\"{a}tze for the radial profiles, a step-like function for the thin wall case $\om\simeq \om_-$ and a Gaussian function for the thick wall case $\om\simeq \om_+$. In both cases we will evaluate $\So$, $Q$, $E_Q$, as well as the conditions for  classical and absolute stability before modifying the ans\"{a}tze. Following that we will repeat the same calculations using our more physically motivated ans\"{a}tze via the Legendre transformation technique described in \eq{easycalc}. Let us comment briefly on the form of the potential. We will see that in  the thin wall limit $\sigma_0\simeq \sigma_+$ with our modified ansatz, although in principle we do not have to restrict ourselves to particular potentials, we will not be able to investigate cases where the effective potential is  extremely flat, hence we will have to limit our investigation to situations where this is not the case. In the thick wall limit $\om\simeq \om_+$, we have to restrict our analysis to the case of polynomial potentials of the form:
\be{thckpot}
U(\sigma)=\half m^2 \sigma^2-A\sigma^n+\sum_{p>n}B_p \sigma^{p}
\ee
where $n\geq 3$, with the nonlinear couplings  $A>0$ and $B_p>0$ to ensure the existence of $Q$-ball solutions. We expect the thin wall approximation to be valid for general $Q$-ball potentials in which the $Q$-ball contains a lot of charge, with $\omega^2 \simeq \omega^2_- \ge 0$. In this limit we can define a positive infinitesimal parameter, $\eo \equiv -\Uo(\sigma_+)\simeq \half (\omega^2-\omega^2_-)\sigma^2_+ \geq 0$, and the effective mass around a "false" vacuum is given by, $\mu^2 \equiv \frac{d^2\Uo}{d\sigma^2}|_{\sigma_+}$. The other extreme case corresponds to the thick wall limit which is valid for $Q$-balls containing a small amount of charge, and it satisfies  $\omega^2 \simeq \omega^2_+=m^2$. For later convenience, in this limit, we define a positive infinitesimal parameter, $m^2_\omega=m^2-\omega^2 \geq 0$.

\subsection{Thin wall approximation for $D\ge 2$}\label{thwapprx}
\paragraph*{Step-like ansatz $\om \simeq \om_-$}
At a first step, we review the standard results in the thin-wall approximation originally obtained by Coleman \cite{Coleman:1985ki}. Adopting a step-like ansatz for the profile we write
\be{equation}
  \sigma(r)=\left\{
    \begin{array}{ll}
    \sigma_0 &\ \ \textrm{for $r<R_Q$} \\
    0 &\ \ \textrm{for $R_Q \leq r$}
    \end{array}
    \right.
\ee
where $R_Q$ and $\sigma_0$ will be defined in terms of the underlying parameters, by minimising the $Q$-ball energy. We can easily calculate $\So,\; Q,$ and $E_Q$:
\be{QEQEGYTHIN}
  \So=\bset{U_0-\half \omega^2 \sigma^2_0}V_D,\ \ Q=\omega \sigma^2_0 V_D, \ \ E_Q = \half \frac{Q^2}{\sigma^2_0
  V_D} + U_0 V_D
\ee
where $U_0 \equiv U(\sigma_0)$ and $V_D=V_D(r=R_Q)$. Note that \eq{QEQEGYTHIN} satisfies the Legendre transformation results \eq{easycalc} as we would have hoped. Since the ansatz \eq{equation} neglects the surface effects, we are working in the regime $\,\mU \gg \mS$ in \eq{virieq}. Therefore we should be able to reproduce the result, $E_Q \simeq \om Q$ with this solution. To see this, we note that the two terms in $E_Q$ are the contributions from the charge and potential energies.. These two contributions virialise since $E_Q$ is extremised with respect to $V_D$ for a fixed charge $Q$, \ie $\partial E_Q/ \partial V_D|_Q=0$, hence $V_D=Q\sqrt{1/(2\sigma^2_0 U_0)}$. This then fixes $R_Q$ because we know, for a $(D-1)$-sphere, $V_D=\frac{R_Q^D}{D} \Omega_{D-1}$ where $\Omega_{D-1} \equiv \int d\Omega_{D-1}=\frac{2\pi^{D/2}}{\Gamma(D/2)}$. Substituting $V_D=Q\sqrt{1/(2\sigma^2_0 U_0)}$ into $E_Q$ [the third equation of \eq{QEQEGYTHIN}] and minimising $E_Q$ with respect to $\sigma_0$, we obtain
\be{QEGYTHIN}
  E_Q=Q \cdot min \bset{\sqrt{\frac{2U_0}{\sigma^2_0}}}= Q\omega_-=\om^2_-\sigma^2_+V_D
\ee
where we have used \eq{EXIST} in which $\omega^2_-=min \left.\bset{\frac{2U_0}{\sigma^2_0}}\right|_{\sigma_0=\sigma_+}$. Thus we recover \eq{virieq} in the limit $\mU \gg \mS$. Finally, we remind the reader that we have obtained the minimised energy $E_Q$ with respect to $V_D (R_Q)$ and $\sigma_0$ in the extreme limit $\om =\om_-$ where we find
\be{sig0}
\sigma_0 = \sigma_+.
\ee
\eq{sig0} implies that the "particle" spends a lot of "time" around
$\sigma_+$ because the effective potential $-\Uo$ around $\sigma_+$ is "flat". Note that $Q$ and $E_Q$ are proportional to the volume $V_D$ in Eqs. (\ref{QEQEGYTHIN}) and (\ref{QEGYTHIN}) just as they are for ordinary matter, in this case Coleman  called it $Q$-matter \cite{Coleman:1985ki}.

\paragraph*{The modified ansatz $\sigma_0\simeq \sigma_+$}
Having seen the effect of an infinitely thin wall, it is natural to ask what happens if we allow for a more realistic case where the wall has a thickness associated with it?  Modifying the previous step-like ansatz to include this possibility \cite{PaccettiCorreia:2001uh, Coleman:1977py} will allow us to include surface effects \cite{Coleman:1985ki, Spector:1987ag, Shiromizu:1998rt} and is applicable for a wider range of $\om$ than in the step-like case. Using the results, we will examine the two different types of potentials, DVPs and NDVPs, which lead to the different cases of \eq{virieq}.

Following \cite{PaccettiCorreia:2001uh}, the modified ansatz is written as
\bea{thindanstz}
\sigma(r)= \begin{cases}
	\sigma_+ -s(r)\; & $for$\; r<R_Q,\\
	\bsig (r)\;  & $for$\; R_Q\leq r \leq R_Q + \delta,\\
    0\; & $for$\; R_Q + \delta < r,
 \end{cases}
\eea
where as before the core size $R_Q$, the core thickness $\delta$, the core profile $s(r)$, and the shell profile $\bsig (r)$ will be obtained in terms of the underlying parameters by extremising $\So$ in terms of a degree of freedom $R_Q$. Continuity of the solution demands that we smoothly continue the profile at $r=R_Q$, namely $\sigma_+-s (R_Q) =\bar{\sigma}(R_Q)$ and $-s^\p(R_Q)=\bar{\sigma}^\p(R_Q)$.

We expand $\Uo$ to leading order around $\sigma _+$, to give  $\Uo(\sigma)\sim -\eo +\half  \mu^2 s^2$ where $s(r)=\sigma_+ - \sigma(r)$. In terms of our mechanical analogy, the "particle" will stay around $\sigma_+$ for a long "time". Once it begins to roll off the top of the potential hill, the damping due to friction ($\propto (D-1)/r$) becomes negligible and the "particle" quickly reaches the origin. Therefore, we can naturally assume
\be{coreQ}
R_Q \gg \delta
\ee
where $\delta$ is the core thickness. We know that $\sigma^\p(0)=-s^\p(0)=0,\; s^\p(R_Q) \neq 0,$ and $s^\p(r)>0$. Using \eq{QBeq}, the core profile $s(r)$ for $r< R_Q$ satisfies the Laplace equation:
\be{eoms}
s^{\p\p}+\frac{D-1}{r}s^\p-\mu^2 s=0
\ee
whose solution is
\be{seq}
s(r)=r^{(1-\frac{D}{2})} \bset{C_1  I_{\frac{D}{2} -1}(\mu r) + C_2 K_{\frac{D}{2} -1}(\mu r)}
\ee
where $I$ and $K$ are, respectively, growing and decaying Bessel functions, $C_1$ and $C_2$ are constants. Since $s(0)$ is finite and $s^\p(r)>0$, it implies that $C_2:=0$. Since $I_\nu(z)\sim z^\nu/2\Gamma(\nu+1)$ for small $z=\mu r$ and $\nu\neq -1,-2,-3 \dots$, thus $s(0)$ is finite:
\be{s0}
s(0)\sim C_1 \frac{\mu^{D/2-1}}{2\Gamma(D/2)}=\sigma_+-\sigma_0
\ee
which gives a relation between $C_1$ and $\sigma_0$. Also the analytic solution is regular at $r=0$: $s^\p(0) \simeq 0$.  For large $r\sim R_Q$, \eq{seq} leads to
\be{sslope}
\frac{s^\p}{s} \simeq \mu - \frac{D-2}{r} \to \mu.
\ee
where we are assuming
\be{app2}
\mu \gg 1/R_Q
\ee
and have used the approximation $I_\nu(z) \sim  \frac{e^{z}}{\sqrt{2 \pi z}}$ for large $z\equiv \mu r$. As already mentioned we note that this result is not strictly valid for extremely flat potentials, \ie $\mu \simeq 1/R_Q$, because the expansion is only valid for $z \equiv \mu r \gg 1$. We will therefore only be applying it to the cases where the effective potential is not very flat.

Turning our attention to the shell regime $R_Q\le r \le R_Q + \delta$. Considering the "friction" term in \eq{QBeq}, we see that it becomes less important for large $r$ compared to the first and third terms in \eq{eoms}, because
\be{frict}
\abs{\frac{D-1}{R_Q}s^\p (R_Q)} \simeq \abs{\frac{D-1}{\mu R_Q} \mu^2 s(R_Q)} \ll \mu^2 s(R_Q) \simeq s^{\p\p}(R_Q) \simeq \abs{\frac{d\Uo}{d s}}_{r=R_Q}
\ee
where we have made use of \eqs{sslope}{app2}. Imposing continuity conditions, namely $\sigma_+-s(R_Q)=\bar{\sigma}(R_Q)$, $-s^\p(R_Q)=\bar{\sigma}^\p(R_Q)$, \eq{QBeq} without the "friction" term becomes
\be{barsig}
\frac{d^2\bar{\sigma}}{dr^2}- \left.\frac{d\Uo}{d\sigma}\right|_{\bar{\sigma}}=0,
\ee
where $\bar{\sigma}(r)$ is defined as being the solution to \eq{barsig}. With
the condition $\bar{\sigma}(R_Q)=\sigma_+-s(R_Q)$ and \eq{s0}, we find 
$\bar{\sigma}(R_Q)=\sigma_+-\sqrt{\frac{2}{\pi}}\frac{\Gamma(D/2)(\sigma_+-\sigma_0) \ e^{\mu R_Q}}{(\mu R_Q)^{(D-1)/2}}$. For the thin wall limit $\sigma_0 \simeq \sigma_+$, we obtain $\bar{\sigma}(R_Q) \sim \sigma_+ - \Delta$ where $\Delta$ is an infinitesimal parameter which satisfies $\sigma_+ \gg \Delta$. Therefore
\be{modthsig}
\bar{\sigma}(R_Q) \gg s(R_Q).
\ee
 Although \eq{nontbdry} does not hold exactly, the "total energy", $\half \bset{\frac{d\bsig}{dr}}^2-\Uo \sim 0$ with \eq{nontbdry}, is effectively conserved with the radial pressure $p_r$ vanishing outside the $Q$-ball core. This fact implies that the surface and effective potential energies virialise with equal contributions, $\mS_{shell} \simeq \mU_{shell}-\half \om Q_{shell}$ where we have introduced shell and core regimes defined by $X_{core}=\Omega_{D-1}\int^{R_Q}_0 dr r^{D-1} \cdots$ and $X_{shell}=\Omega_{D-1} \int^{R_Q+\delta}_{R_Q} dr r^{D-1} \cdots$ for some quantity $X$. Using $\sigma^\p<0$  and the condition $\bar{\sigma}(R_Q+\delta)=0$, the thickness of the $Q$-ball can be written as $\delta(\om)=\int^{\bar{\sigma}(R_Q)}_{0}\frac{d\sigma}{\sqrt{2\Uo}}$. Since $\delta$ is real and positive, we have to impose
\be{const}
\bar{\sigma}(R_Q) < \sigma_-,
\ee
recalling $\Uo(\sigma_-)=0$ for $\sigma_-\neq 0$.

With the use of \eq{easycalc}, we turn our attention to extremising the Euclidean action $\So$ in \eq{Uo} for the degree of freedom $R_Q$. Using the obtained value $R_Q$, we will differentiate $\So$ with respect to  $\om$ to obtain $Q$ as in \eq{legendre} which leads us to the $Q$-ball energy $E_Q$ as in \eq{legtrns} and the characteristic slope $E_Q/\om Q$. For convenience we split $\So$ into the core part $\So^{core}$ for $r<R_Q$ and the shell part $\So^{shell}$ for $R_Q \leq r \leq R_Q + \delta$ using \eq{thindanstz}. Using $V_D=\frac{R_Q^D}{D} \Omega_{D-1} \gg \partial V_D \equiv R_Q^{D-1}\Omega_{D-1} \gg \partial^2 V_D \equiv R_Q^{D-2}\Omega_{D-2}$ and \eqs{eoms}{sslope}, we find,
\be{coresw}
\So^{core}=-V_D \cdot \eo + \pa V_D \cdot \bset{\half \mu s^2(R_Q)}  - \pa^2 V_D \cdot \bset{ \frac{\Omega_{D-1}}{\Omega_{D-2}} \frac{(D-2)}{\mu} \half \mu s^2(R_Q)}
\ee
where the first term, $\eo$, in \eq{coresw} comes from the effective potential energy, while the second and third terms arise from the surface energy. Since $\eo$ is an infinitesimal parameter in the other thin wall limit $\om\simeq \om_-$, it gives
\be{ratcore}
\mU_{core}\simeq \half \om Q_{core}.
\ee
The effective potential energy balances the surface energy in the shell (see \eq{barsig}), therefore by introducing the definition $T\equiv \int^{\bsig(R_Q)}_0 d\sigma \sqrt{2\Uo}$, we see
\bea{shellsw}
\So^{shell} &=&\Omega_{D-1}\int^{\bsig(R_Q)}_0 d\sigma r^{D-1}\sqrt{2\Uo(\sigma)} \lsim \Omega_{D-1} (R_Q+\delta)^{D-1} T \\
\label{shllsw2} &\to& \pa V_D \cdot T + \pa^2 V_D \cdot \bset{\frac{\Omega_{D-1} }{\Omega_{D-2}}(D-1)\delta \cdot T} + \order{R^{D-1}_Q,\frac{R^2_Q}{\delta^2}} \cdot T,
\eea
where we have used the fact that the integrand has a peak at $r=R_Q+\delta$ in the second relation of \eq{shellsw} \cite{Hong:1987ur} and Taylor-expanded $ (R_Q+\delta)^{D-1}$ in going from \eq{shellsw} to \eq{shllsw2} because of our approximation \eq{coreQ}. Combining both expressions \eqs{coresw}{shllsw2}, we obtain
\bea{}
 \So &=& \So^{core} + \So^{shell} \\
\label{swall} &\simeq & - \eo \cdot V_D  + \tau \cdot \pa V_D  +h\cdot \partial^2 V_D
\eea
where $\tau \equiv T+ \half \mu s^2(R_Q)$. Note that whilst in $\tau$, $T$ contains the equally virialised surface and effective potential energies from the shell, the second term $\half \mu s^2(R_Q)$ contains a surface energy term from the core. Moreover we have defined
$h\equiv \frac{\Omega_{D-1}}{\Omega_{D-2}}\sbset{ (D-1) \delta \cdot T - \frac{(D-2)}{\mu} \half \mu s^2(R_Q)}$ which is negligible compared to $\tau$ because of the assumptions \eqs{coreQ}{app2}. Therefore, we will take into account only the first two terms in $\So$, \eq{swall}. It is also important to realise that
\be{tens}
\tau = \int^{\bsig(R_Q)}_0 d\sigma \sqrt{2\Uo} + \int^{\sigma_+}_{\bsig(R_Q)} d\sigma \sqrt{2 U_{\om_-}}\to \int^{\sigma_+}_0 d\sigma \sqrt{2U_{\om_-}}=const.
\ee
which is independent of $\om$ and $D$ in the limit of $\om\to \om_-$, where we have used the other thin wall limit $\om\simeq \om_-$. Our modified ansatz is not only valid in the extreme limit $\om=\om_-$ but also in the limit $\om\sim \om_-$ as long as $\tau$ depends on $\om$ "weakly". Note that the condition of \eq{const} also ensures that $\tau$ is positive and real. In addition, the second term in \eq{tens} is negligible compared to the first term, \ie
\be{ratshell}
\mS_{shell} \simeq \mU_{shell} - \half \om Q_{shell} \gg  \mS_{core}
\ee
because of $\sigma_+ \sim \bar{\sigma}(R_Q)$, see \eq{modthsig}.

We can make progress by using the Legendre transformation  of \eq{easycalc}, which implies that we need to find the extrema of $\So$ with fixed $\omega$, \ie $\frac{\partial \So}{\partial R_Q}=0$. This is equivalent to the virialsation between $\eo$ and $\tau$. Then one can compute the core radius,
\be{RQ}
	R_Q=(D-1)\frac{\tau}{\eo}.
\ee
Note that this implies that one-dimensional thin wall $Q$-balls do not exist due to the positivity of $R_Q$. By using \eqs{swall}{RQ} and \eq{easycalc}, we can compute the desired quantities to compare with the results we obtained using the step-like ansatz, in particular \eqs{QEQEGYTHIN}{QEGYTHIN}, and we can confirm that the classical stability condition \eq{CLS} is satisfied:
\bea{}
\label{soq1} \So&\simeq & \frac{\tau}{D}\; \partial V_D =
\frac{\eo}{D-1}\; V_D>0,\ \ Q(\omega)\simeq \omega \sigma^2_+ V_D \\
\label{linearthin} E_Q &\simeq& \omega^2 \sigma^2_+ V_D  + \frac{\tau}{D}\; \partial V_D\\
\label{surfthin} &\simeq& \omega Q \sbset{\frac{2D-1}{2(D-1)}-\frac{\om^2_-}{2(D-1)\om^2}} \\
\label{q1class}\frac{\omega}{Q}\frac{dQ}{d\omega}&\simeq&1- \frac{D\omega^2 \sigma^2_+}{\eo}  \simeq - \frac{D\omega^2 \sigma^2_+}{\eo} <0.
\eea
We can see the virialisation between $\tau$ and $\eo$ for the second and third terms in \eq{soq1}. As in \eq{QEGYTHIN}, the first term of $E_Q$, in \eq{linearthin}, is a combination of an energy from the charge and potential energy from the core throughout the volume, while the new second term $\frac{\tau}{D}$, called the surface tension,
represents the equally virialised surface and effective potential energies from the shell as in \eq{ratshell}. In the limit $\om\simeq \om_-$, $\eo$ becomes asymptotically zero which implies \eq{ratcore} as we saw. We have also seen $\mS_{shell}\gg \mS_{core}$. Using $\mU=\mU_{core}+ \mU_{shell}$, $\mS=\mS_{core}+\mS_{shell} \sim \mS_{shell}$, and \eqs{ratcore}{ratshell}, we obtain
\be
{UoverS}
\mU \sim \mS + \om_- Q
\ee
which we will use shortly. Since the characteristic function, $E_Q/Q$, increases monotonically as a function of $\om$ and $\So>0$, \ie $\frac{d}{d\om}\bset{\frac{E_Q}{Q}}>0$ or \eq{q1class}, the classical stability condition \eqs{CLS}{chslope} is satisfied without specifying any detailed potential forms. However, the physical properties of the finite thickness thin wall $Q$-balls do depend on the vacuum structures of the underlying potential. To demonstrate this we consider two cases of non-degenerate vacuum potentials (NDVPs) with $\om_- \neq 0$ and degenerate vacuum potentials (DVPs) with $\om_-=0$ (see red-solid lines in \fig{fig:twomdl}). Suppose that the thin wall $Q$-balls have identical features over a large range of $\om$, we can find the approximate threshold frequency $\om_a$ using \eqs{freeengy}{virieq} as we assumed when we obtained \eq{viriwa}.

\paragraph*{NDVPs}

This type of potential reproduces the results we obtained in \eq{QEGYTHIN} corresponding to the regime $\mU\gg \mS$ which corresponds to the existence of $Q$-matter in that the charge and energy are proportional to the volume $V_D$ due to the negligible surface tension in \eq{linearthin}. Hence, the modified ansatz \eq{thindanstz} can be simplified into the original step-like ansatz \eq{QEQEGYTHIN} with negligible surface effects in the extreme limit $\om=\om_-$. To see that, we need to recall the definition of $\om_-$ in \eq{LEFT}.
We can realise that $\mu$ is the same order as $\om_-$ except the case of $\om_-=0$. Using $\mu\sim \om_-$, we can show that $\half \om Q \gg \mS_{core}\sim \half \mu s^2(R_Q) \partial V_D$ where we have used \eqs{app2}{modthsig}. Using \eqs{ratcore}{ratshell} and $\half \om Q \gg \mS_{core}$ which we just showed, we can obtain the desired result $\mU\gg \mS$. Similarly \eq{surfthin} in the limit $\om \simeq \om_-$ simplifies to give $\frac{E_Q}{\om Q} \sim 1$ which is the result of \eq{virieq} with $\mU\gg \mS$. Using \eq{surfthin} and \eqs{freeengy}{virieq},
we can also find the critical value $\om_a$ for absolute stability
\be{ndvpwa}
\frac{\om_a}{m}=\frac{D-1}{2D-1}
\bset{1+\sqrt{1+\frac{(2D-1)}{(D-1)^2}\frac{\om^2_-}{m^2} }}.
\ee
Finally, thin wall $Q$-balls in NDVPs are classically stable without the need for the detailed potential forms, however the absolute stability condition for $\om \sim \om_-$ depends on the spatial dimensions $D$ and on the mass $m$.

\paragraph*{DVPs}
For the case of the presence of degenerate minima where $\om_-=0$, since  $\eo=\half \omega^2 \sigma^2$, we immediately see from \eq{linearthin} that
\be{eqqdeg}
  \frac{E_Q}{\om Q}\simeq \frac{2D-1}{2(D-1)}
\ee
which reproduces \eq{virieq} for the case of $\mS\sim \mU$. As in NDVPs, we know \eq{UoverS} in the limit $\om \simeq \om_-$, but the second term $\om_- Q$ becomes zero in the present potentials. It follows that $\mU_{core} \simeq 0$ and $\mU_{shell} \simeq \mS_{shell}\gg \mS_{core}$, hence $\mS \sim \mU$. In other words, most of the $Q$-ball energy is stored within the shell. In addition the charge $Q$ and energy $E_Q$ are not scaled by the volume, which implies that the modified ansatz does not recover the simple ansatz as opposed to NDVPs. In particular we find that $Q=\frac{\Omega_{D-1} \sqrt{2(D-1)\tau}}{D\sigma_+}\; R^{(D-1/2)}_Q \propto R^{(D-1/2)}_Q,\hspace*{10pt} E_Q= \frac{2D-1}{D}\Omega_{D-1} \tau\; R^{D-1}_Q  \propto R^{D-1}_Q$. A nice check of our general results follows by writing $E_Q$ in terms of the charge $Q$ by eliminating $R_Q$ between the two expressions. This gives $E_Q\propto Q^{2(D-1)/(2D-1)}$ which reproduces the three dimensional results obtained in \cite{PaccettiCorreia:2001uh}.

Finally, let us recap, the key approximations and conditions we have made in this modified ansatz. They are Eqs. (\ref{coreQ}, \ref{app2}, \ref{const}), and \eq{tens}. We believe that the estimates we have arrived at for the thin wall $Q$-balls are valid as long as: the core size is much larger than the shell thickness; the effective potential is not too flat around $\sigma_+$; the core thickness $\delta$ and surface tension $\tau/D$ are positive and real; $\tau$ is insensitive to both $\om$ and $D$.  With the extreme limit $\om\to \om_-$, the $Q$-balls in DVPs recover the simple step-like ansatz, while the ones in NDVPs do not. One-dimensional $Q$-balls do not support thin wall approximation due to the absence of the friction term in \eq{QBeq}.
\subsection{Thick wall approximation for $\omega^2 \simeq \omega^2_+$, \ie $m_\omega\to 0$}
\paragraph*{Gaussian ansatz}

As we have started with the simple step-like ansatz in the thin wall approximation, a Gaussian function is a simple approximate profile to describe the thick wall $Q$-balls \cite{Gleiser:2005iq}. Using a Gaussian ansatz
\be{gaussansatz}
\sigma(r)=\sigma_0\exp\bset{-\frac{r^2}{R^2}},
\ee
we will extremise $\So$ with respect to $\sigma_0$ and $R$ with fixed $\om$, instead of minimising $E_Q$ with fixed $Q$. Notice that the slope $-\sigma^\p/\sigma$ becomes $2r/R^2$ which is linearly proportional to $r$ and the solution is regular at $r=0$: $\sigma^\p(0)=0$. Neglecting higher order terms $B_p$ in \eq{thckpot} with \eq{gaussansatz}, one can obtain straightforwardly
\bea{THCKQ}
    Q&=&\bset{\frac{\pi}{2}}^{D/2} \omega \sigma^2_0 R^D,\hspace*{5pt} \So\simeq \bset{\half \mo^2+\frac{D}{R^2}-A\sigma^{n-2}_0\bset{\frac{2}{n}}^{D/2}}\frac{Q}{\omega},\\
\label{eqthick1}    E_Q&\simeq&\sbset{\half \bset{m^2 + \omega^2} +
    \frac{D}{R^2}- A\sigma^{n-2}_0 \bset{\frac{2}{n}}^{D/2}} \frac{Q}{\omega}.
\eea
\eq{easycalc} can be easily checked in \eqs{THCKQ}{eqthick1}. The first and last terms in \eq{eqthick1} are the potential energy terms. The second energy term
comes from the charge energy, and the surface energy term appears in the third
term. By finding the extrema of $\So$ with respect to $\sigma_0$ where $\dd{\So}{\sigma_0}=0$, it defines the underlying parameter $\sigma_0$ as
\be{apprxsigma0}
\sigma_0=\sbset{\bset{m^2_\omega+\frac{2D}{R^2}}\frac{1}{nA}\bset{\frac{n}{2}}^{D/2}}^{1/n-2}
\tol \bset{\frac{\mo^2}{2A}}^{1/n-2}\sim \sigma_-
\ee
where we have neglected the surface term and used the approximation $D/2\simeq \order{1}$ in the second relation of \eq{apprxsigma0}. Then we can check that the Gaussian ansatz naturally satisfies the other thick wall limit $\sigma_0\simeq \sigma_-$ and that the higher order terms $B_p$ in \eq{thckpot} are negligible. Using the first relation of \eq{apprxsigma0}, one needs to extremise $\So$ with respect to  another degree of freedom $R$ with $\dd{\So}{R}=0$ which determines $R$:
\be{gausscore}
R=\sqrt{\frac{2(2-D)}{\mo^2}}\ge 0.
\ee
The reality condition on $R$ implies that the Gaussian ansatz is valid only for $D=1$. The width of the gaussian function $R$ in \eq{gausscore} becomes very large in the thick wall limit $\mo \to 0$, thus we can justify that the surface term in \eqs{eqthick1}{apprxsigma0} are negligible. Therefore we are looking at the regime $\mU \gg \mS$ which should lead us to $E_Q\simeq \om Q$ as in the first case of \eq{virieq}. To do this for $D=1$ we substitute \eq{apprxsigma0} into $Q,\; E_Q,\; \So$:
\bea{}
Q&=&\sqrt{\frac{\pi}{2}}\omega \sigma^2_0 R,\hspace{10pt} \So=\bset{\half- \frac{1}{n}}\frac{2\mo^2 Q}{\omega}>0,\\
\label{gausseq}\frac{E_Q}{\om Q}&=&\bset{\half+\frac{1}{n}} +\bset{\half-\frac{1}{n}}\bset{\frac{2m^2}{\om^2}-1}\tol 1
\eea
where we have considered the thick wall limit $\omega\simeq m$ in the second relation of \eq{gausseq}. We can check \eq{virieq} and the analytic continuation \eq{freeengy}. In the same limit, the Euclidian action becomes an infinitesimally small positive value: $\So \to 0^+$.

Using the second relation $\sigma_0$ in \eq{apprxsigma0} and \eq{gausscore}, one can find
\be{gausscls}
\frac{\om}{Q}\frac{dQ}{d\om} \simeq 1-\frac{\omega^2}{\mo^2}\bset{\frac{4}{n-2}-1}
\to  -\frac{\omega^2}{\mo^2}\bset{\frac{4}{n-2}-1}\le 0
\ee
where we have used the fact that $\mo$ is a positive infinitesimal parameter in the limit, $\om\simeq \om_+$ going from the first relation to the second one. \eq{gausscls} shows that the classical stability condition clearly depends on the non-linear power $n$ in the potential \eq{thckpot}: $n \le 6$. This is contradictory because \eq{gausseq} gives $\frac{d}{d\om}\bset{\frac{E_Q}{Q}}\to -1+\frac{4}{n}$ which implies $n \le 4$ for the other classical stability condition using \eq{chslope}. We will shortly see that this contradiction between \eq{CLS} and \eq{chslope} is an artefact of the Gaussian ansatz. Moreover, our conclusion should state that the Gaussian approximation is valid only for $D=1$. These awkward consequences are improved with the following physically motivated ansatz.
\paragraph*{The modified ansatz}
Having considered the case of the simple Gaussian ansatz following the spirit of \cite{Gleiser:2005iq}, we found some problems for the classical stability. To fix these, we need a more realistic ansatz \cite{Kusenko:1997ad, PaccettiCorreia:2001uh, Shatah:1985vp, Blanchard:87, Multamaki:1999an}. To do this we drop an explicit detailed profile to describe thick wall $Q$-balls and rescale the field profile so as to work in dimensionless units whilst extracting out the explicit dependence on $\omega$ from the integral in $\So$. As in the thin wall approximation with the modified ansatz, we will again make use of the technique \eq{easycalc} to obtain other physical quantities from $\So$.

We begin by defining  $\sigma=a\wt{\sigma}$ and $r=b\wt{r}$ with $a$ and $b$ which will depend on $\om$. Substituting them into \eq{Uo} with the potential \eq{thckpot} we obtain:
\bea{so1}
\So &=& \int d\Omega_{D-1} \int d\tilde{r} \tilde{r}^{D-1}b^D
\set{\half \bset{\frac{a}{b}}^2 \wt{\sigma}^{\p 2}+\half a^2 \mo^2 \tilde{\sigma}^2-A a^n \tilde{\sigma}^n +
	\sum_{p>n}B_p a^{p}\wt{\sigma}^{p}}, \\
\label{so2}&=& b^D\bset{\frac{a}{b}}^2 \Omega_{D-1} \int d\tilde{r} \tilde{r}^{D-1} \half \set{\wt{\sigma}^{\p 2}+\tilde{\sigma}^2-\tilde{\sigma}^n + \sum_{p>n}B_p b^2 a^{p-2}\wt{\sigma}^{p}},\\
\label{so3}&\simeq&  \mo^{4/(n-2) -D+2} A^{-2/(n-2)}\Omega_{D-1} S_n
\eea
with the rescaled action $S_n=\int d\tilde{r} \tilde{r}^{D-1} \bset{\half \wt{\sigma}^{\p 2}+\wt{U}}$ with $\wt{U}=\half \wt{\sigma}^2-\half \wt{\sigma}^n$, and we have neglected the higher order terms involving $B_p$.
In going from \eq{so1} to \eq{so2} we have set the coefficients of the first three terms in the brackets to be unity in order to explicitly remove the $\om$ dependence from the integral in $\So$. In other words we have set
$\half \bset{\frac{a}{b}}^2=\half a^2 \mo^2=Aa^n$. This implies, $a=\bset{\frac{\mo^2}{2A}}^{1/(n-2)}=\sigma_-$ and $b=\mo^{-1}$. Crucially $S_n$ is independent of $\omega$, and is positive definite\cite{Kusenko:1997ad, Multamaki:1999an, PaccettiCorreia:2001uh}. Adopting the powerful approach developed in \eq{easycalc}, given $S_\om$ we can differentiate it to obtain $Q$ and then use the Legendre transformation to obtain $E_Q$. This is straightforward and yields
\bea{qthck}
Q(\omega)&=& \om m^{4/(n-2)-D}_\omega \bset{\frac{4}{n-2}-D+2}A^{-2/(n-2)}\Omega_{D-1} S_n \propto \mo^{4/(n-2)-D},\\
\label{eqthick2} E_Q      &=& m^{4/(n-2)-D}_\omega \sbset{\mo^2+  \omega^2\bset{\frac{4}{n-2}-D+2}}A^{-2/n-2}\Omega_{D-1} S_n,\\
\label{modslope}          &=& \om Q \sbset{1+\frac{\mo^2}{\om^2}\bset{\frac{4}{n-2}-D+2}^{-1}  }\to \om Q.
\eea
The first term in \eq{modslope} is the energy contributed by the charge, while the second term is dominated by the effective potential energy, hence $\mU\gg \mS$. Therefore, we can also recover the result $E_Q\simeq \om Q$ in the thick wall limit $\om\simeq \om_+$ as we would expect from \eq{virieq} when $\mU\gg \mS$. Since $Q$ and $E_Q$ should be positive definite, it places the constraint \cite{Multamaki:1999an}
\be{validthck}
D < \frac{4}{n-2}+2.
\ee
With the condition \eq{validthck}, it is easy to see that $\So \to 0^+$ in the thick wall limit, $\om\simeq \om_+$ where $\mo^2 \to 0^+$. There is another constraint emerging from the need for the solution to be classically stable:
\bea{moddiffchg}
\frac{\om}{Q}\frac{dQ}{d\om} \simeq 1-\frac{\omega^2}{\mo^2}\bset{\frac{4}{n-2}-D}
&\to&  -\frac{\omega^2}{\mo^2}\bset{\frac{4}{n-2}-D}\le 0, \\
\label{modcls} &\lr& D \le \frac{4}{n-2}
\eea
which coincides with \eq{gausscls} in the case of $D=1$. Notice that the modified ansatz is valid not only for $D=1$ but also $D<\frac{4}{n-2}+2$ in \eq{validthck}. For $D=3$ this result matches that of \cite{PaccettiCorreia:2001uh}. The classical stability condition, \eq{modcls}, is consistent with the need for $Q$ and $E_Q$ to be finite. \eq{modcls} is more restrictive than that given in \eq{validthck}. Furthermore, we should check the relation \eq{chslope} for the characteristic function $E_Q/Q$ in terms of $\om$. It follows that $\frac{d}{d\om}\bset{\frac{E_Q}{Q}} \simeq 1- 2\bset{\frac{4}{n-2}-D+2 }^{-1} \ge 0$, which requires the same condition as \eq{modcls}. With this fact and \eq{modslope}, it implies that thick wall $Q$-balls satisfy the conditions for both classical and absolute stability. Moreover it also reproduces the previous results  in \cite{Kim:1992mm}, for the case of $D=2$ and $n=4,\; p=6$ (6-th order potential). Unlike the Gaussian ansatz \eq{gaussansatz}, our modified ansatz now shows consistent results between \eq{CLS} and \eq{chslope}.

Let us remark on the validity of our analysis following \cite{Multamaki:1999an}. In this section we have used a modified ansatz which has involved a re-scaling of $\sigma$ and $r$ in such a way as to leave us with a dimensionless action $S_n$. There are restrictions on our ability to do this as first pointed out in  \cite{Multamaki:1999an} for the case of $D=3$. We can generalise this to our $D$ dimensional case. Given that the $Q$-ball solutions extremise $S_n$, we may rescale $r$ or $\sigma$ introducing a one-parameter rescaling,  $r \to \al r$ or $\sigma \to \lambda \sigma$ which will deform the original solution. Defining $X(\al)\equiv S_n[\al r,\sigma(\al r)]$ and $Y(\lambda)\equiv S_n[\lambda \sigma(r)]$, we impose the condition that the action $S_n$ is extremised when $\al=\lambda=1$, which implies $\frac{dX}{d\al} |_{\al=1}=0=\frac{dY}{d\lambda}|_{\lambda=1}$. It is possible to show that these conditions imply that consistent solutions require the same condition as \eq{validthck}. The three dimensional case leads to the result, $n<6$, as originally obtained in \cite{PaccettiCorreia:2001uh}. The particular choice  of $n=4$ which we will investigate shortly implies $D < 4$ for the validity of our thick wall approximation with the modified ansatz. Moreover, thick wall $Q$-balls become classically unstable for $D \geq 3$ as can be seen from \eq{modcls}.

What have we learnt from extending the ansatz beyond the Gaussian one? We have seen that they have lead to different results. For instance, the Gaussian ansatz essentially has a contradiction for the classical stability analysis even for $D=1$, whereas the solutions based on the modified ansatz are valid for $D$, which satisfies \eq{validthck}, and give consistent results, \eq{modcls}, for classical stability.

As we will see in the next section, our numerical results in which we obtain the full $Q$-ball solution support the modified ans\"{a}tze for both thin and thick wall cases.

\section{Numerical results}
\label{numerical}
In this section we obtain numerical solutions for $Q$-balls using the polynomial potential in \eq{thckpot} including only one higher order term:
\be{usig}
U(\sigma)=\half m^2 \sigma^2-A \sigma^n+ B \sigma^p,
\ee
where $A>0,\,B>0, p>n>2$. We shall confirm the results obtained analytically using the modified ans\"{a}tze for both the thin and thick wall $Q$-balls. Recall that $U_\om(\sigma)=U(\sigma) - \frac12 \om^2\sigma^2$, with $U_\om(\sigma_-)=0$ and $\sigma_+$ marks the maximum of the effective potential $-U_\om$ where $\sigma_+\neq 0$. For a particular case  $p=2(n-1)$, we find
\be{sig-}
\sigma_-(\om)=\bset{\frac{A-\sqrt{A^2-2 B m^2_\omega}}{2B}}^{1/(n-2)},\hspace{5pt}
\sigma_+(\om)=\bset{\frac{An+\sqrt{(An)^2-4Bp m^2_\omega}}{2Bp}}^{1/(n-2)}.
\ee
Also, for convenience, we set
\be{set}
\omega_+=m=1,\hspace{5pt} \omega_-=\sqrt{1-\frac{A^2}{2B}}\ge 0 \lr A \le \sqrt{2B},
\ee
where we recall the definitions of $\om_+$ and $\om_-$ are that $\om_+^2 \equiv \frac{d^2U}{d\sigma^2}|_{\sigma=0} = m^2$ and $U_{\om_-} (\sigma_+) \equiv 0$.
Setting  $\omega_-=0$ in  \eq{set} implies that $U(\sigma)$ in \eq{usig} has degenerate vacua at $\sigma=0,\; \pm \sigma_+$, whilst the original potential $U$ does not have degenerate vacua with $\om_-\neq 0$. In this section, we shall consider two examples of the potential $U$,  which can be seen as the red (solid) lines in \fig{fig:twomdl}. The degenerate vacua potential (DVP) on the left has $\omega_-=0$ ($A=\sqrt{2B}$) and the non-degenerate vacua potential (NDVP) on the right has $\omega_-=0.5$ ($A=\sqrt{3B/2}$).  In order to determine actual values for $A$ and $B$, we define $\sigma_+(\om_+) =1$ and set $n=4,\; p=6$ for both cases, hence $A=\frac43,\,B=\frac89$ in DVP and $A=1,\,B=\frac23$ in NDVP. \fig{fig:twomdl} also includes plots of the effective potentials for various values of $\om$.
\subsection{Numerical techniques and parameters}

To obtain the $Q$-ball profile we need to know the initial "position" $\sigma_0=\sigma(r=0)$. This is done using a shooting method, whereby we initially guess at a value of $\sigma_0$, then solve \eq{QBeq} for the $Q$-ball profile, and depending on whether we overshoot or undershoot the required final value of $\sigma$, we modify our guess for $\sigma_0$ and try again. Throughout our simulations, we need to specify the following three small parameters, $\epsilon,\; \xi,\; \eta$ which, respectively, determine our simulation size, $r_{max}$, the radius at which we can match the analytic and numerical solutions ($R_{ana}$), and the core size $R_Q$. The smoothly continued profile is computed up to $r=R_{max}$.

\paragraph*{Shooting method}

Let us consider an effective potential $-\Uo$ which satisfies the $Q$-ball \textit{existence condition}, \eq{EXIST}. We have to initially guess $\sigma_0$ subject to it be being in the appropriate region $\sigma_- \le \sigma_0 <\sigma_+$. For example it might be $\sigma^0_G=\frac{\sigma_+ + \sigma_-}{2}$. There are then three possibilities, the particle could overshoot, undershoot, or shoot properly. The last case is unlikely unless we are really "lucky". If it overshoots then we would find $\sigma(r_O)<0$ at some  "time" $r_O$. If that were to happen we could update $\sigma^0_G$ to  $\sigma^1_G = \frac{\sigma^0_G + \sigma_-}{2}$ as our next  guess. On the other hand if it undershoots, the "velocity" of the "particle" might be positive at some "time" $r_U$, $\sigma^\p(r_U)>0$. If that were to happen we might update  $\sigma^0_G$ to  $\sigma^1_G=\frac{\sigma_+ + \sigma^0_G}{2}$ as our next guess. After repeating the same procedures say $N$ times, we obtain the finely-tuned initial "position" $\sigma_0\simeq \sigma^N_G$ as our true value. To be compatible with numerical errors, our numerical simulation should be stopped with an appropriate accuracy parameterised by $\epsilon$:
\be{eps}
 \epsilon > \sigma(r_U=r_{max}) > 0
\ee
where $r_{max}$ is the size of our simulations, and $\epsilon$ measures the numerical accuracy where a small value of $\epsilon$ corresponds to good numerical accuracy. Unfortunately the final profiles still have small numerical errors for large $r$. To compensate for these errors, the profiles should continue to the analytical ones smoothly at some point $r=R_{ana}$.

\paragraph*{Matching analytic and numerical solutions at $R_{ana}$}
For large $r$, the $Q$-ball \eq{QBeq} can be reduced to
\be{LQBeq}
\sigma^{\p\p}+\frac{D-1}{r}\sigma^\p-m^2_\omega\ \sigma=0.
\ee
The analytic solution becomes
\be{slope}
\sigma(r)\sim E \sqrt{\frac{\pi}{2m_\omega}}\ r^{-\frac{D-1}{2}}e^{-m_\omega r} \hspace{10pt} \lr \hspace{10pt} -\frac{\sigma^\p}{\sigma}\sim\frac{D-1}{2r}+m_\om
\ee
where $E$ is a constant which is determined later. Note that we have used the fact that the modified Bessel function of the second kind has the relation $K_\mu(r)\simeq \sqrt{\frac{\pi}{2r}}e^{-r}$ for large $r$ and any real number $\mu$. In order to smoothly continue to the analytic profile \eq{slope} at the continuing point $R_{ana}$, the following condition is required using the second relation of \eq{slope}:
\be{xi}
\abs{\frac{D-1}{2r}+m_\om+\frac{\sigma^\p_{num}}{\sigma_{num}}}<\xi
\ee
where a parameter $\xi$ should be relatively small. Hence we can find the appropriate profile in the whole space
\be{mixedprof}
    \sigma(r)=
    \left\{
    \begin{array}{ll}
    \sigma_{num}(r)&\ \ \textrm{for $r<R_{ana}$}, \\
    \sigma_{num}(R_{ana})\bset{\frac{R_{ana}}{r}}^{(D-1)/2}e^{-m_{\omega}(r-R_{ana})} &\ \ \textrm{for $R_{ana} \le r \le R_{max}$},
    \end{array}
    \right.
\ee
where we have computed $E$ using \eq{slope} and our simulations are carried out up to $r=R_{max}$.
\paragraph*{Core size and thickness of thin-wall $Q$-ball}
Using \eq{sslope}, we can define the core size $r=R_Q$ and the numerical thickness $\delta_{num}$ by the slope $-\sigma^\p/\sigma$ with the following condition
\bea{core}
\abs{\bset{\frac{D-2}{r}-\mu}\bset{\frac{\sigma_+-\sigma}{\sigma}} +\frac{\sigma^\p_{num}}{\sigma_{num}}}&<&\eta,\\
\delta_{num} \equiv R_{ana}-R_Q. & &
\eea
Notice that the definition of $\delta_{num}$ is different from the definition in \eq{coreQ} where $\delta(\om)=\int^{\bar{\sigma}(R_Q)}_0 \frac{d\sigma}{\sqrt{2\Uo}}$.
\paragraph*{Numerical parameters}

We have run our code in two different regimes of $\om$ for both DVP and NDVP because the profiles for large $\om$ are needed to look into larger simulation size $r_{max}$ compared to the ones for small $\om$. Due to numerical complications, we do not conduct our simulations near the extreme thin wall limit, \ie $\om \simeq \om_-$. However, by solving close to the thin wall limit, our numerical results for $\sigma_0\simeq \sigma_+$ and $R_Q \gg \delta_{num}$ allow us to recover the expected properties of thin wall $Q$-balls with the modified ansatz \eq{thindanstz}. Finally, our results presented here correspond to the particular sets of parameters summarised in \tbl{tbl:nondeg}.
\begin{figure}[!ht]
  \def\@captype{table}
    \begin{center}
      \begin{tabular}{|c||c|c|c|c|c|}
	\hline
	\multicolumn{6}{|c|}{DVP}\\
	\hline
	$\omega$ & $\epsilon$ & $r_{max}$ & $R_{max}$ & $\xi$ & $\eta$ \\
    	\hline
0.38-0.73 & 4.0 $\times 10^{-2}$ & 30 & 200 & 8.0 $\times 10^{-3}$ & 1.0 $\times 10^{-1}$ \\
0.73-0.99999 & 1.0$\times 10^{-5}$ & 40 & 200 & 8.0 $\times 10^{-3}$ & 1.0 $\times 10^{-1}$ \\
\hline
      \end{tabular}
    \end{center}
  %
  \hfill
    \begin{center}
      \begin{tabular}{|c||c|c|c|c|c|}
	\hline
	\multicolumn{6}{|c|}{NDVP}\\
	\hline
	$\omega$  & $\epsilon$ & $r_{max}$ & $R_{max}$ & $\xi$ & $\eta$ \\
    	\hline
0.60-0.85 & 3.0 $\times 10^{-3}$ & 30 & 200 & 8.0 $\times 10^{-3}$ & 1.0 $\times 10^{-1}$ \\
0.85-0.99999 & 1.0$\times 10^{-5}$ & 50 & 200 & 8.0 $\times 10^{-3}$ & 1.0 $\times 10^{-1}$ \\
\hline
      \end{tabular}
	\end{center}    \tblcaption{The numerical parameters in DVP (top) and in NDVP (bottom).}
    \label{tbl:nondeg}
\end{figure}
\subsection{Stationary properties in DVP and NDVP}

We devote a large part of this section to justifying the previously obtained analytical results in the thin and thick wall approximations by obtaining the appropriate numerical solutions.

\paragraph*{Profiles with our numerical algorithm}

\begin{figure}[!ht]
  \begin{center}
   \subfigure{\label{fig:deggrad}
	\includegraphics[angle=-90, scale=0.32]{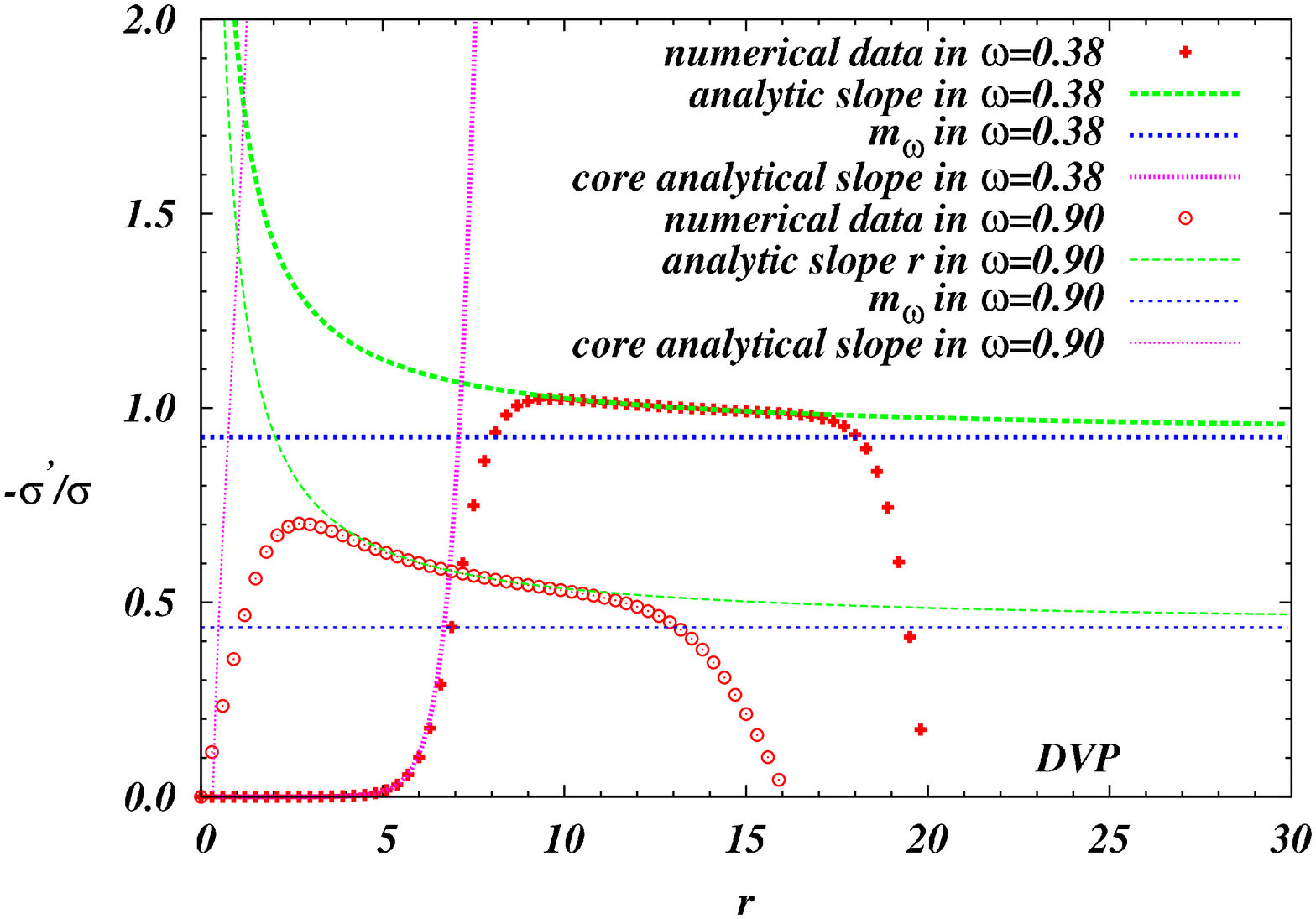}}
   \subfigure{\label{fig:nondeggrad}
	\includegraphics[angle=-90, scale=0.32]{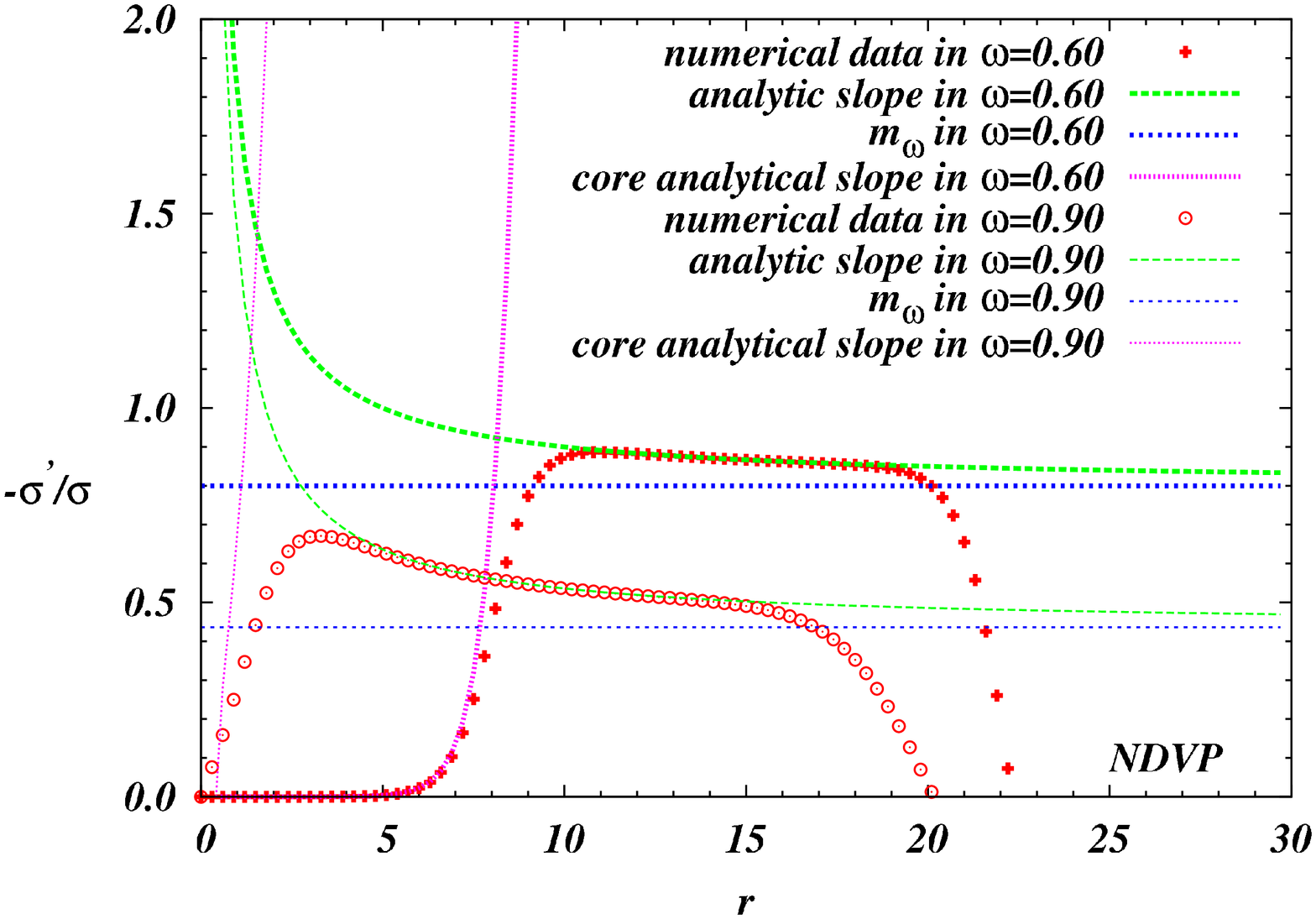}}\\
  \subfigure{\label{fig:degprof}  \includegraphics[angle=-90,scale=0.32]{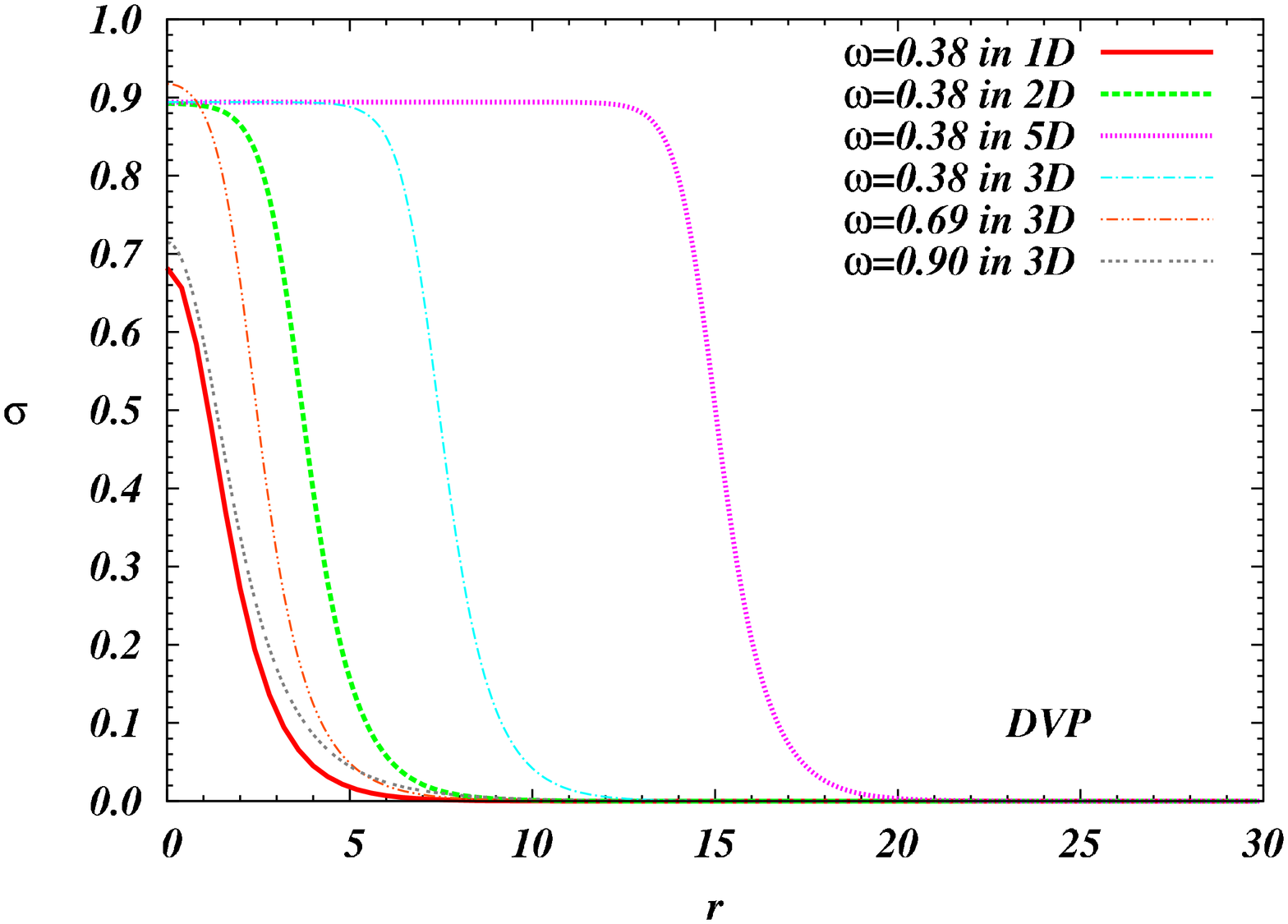}}
  \subfigure{\label{fig:nondegprof}\includegraphics[angle=-90,scale=0.32]{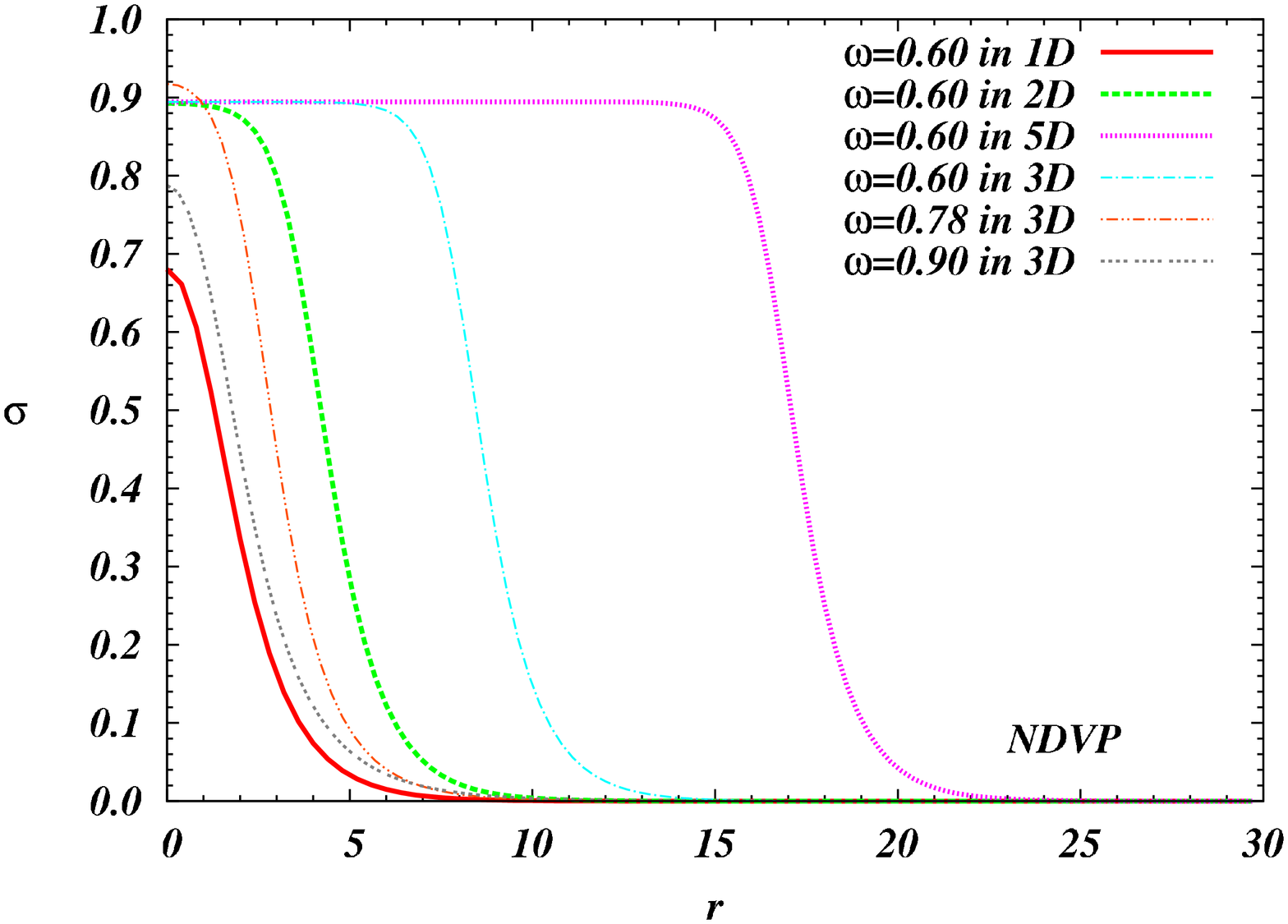}}
  \end{center}
  \figcaption{The top two panels show the numerical slopes $-\sigma^\p/\sigma$ for the case of $D=3$ for two values of $\om$ for both DVP (left) and NDVP (right). The red (one-dotted and one with circles) lines show the numerical slopes and the green dotted lines with two different widths the corresponding analytic solutions. The purple dot-dashed lines with two different widths show the analytic fits for the core profiles. The bottom two panels show the full $Q$-ball profile as described in \eq{mixedprof} for a number of values of $\om$ and $D$. Note how the core size increases with $D$.
}
  \label{fig:degexp}
\end{figure}

In the top two panels of \fig{fig:degexp} the two red lines (one dotted and one with circles) show the numerical slopes $-\sigma^\p/\sigma$ for the case of $D=3$ for two values of $\om$. These are then matched to the analytic profiles (green dotted lines) in order to achieve the full profile as given in  \eq{mixedprof}. Recall that we expect in general for all values of $\om$, the analytic fits to be accurate for large $r$, the numerical fits to be most accurate for small $r$ and there to be an overlap region where they are both consistent with each other as seen in \fig{fig:degexp}.  We have also plotted in dot-dashed purple lines our analytic fits,  \eq{core}, for the slopes of the thin wall cores from $r=0.5$. We should remind the reader that this fit only really works for the case of small $\om$ because we are dealing with thin wall $Q$-balls.  Notice, it is clear from the purple lines that the core sizes cannot be determined by this technique for the case $\om=0.9\simeq \om_+$.

\begin{figure}[!ht]
  \begin{center}
   \subfigure{\label{fig:degini}\includegraphics[angle=-90, scale=0.32]{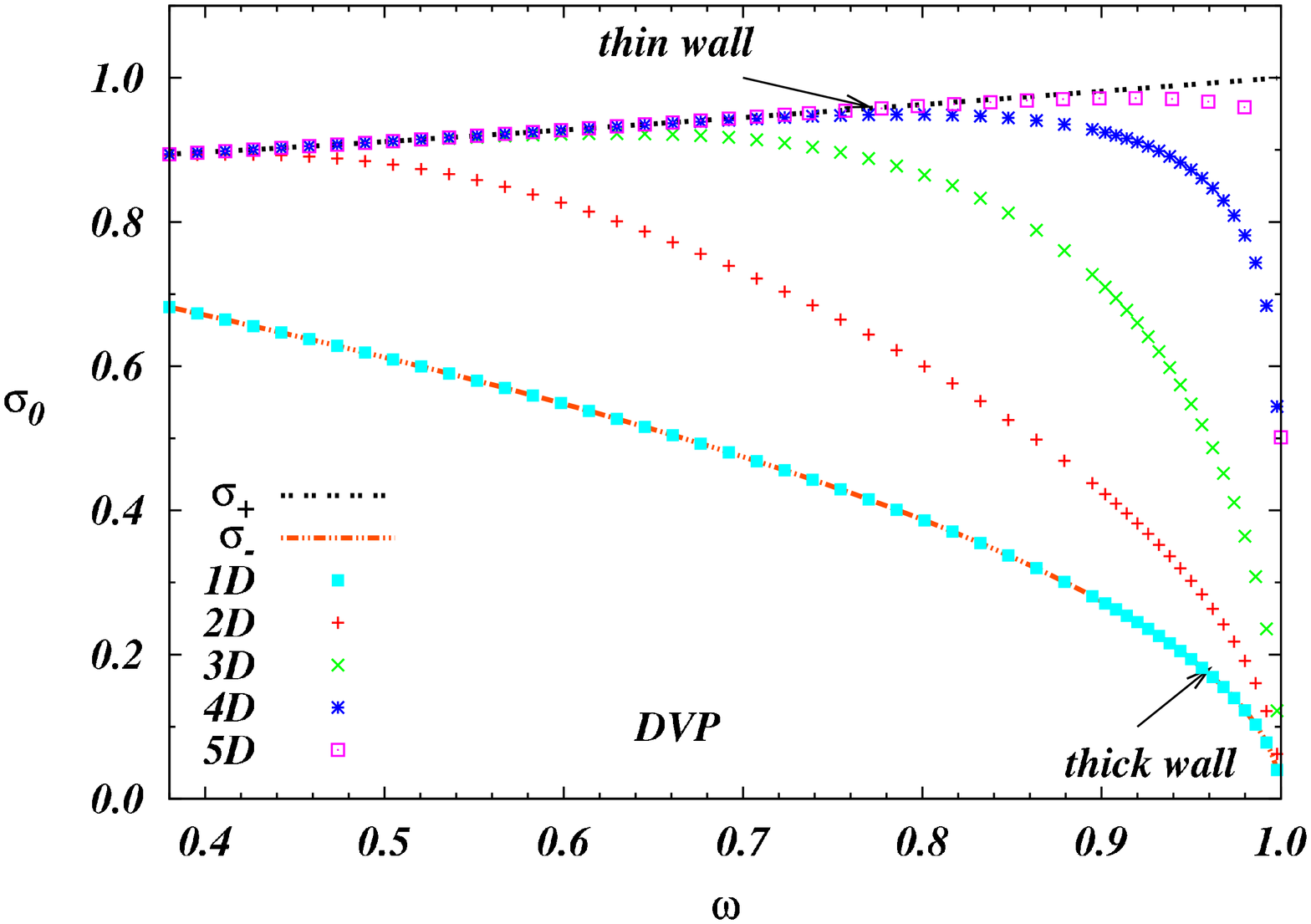}}
   \subfigure{\label{fig:nondegini}\includegraphics[angle=-90, scale=0.32]{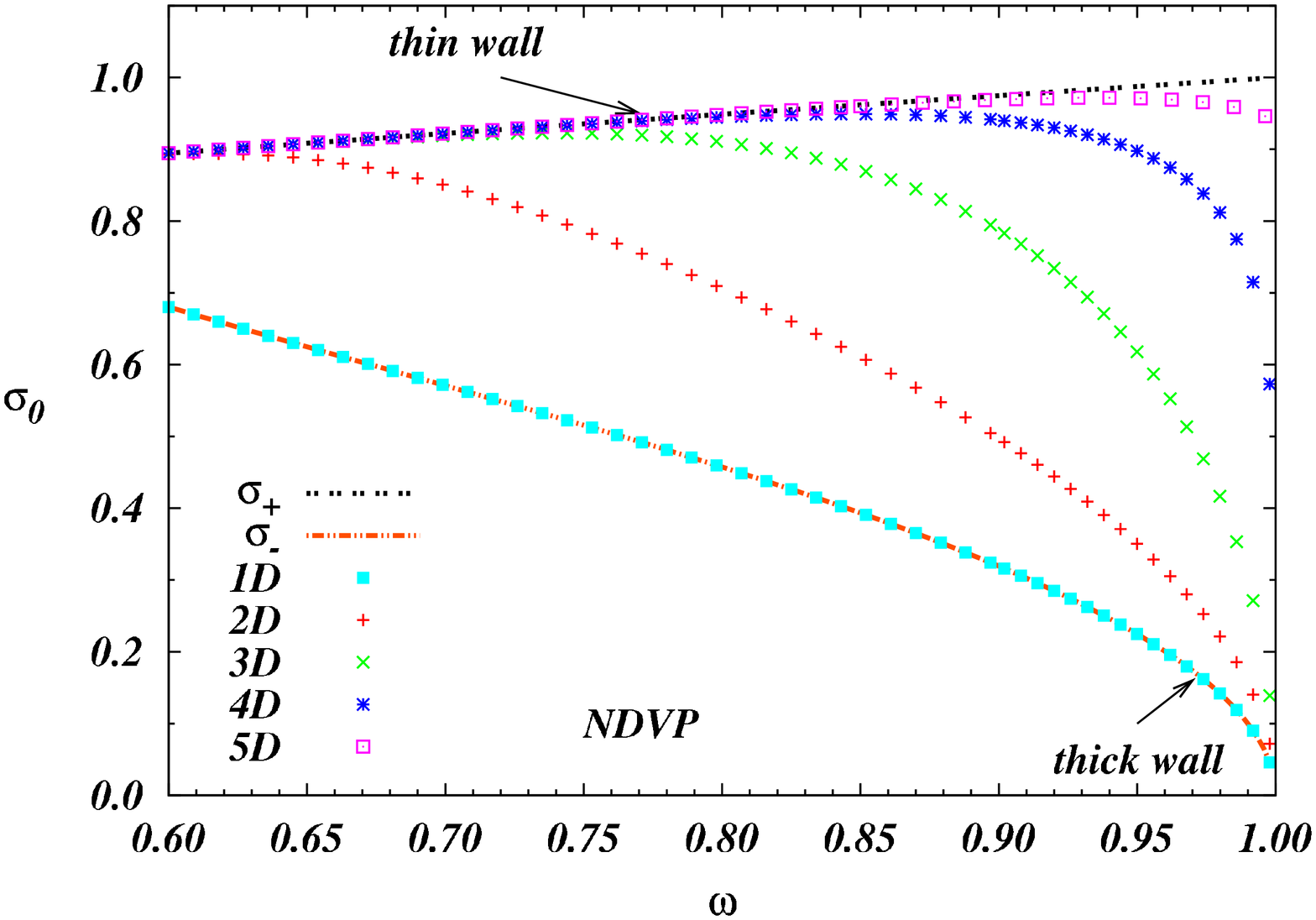}}\\
   \subfigure{\label{fig:dgthck}\includegraphics[angle=-90, scale=0.32]{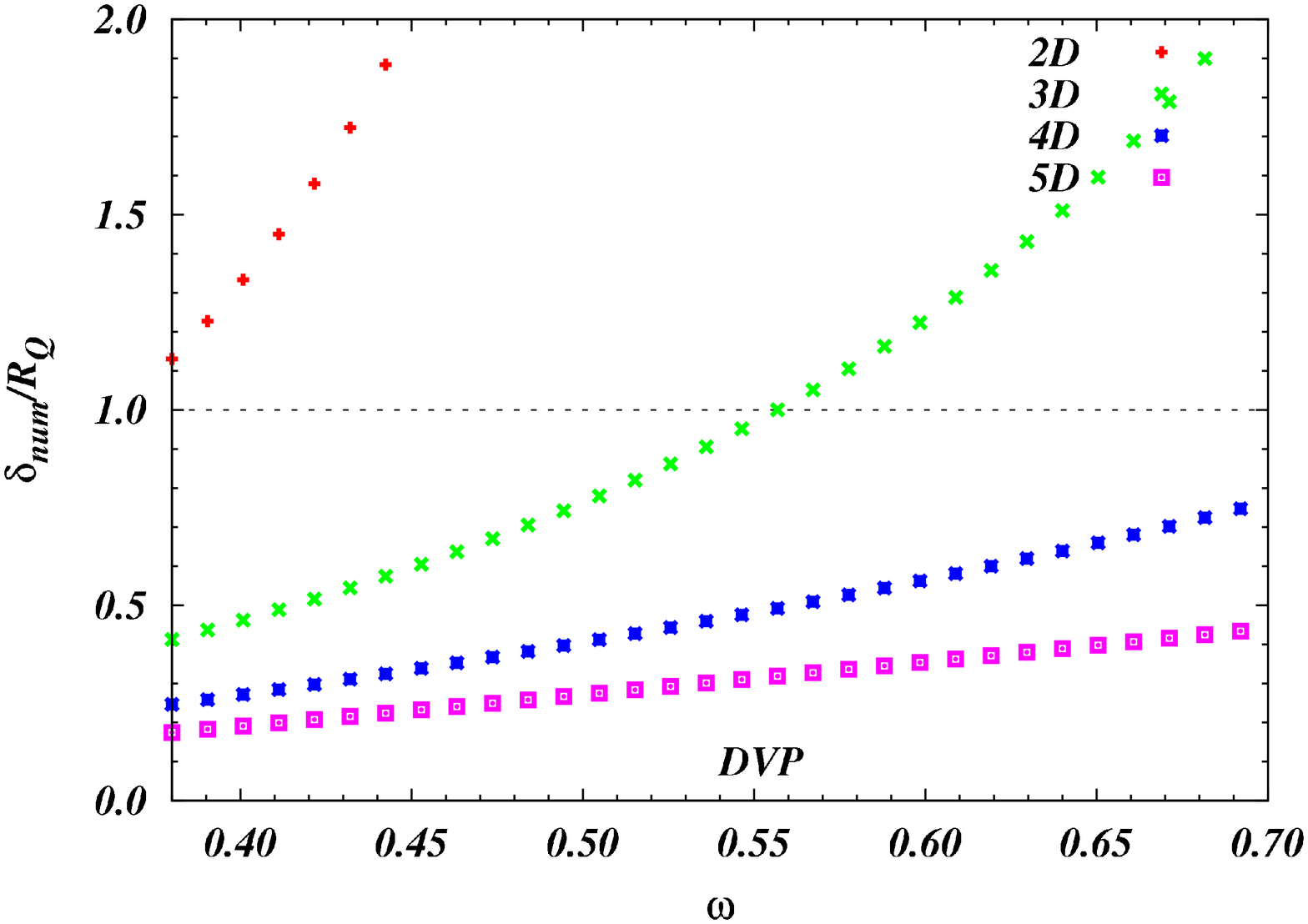}}
   \subfigure{\label{fig:ndgthck}\includegraphics[angle=-90, scale=0.32]{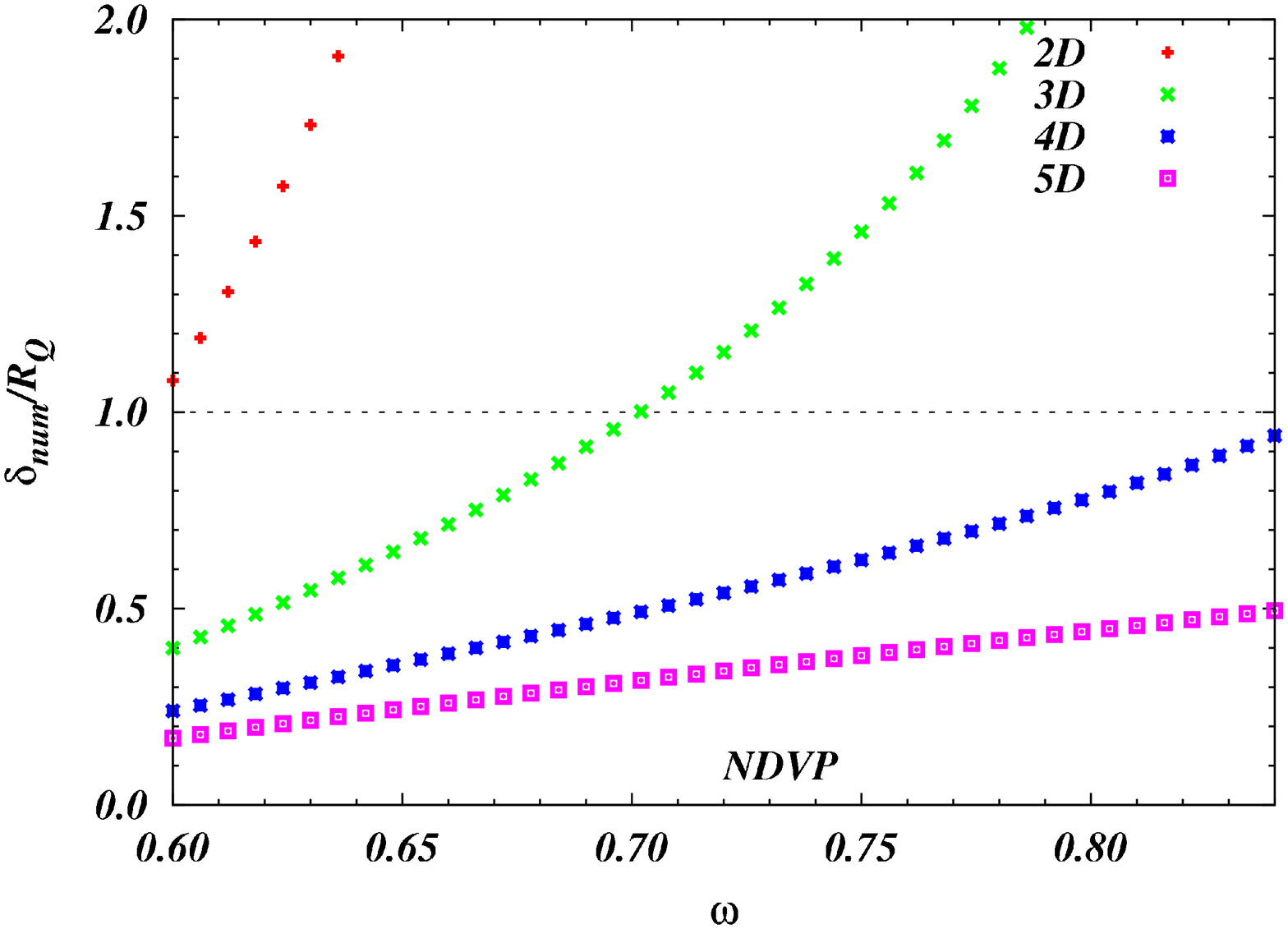}}\\
   \subfigure{\label{fig:degrq}\includegraphics[angle=-90, scale=0.32]{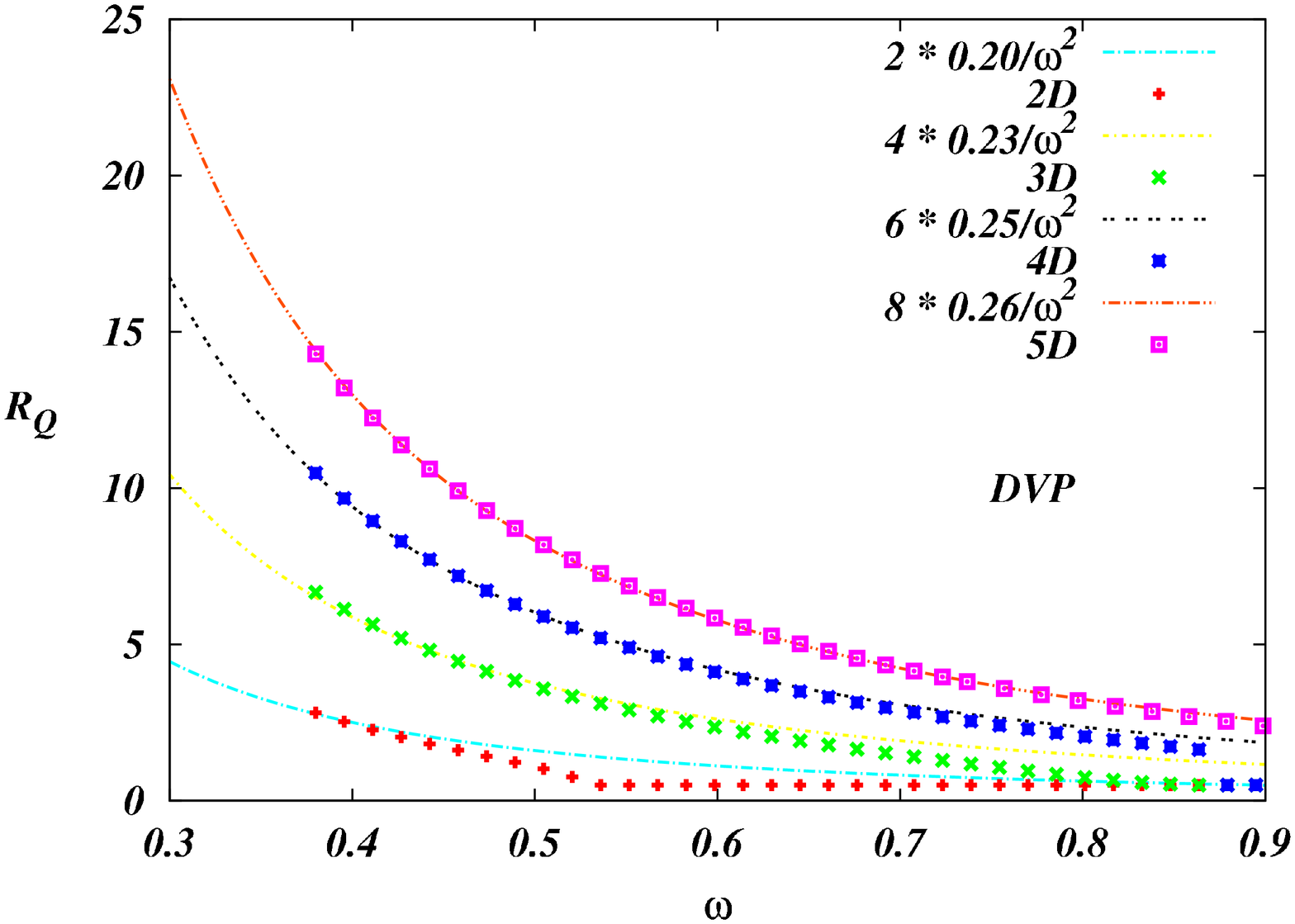}}
   \subfigure{\label{fig:nondegrq}\includegraphics[angle=-90, scale=0.32]{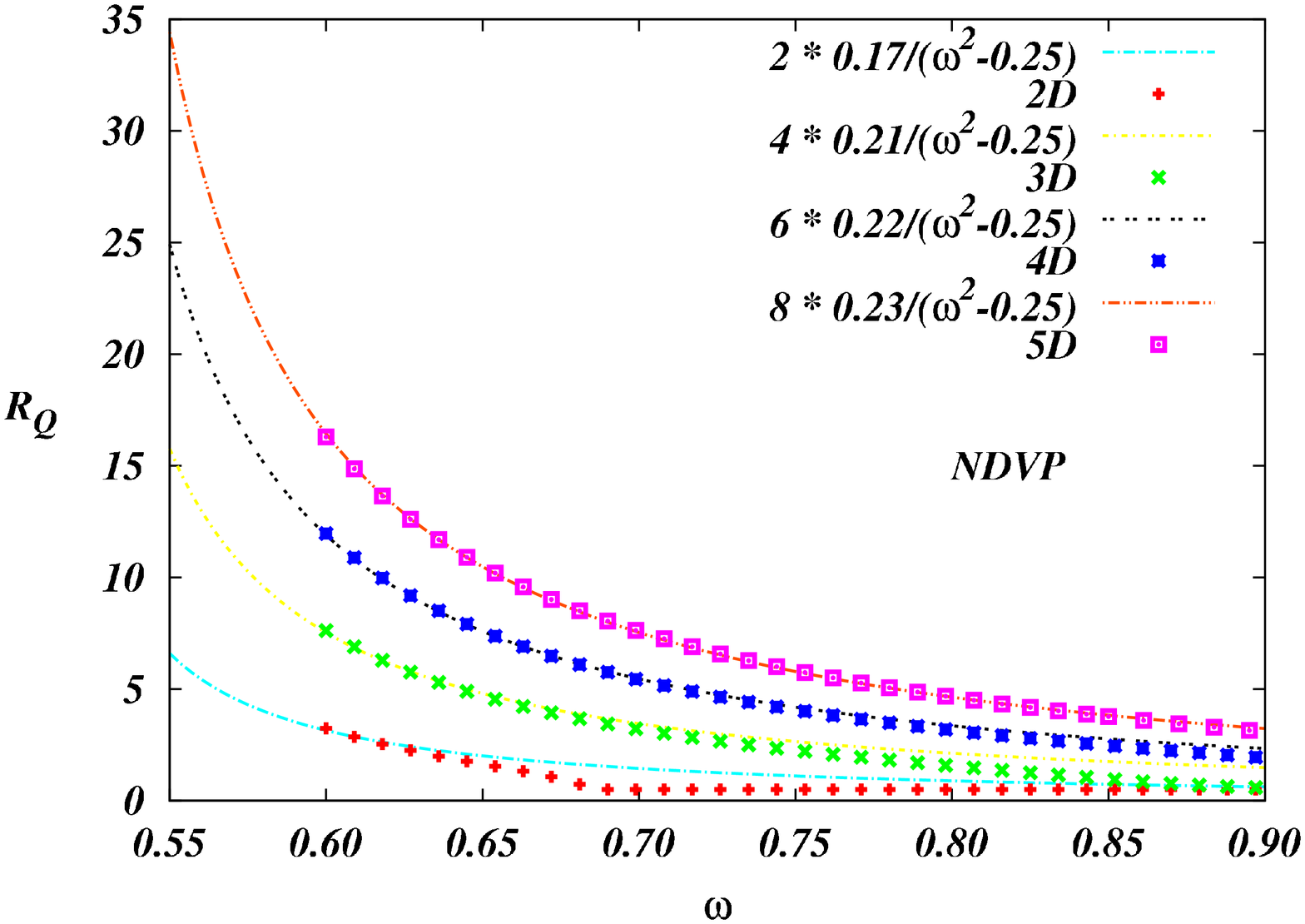}}
  \end{center}
  \figcaption{The initial "positions" $\sigma_0$ (top), $\delta_{num}/R_Q$ (middle), and the core sizes $R_Q(\om)$ (bottom). The top panels show $\sigma_\pm$, \eq{sig-} as black and orange dot-dashed lines respectively. The middle panels show the range of values of $\om$ for a given value of $D$ in which the core thickness is smaller than the core size, a crucial assumption we have to make. In the bottom panel, the analytical core sizes in \eq{RQ-DVP} are plotted with the numerical ones for the following $\om$ ranges: $[0.38-0.40],\ [0.38-0.55],\ [0.38-0.60],\ [0.38-0.70]$ in DVP, and $[0.60-0.62],\ [0.60-0.65],\ [0.60-0.75],\ [0.60-0.85]$ in NDVP and for $D=2,\ 3,\ 4,\ 5$, respectively.  As can be seen, the fits are excellent. The range of $\om$ values chosen have been based on the results shown in the top two panels and correspond to that range where the thin-wall $Q$-balls are solutions (except for $D=2$).}
  \label{fig:sig}
\end{figure}

The bottom two panels show the full profiles satisfying  \eq{mixedprof} for arbitrary $D$ up to $D=5$.  We have been able to obtain the $Q$-ball profiles in the whole parameter space $\om$ except for the extreme thin wall region $\om\simeq \om_-$. Both DVP and NDVP $Q$-balls have profiles with similar behaviours in that as the spatial dimension increases, so does their core size.
\paragraph*{Criteria for the existence of a thin wall $Q$-ball with core size $R_Q$}
The top and middle panels of \fig{fig:sig} show the numerical results for $\sigma_0(\om)$ and $\delta_{num}/R_Q$ against $\om$  for a number of spatial dimensions $D$. For the case of $D \geq 3$ it is clear from the top panels that the $Q$-balls are well described by the thin wall result \eq{sig-} for most values of $\om$, with the range increasing as $D$ increases. The case of $D=2$ is less clear, it appears to asymptote onto the line. We believe there is a solution that exists for that case for small values of $\om$. An important point is that for the approximation to be valid we are working in the regime  $\delta_{num}/R_Q <1$ which can be seen to be true from the middle panels (again we believe the case of $D=2$ is heading below the line  $\delta_{num}/R_Q =1$ for small $\om$.

These results are consistent with our analytic solutions for finite thickness $Q$-balls given by \eq{thindanstz}, subject to the criteria $\sigma_0\simeq \sigma_+$ and $R_Q \gtrsim \delta_{num}$, even though $\om \sim \om_+$.

For $D=1$ we see in the top panels that  $\sigma_0$ exactly matches $\sigma_-$, (the orange dot-dashed lines). The bottom two panels in \fig{fig:sig} show the core sizes $R_Q$ of thin-wall $Q$-balls which satisfy our criterion \eq{core}. Recall that $R_Q$ in \eq{RQ} is a function of $\om$ assuming $\tau$ depends on $\om$ weakly, thus we plot the numerical core sizes comparing them with our analytical approximation for DVP and NDVP, respectively
\be
{RQ-DVP}
R^{DVP}_Q \simeq \frac{2(D-1) \tau_{num}}{\omega^2};\hspace*{10pt} R^{NDVP}_Q \simeq \frac{2(D-1)\tau_{num}}{(\omega^2-\om^2_-)}
\ee
where the parameter $\tau_{num}$ is
computed numerically (see \tbl{tbl:ndrq}). The presented numerical core sizes match excellently with the
analytical fittings over a wide range of $\om$. Some numerical errors
appear around $\om\simeq \om_+$ since we cannot determine the thick wall cores with this technique, see the top two panels in \fig{fig:degexp}. \tbl{tbl:ndrq} shows analytical and numerical values of $\tau$ using \eq{tens} and the above fitting technique. We confirm that the values of $\tau$ (a part of the surface tension $\tau/D$ in \eq{linearthin}) are nearly constant, depending slightly on $D$. Therefore, the assumptions we made for thin wall $Q$-balls are valid as long as $\sigma_0\simeq \sigma_+$ and $R_Q \gtrsim \delta_{num}$. 

\begin{figure}[!ht]
  \def\@captype{table}
  \begin{minipage}[t]{\textwidth}
    \begin{center}
      \begin{tabular}{|c||c|c|c|c|c|}
	\hline
	$\tau$ & $\tau_{ana}$ & $2D$ & $3D$ & $4D$  & $5D$ \\
    	\hline
		DVP & 0.19 & 0.20 & 0.23 & 0.25 & 0.26 \\
		NDVP & 0.16 & 0.17 & 0.21 & 0.22 & 0.23 \\
\hline
      \end{tabular}
	\end{center}
	\tblcaption{The values of $\tau_{ana}$ and $\tau_{num}$ in terms of $D$ in DVP and NDVP, see \eqs{tens}{RQ}.}
    \label{tbl:ndrq}
  \end{minipage}
\end{figure}

\paragraph*{Configurations}

\begin{figure}[htp]
    \begin{center}
    \subfigure{\label{fig:degrhoq}\includegraphics[angle=-90,scale=0.32]{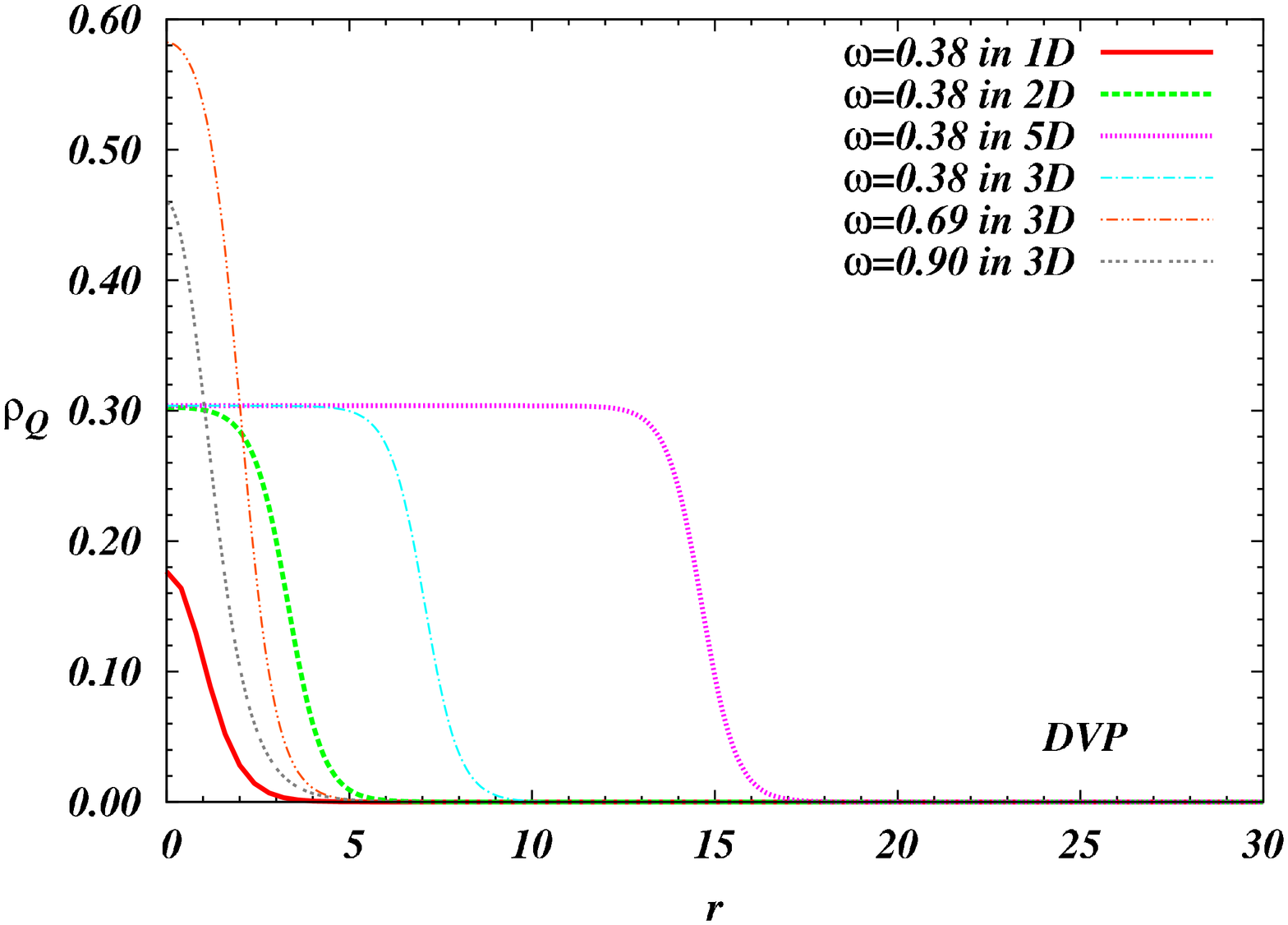}}
    \subfigure{\label{fig:nondegrhoq}\includegraphics[angle=-90,scale=0.32]{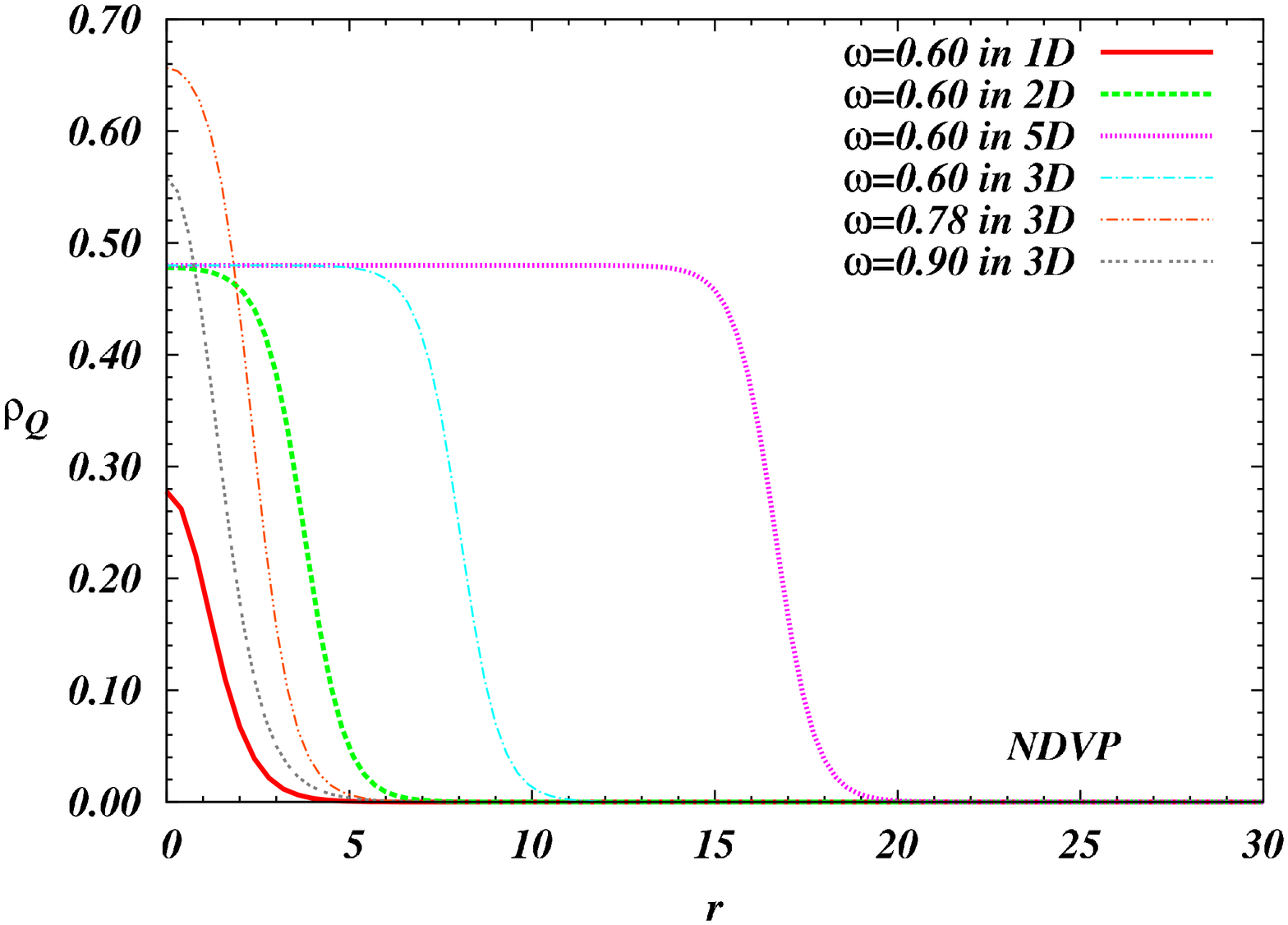}}\\
        \subfigure{\label{fig:degrhoe}\includegraphics[angle=-90,scale=0.32]{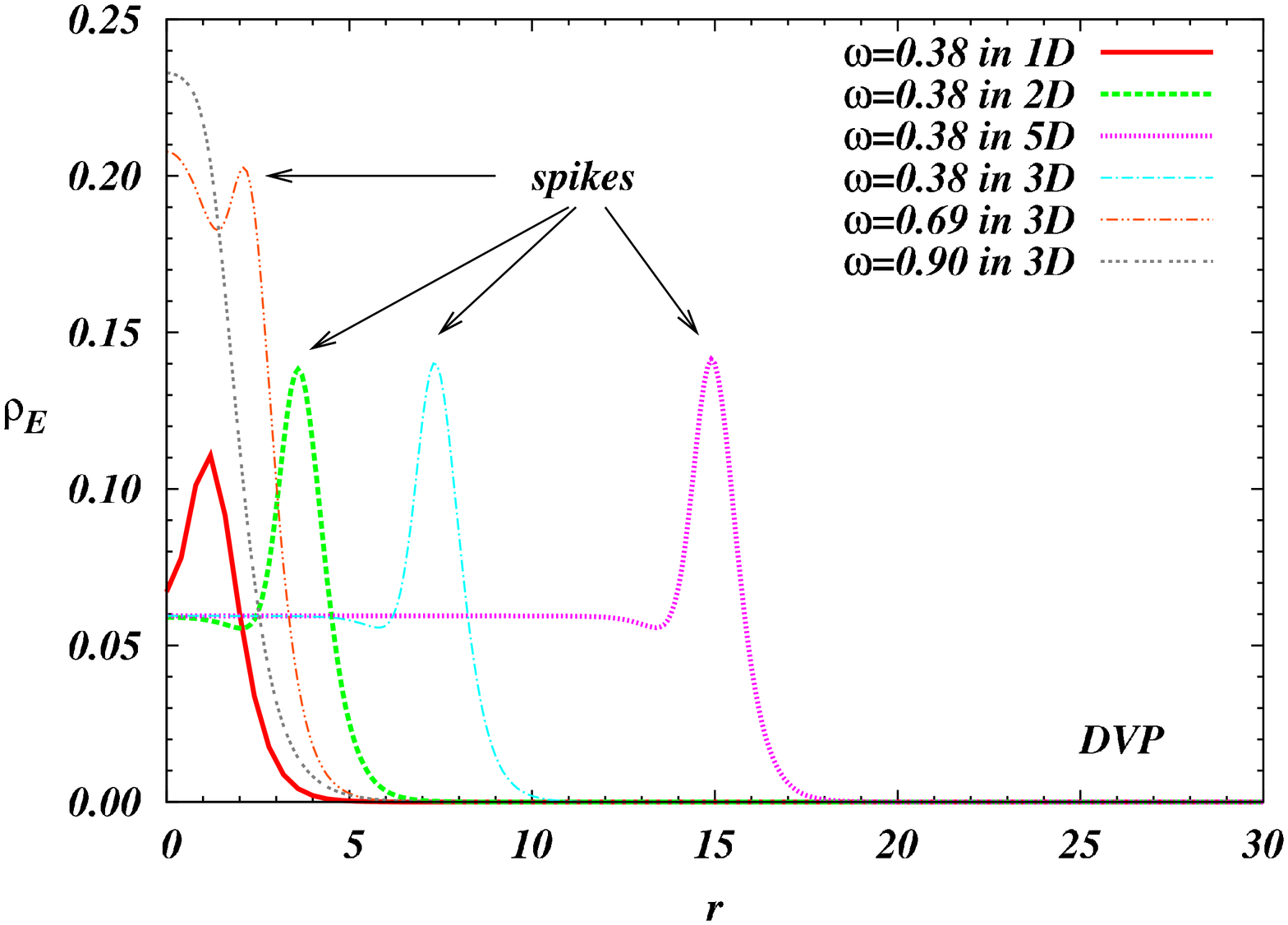}}
    \subfigure{\label{fig:nondegrhoe}\includegraphics[angle=-90,scale=0.32]{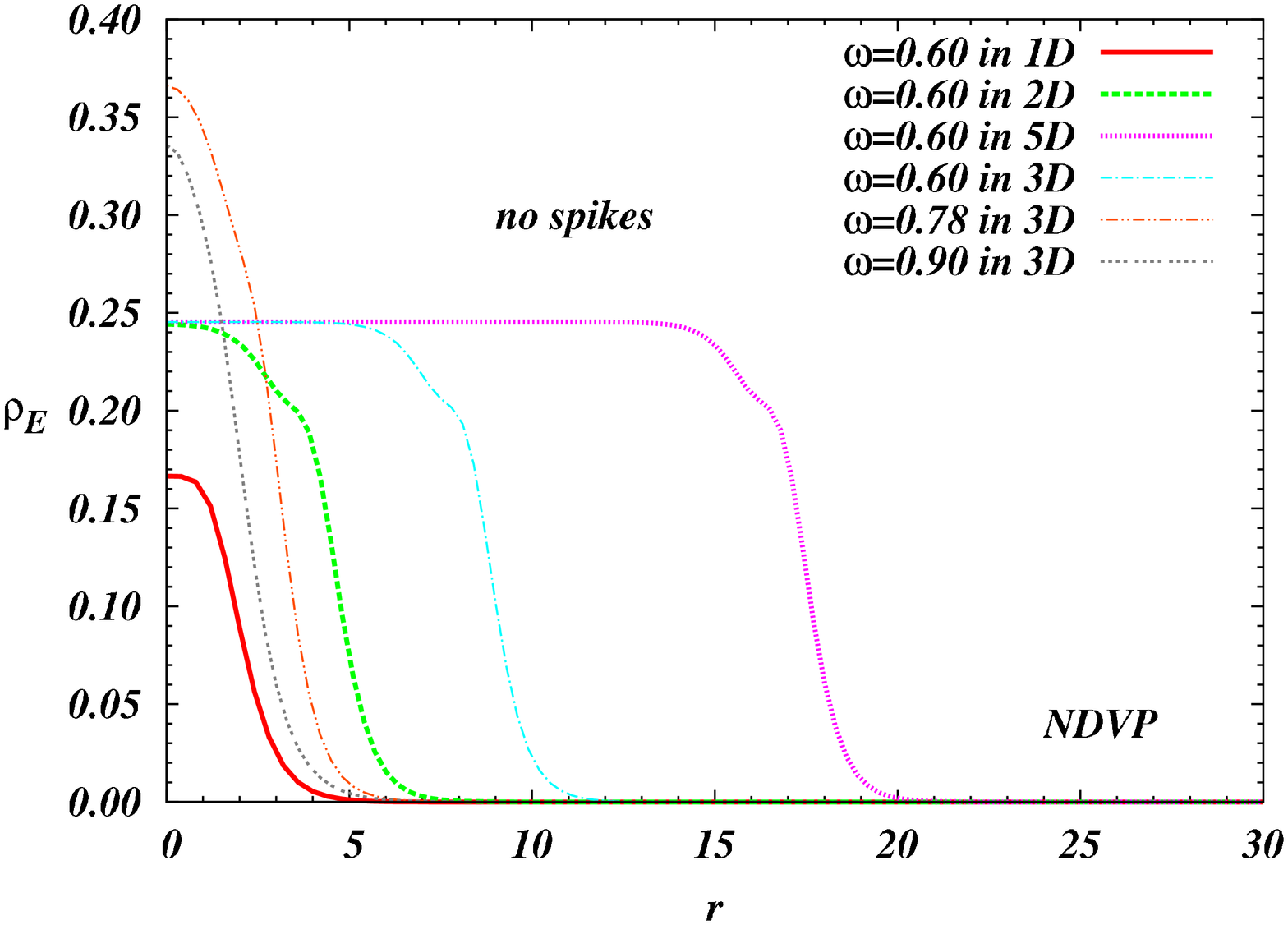}}
  \end{center}
  \figcaption{The configurations for charge density $\rho_Q$ (top) and energy density $\rho_E$ (bottom) computed using \eq{rho_Q} for both DVP (left) and NDVP (right). The presence of spikes of $\rho_E$ in DVPs contributes to their increased surface energies.}
  \label{fig:conf}
\end{figure}

\fig{fig:conf} illustrates the configurations of charge density $\rho_Q$ (top) and energy density $\rho_E$ (bottom), in both DVP (left) and NDVP (right). Each of the DVP energy densities around $\om\sim \om_-$ has a spike within the shells, while those spikes are not present in NDVP. The presence of spikes can contribute to the increase in surface energy $\mS$, which accounts for the different observed ratio for  $\mS/\mU$ in the two cases, where $\mU$ is the potential energies. Otherwise DVP and NDVP models have similar profiles in \fig{fig:degexp}. Moreover, we have numerically checked that $Q$-balls for $D\ge 2$ generally have positive radial pressures, whereas the $1D$ radial pressures are always zero, \ie $\half \sigma^{\p2}=\Uo$ due to \eq{radialp}.

\paragraph*{Virialisation and characteristic slope $E_Q/\om Q$}

\begin{figure}[htp]
\begin{center}
    \subfigure{\label{fig:degsu}\includegraphics[angle=-90,scale=0.32]{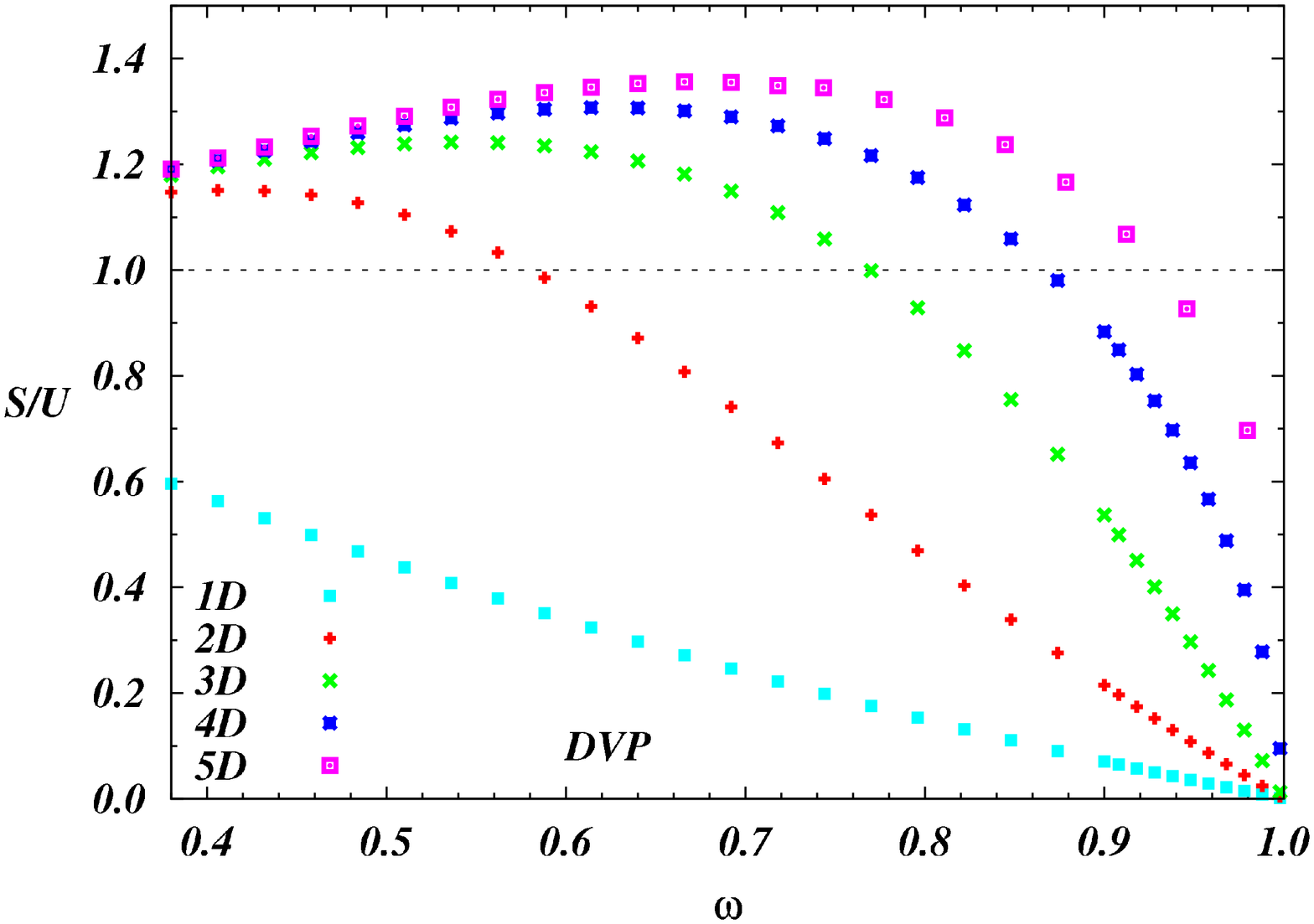}}
    \subfigure{\label{fig:nondegsu}\includegraphics[angle=-90,scale=0.32]{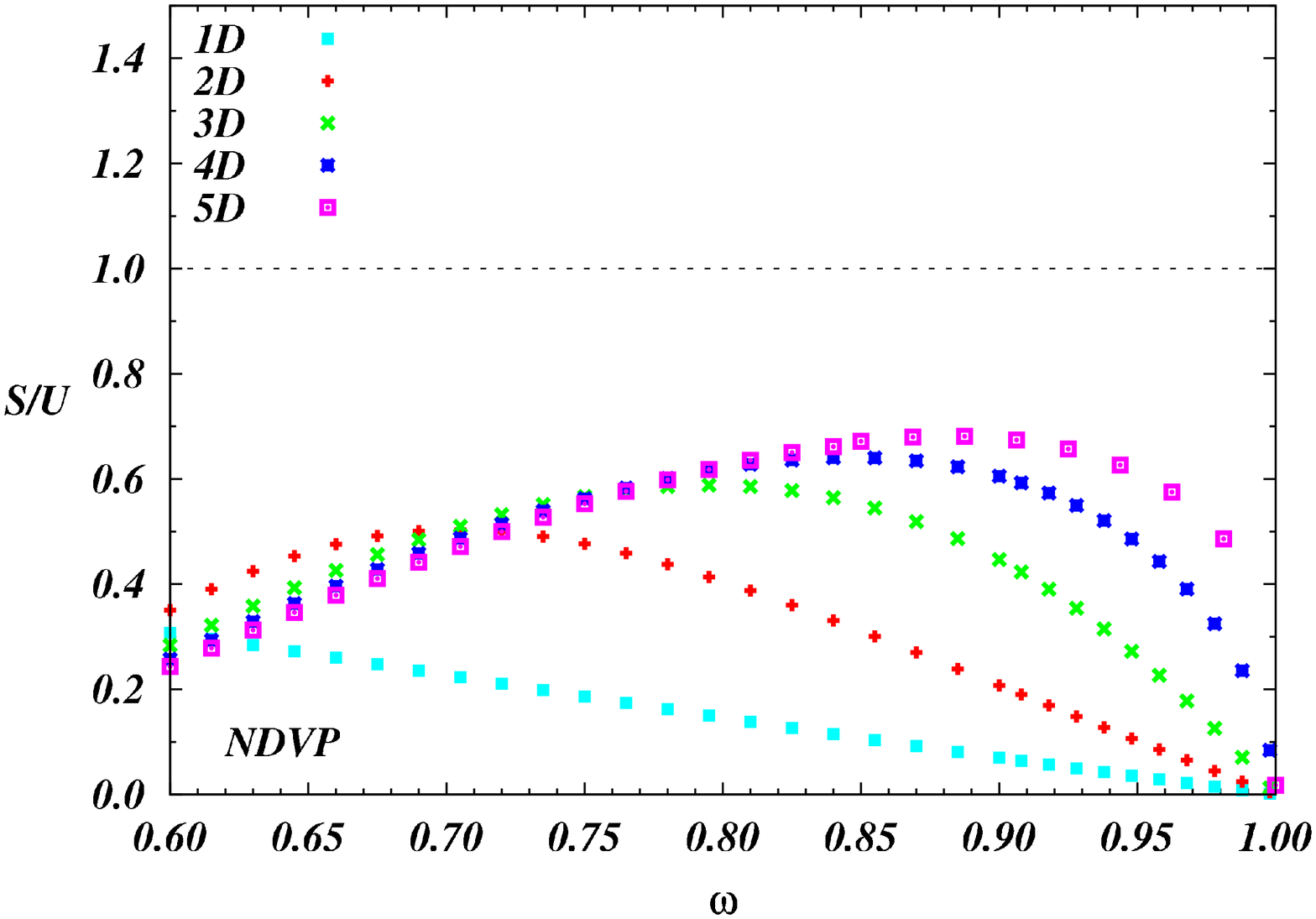}}\\
    \subfigure{\label{fig:degslope}\includegraphics[angle=-90,scale=0.32]{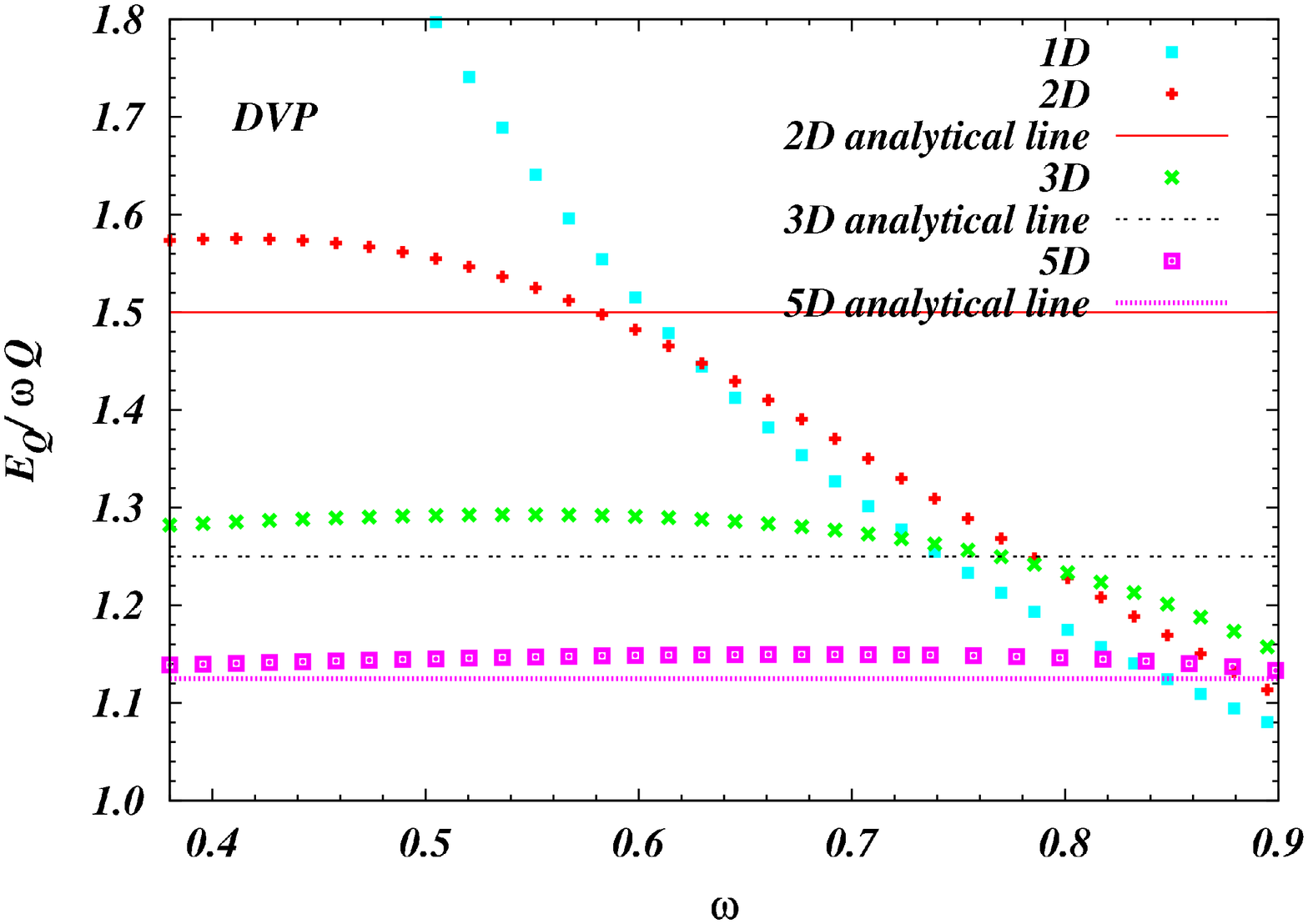}}
    \subfigure{\label{fig:nondegslope}\includegraphics[angle=-90,scale=0.32]{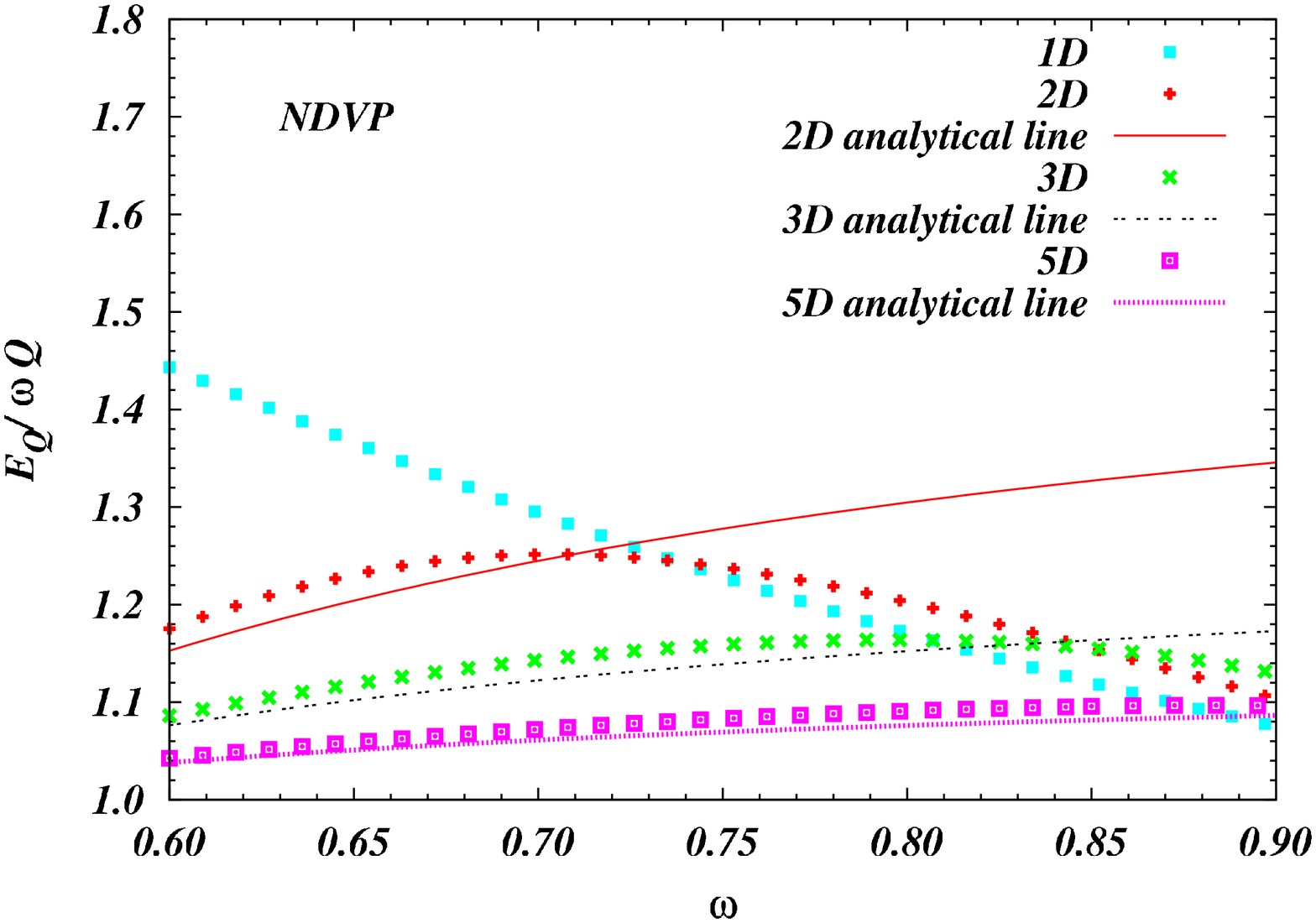}}\\
    \subfigure{\label{fig:degslopethck}\includegraphics[angle=-90,scale=0.32]{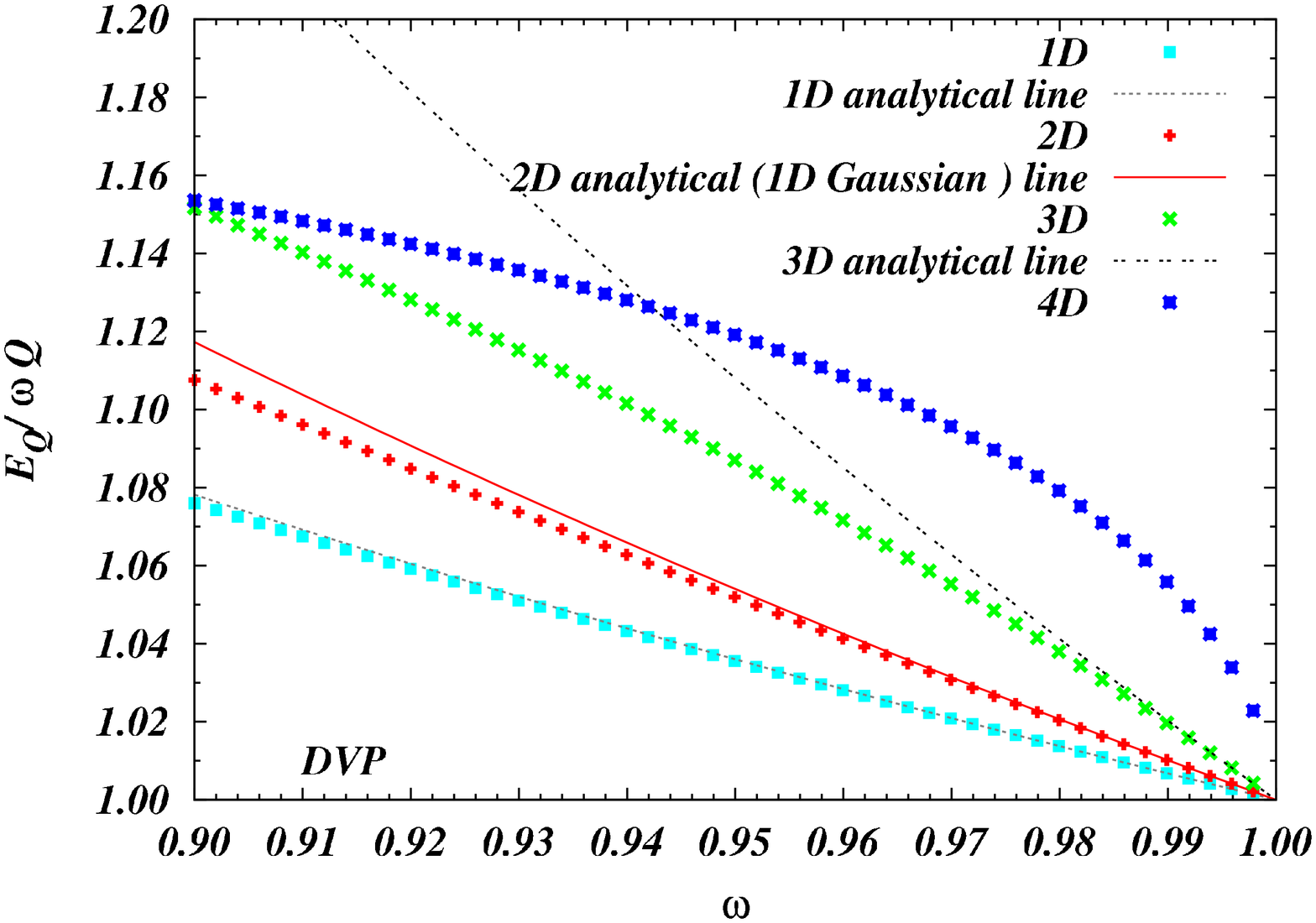}}
    \subfigure{\label{fig:nondegslopethck}\includegraphics[angle=-90,scale=0.32]{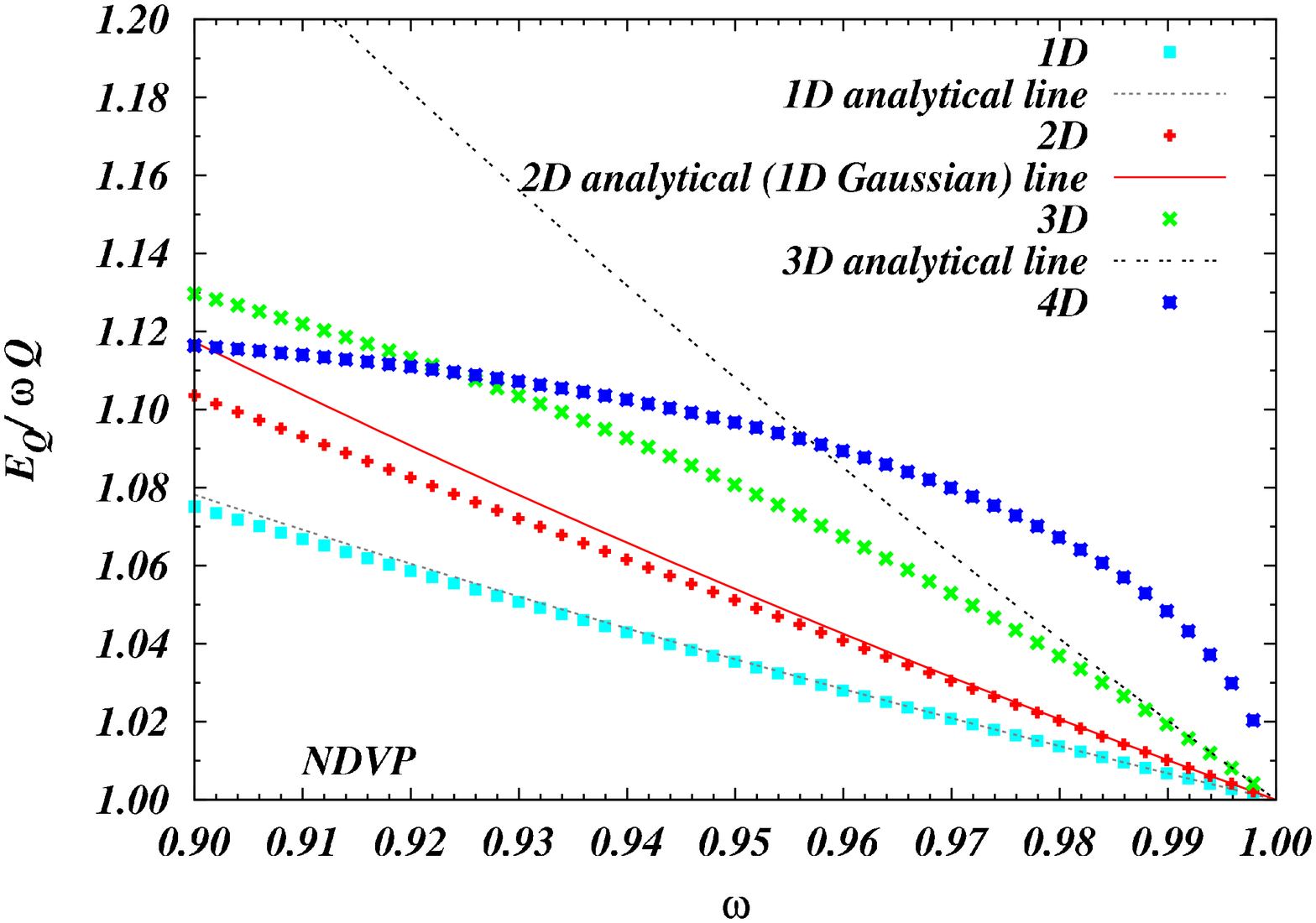}}\\
    \end{center}
  \figcaption{The ratio of $\mathcal{S/U}$ where $\mS$ and $\mU$ are surface and potential energies (top panels), the characteristic slope $E_Q/\om Q$ in the thin-wall-like limit, $\om\sim \om_-$, with the analytic lines \eq{surfthin} (middle panels),
and in the thick-wall-like limit, $\om \simeq \om_+$, (bottom panels), with the analytic lines \eqs{gausseq}{modslope}.}
  \label{fig:egy}
\end{figure}

The top panels in \fig{fig:egy} illustrate the ratios $\mS/\mU$ and the four bottom ones show the characteristic slopes of $E_Q/\om Q$ against $\om$ in both the thin-wall (middle-panels) and thick-wall (bottom-panels) limits. According to our analytic arguments \eq{UoverS}, we expect $\mS/\mU \simeq 1$ in the extreme limit $\om \simeq \om_-=0$ in DVP. Similarly, we expect
$\mS/\mU \sim 0$ in the same extreme thin wall  limit $\om=\om_-=0.5$ for NDVP. The latter case corresponds to the existence of $Q$-matter with the simple step-like ansatz \eq{equation}. Although we are unable to probe these precise regimes, we believe the slopes of the curves indicate they are heading in the right direction.
The thin wall slopes $E_Q/\om Q$ in the two middle panels lie nearby the analytical ones, \eqs{surfthin}{eqqdeg}, as long as $\sigma_0\simeq\sigma_+$ (see \fig{fig:sig}) except for the $2D$ cases because for $D\le 2$ the profiles are not well fitted by thin wall predictions.  Similarly, the thick wall slopes $E_Q/\om Q$ in the bottom two panels agree with our analytical predictions \eq{modslope} using the modified ansatz rather than with \eq{gausseq} using the simple Gaussian ansatz. We have confirmed that the analytic thick wall slopes with \eq{modslope} can not apply to higher dimensions $D \ge 4$, see \eq{validthck}. Around the thick wall limit $\om\simeq \om_+$, the behaviours in both potentials are $\mS \ll \mU$ (see top panel), which implies $E_Q \simeq \om Q$ as predicted in \eqs{gausseq}{eqthick2}; hence we can verify that the solutions are continued to the free particle solutions, see \eq{freeengy}. Our physically motivated modified ans\"{a}tze in both the thin and thick wall limits therefore have clear advantages over the simple ans\"{a}tze in \eqs{equation}{gaussansatz}.
\paragraph*{$Q$-ball stability}

\begin{figure}[htp]
  \begin{center}
    \subfigure{\label{fig:dcls}\includegraphics[angle=-90,scale=0.32]{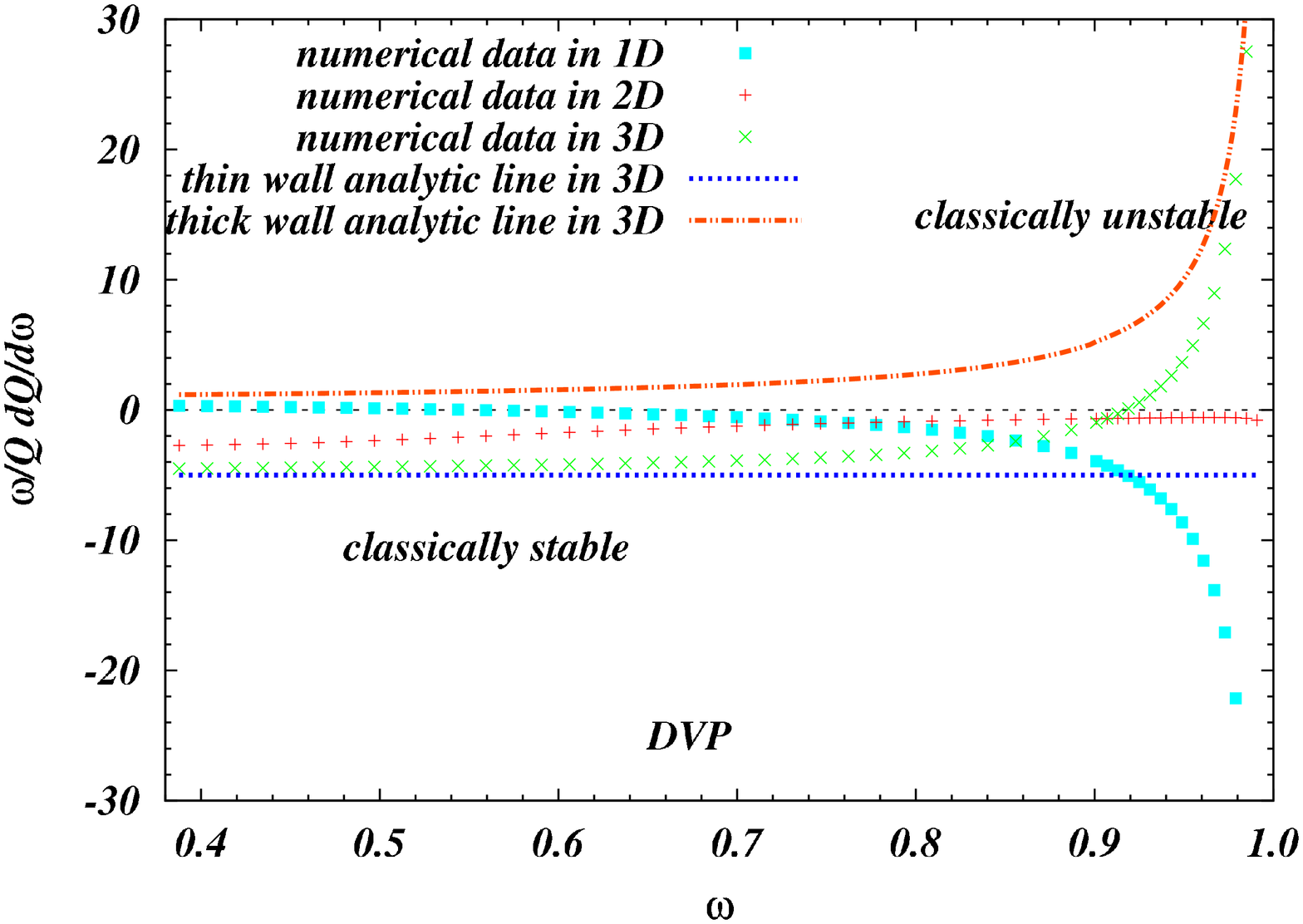}}
    \subfigure{\label{fig:ndcls}\includegraphics[angle=-90,scale=0.32]{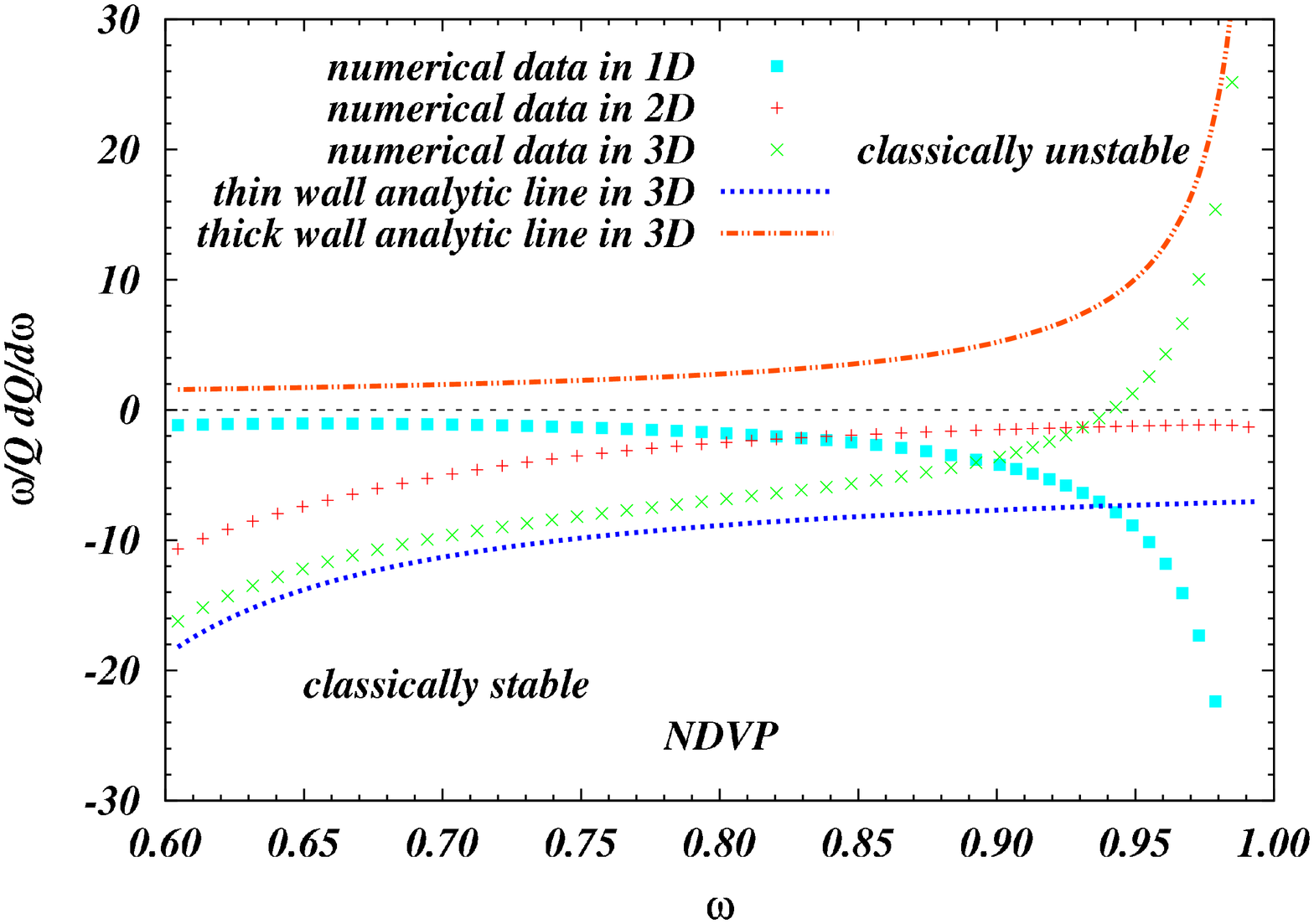}}\\
    \subfigure{\label{fig:degabs}\includegraphics[angle=-90,scale=0.32]{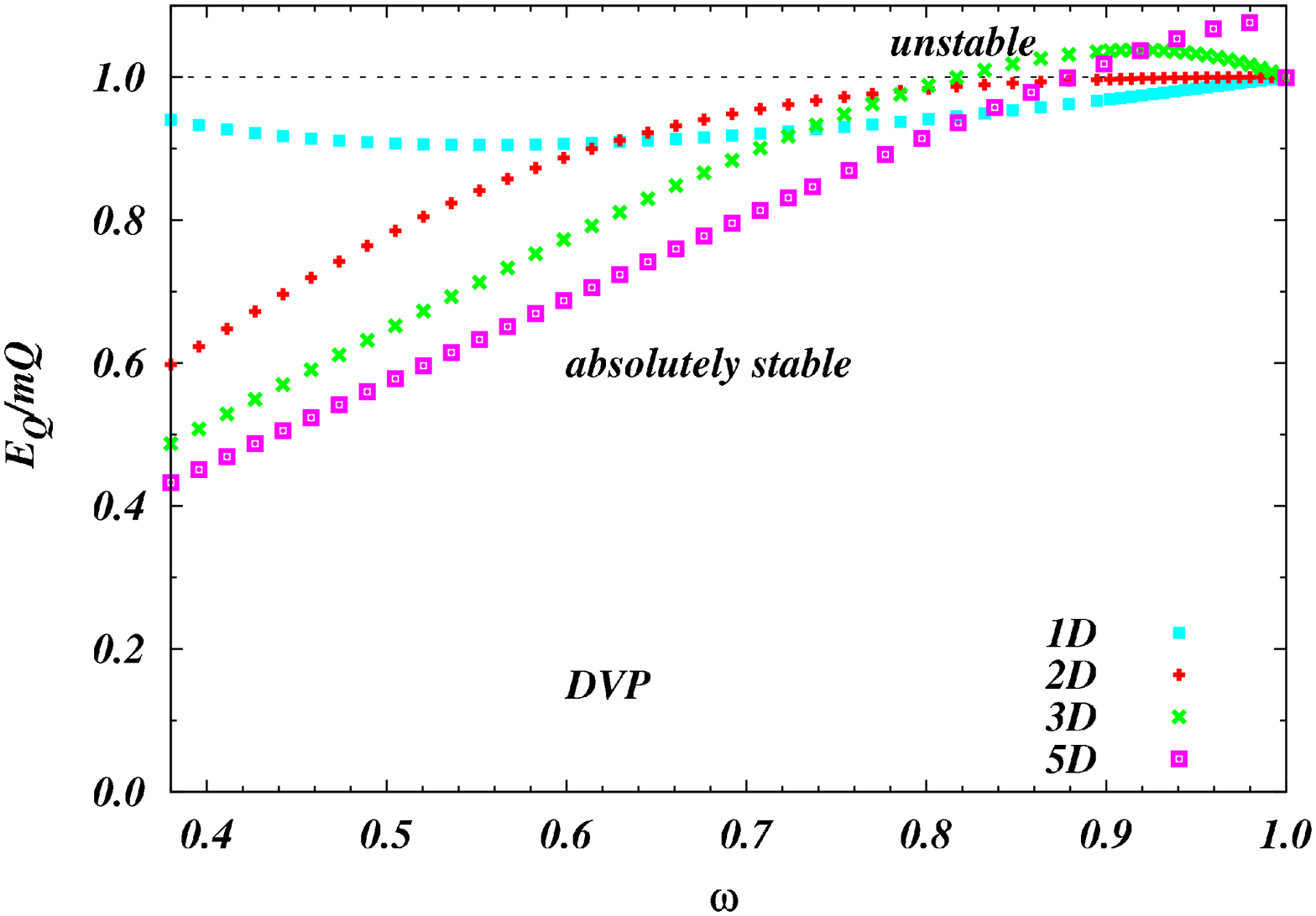}}
    \subfigure{\label{fig:nondegabs}\includegraphics[angle=-90,scale=0.32]{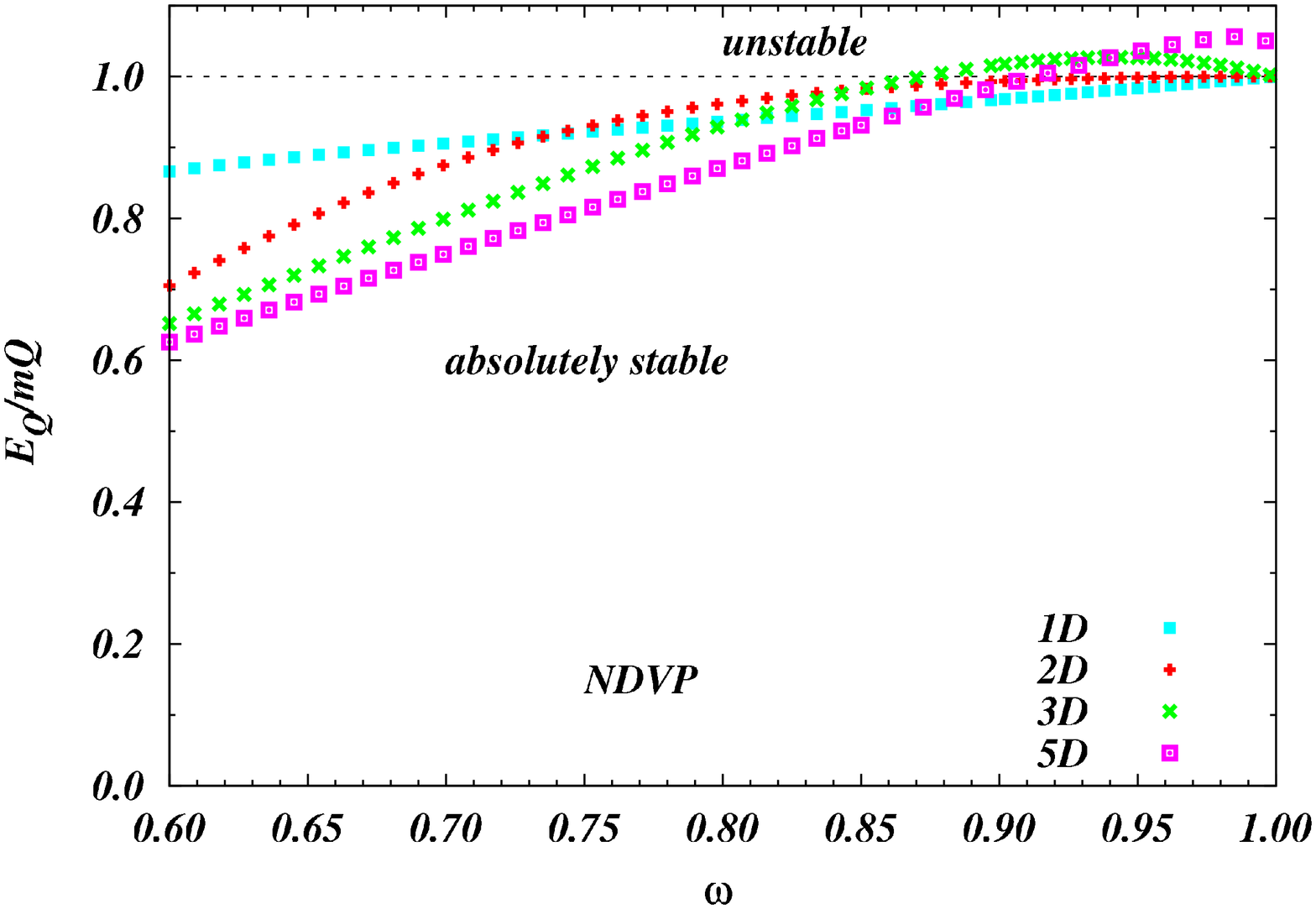}}
  \figcaption{Classical stability using \eq{CLS} for the top panels and absolute stability using \eq{ABSCOND} for the bottom panels. The $3D$ analytical lines of \eqs{q1class}{modcls} for classical stability agree with the corresponding numerical data. Above the zero-horizontal axes in the top panels, the $Q$-balls are classically unstable. Similarly, $Q$-balls above the horizontal axis, $E_Q=mQ$, are absolutely unstable. The one dimensional $Q$-balls are always classically stable. The $1D$ slopes $E_Q/m Q$ have different behaviours depending on DVP and NDVP unlike the other dimensional cases.}
  \label{fig:clsabs}
   \end{center}
\end{figure}

\fig{fig:clsabs} shows the classical and absolute stability lines for $Q$-balls. \tbl{tbl:eqwa} indicates the approximate analytical values of $\om_a$ derived by \eqs{viriwa}{ndvpwa}, which can be compared to the numerically obtained critical values $\om$ for the stabilities denoted by $\om_c,\; \om_s,\; \om_{ch},\; \om_a,$ and $\om_f$ in \tbl{tbl:dwmc}. Each of these are defined by  $\left.\frac{dQ}{d\om}\right|_{\om_c}=\left.\frac{d^2S_\om}{d\om^2}\right|_{\om_s}=\left.\frac{d}{d\om}\bset{\frac{E_Q}{Q}}\right|_{\om_{ch}}=0$, $E_Q/Q|_{\om_a}=m$,
 and $\left.\frac{d\om}{dQ}\right|_{\om_f}=0$ respectively. The $3D$ analytical plots of $\frac{\om}{Q} \bset{\frac{dQ}{d\om}}$
in the thin and thick wall limits, \eqs{q1class}{modcls}, can be seen to match the corresponding numerical data in the appropriate limits of $\om$. We have confirmed numerically that  for both DVP and NDVP cases $\om_c=\om_f \simeq \om_s\simeq \om_{ch}$, see \tbl{tbl:dwmc}. This can be easily understood from \eqs{CLS}{chslope} and \eq{SAF}.

Recall \eq{modcls}, the polynomial potential \eq{usig} with $n=4$, leads to the classical stability condition $D\le 2$ for the thick wall case. The top panels in \fig{fig:clsabs} demonstrate that
thick wall $Q$-balls in $D \ge 3$ are classically unstable.
In \tbl{tbl:dwmc}, one can check that the absolute stability condition is more severe than the classical one, then there are three types of $Q$-ball \cite{Friedberg:1976me} as before: \emph{absolutely stable $Q$-balls} for $\om < \om_a$, \emph{meta-stable $Q$-balls} for $\om_a \leq \om \leq \om_c$, which are not quantum-mechanically stable but classically stable, and can decay into multiple $Q$-balls, or \emph{unstable $Q$-balls} for $\om_c < \om$.

Both DVP and NDVP analytical values $\om_a$ in \tbl{tbl:eqwa} agree well with the numerical ones in \tbl{tbl:dwmc}. Generally speaking, the higher dimensional $Q$-balls are more stable classically as well as quantum mechanically. Moreover, thin wall $Q$-balls are always classically stable as demonstrated in \eq{q1class}, but the classical stability of thick wall $Q$-balls is model- and $D$- dependent as in \eq{modcls}. The one- and two- dimensional $Q$-balls have a much richer structure than the thin and thick wall $Q$-balls. It is a challenging task to understand their intermediate profiles \cite{Sakai:2007ft}.

\begin{figure}[ht]
  \def\@captype{table}
    \begin{minipage}[t]{\textwidth}
    \begin{center}
      \begin{tabular}{|c||c|c|c|c|}
	\hline
		& \multicolumn{4}{|c|}{$\om_a$} \\
        \hline
$D$ &  $\mathcal{S} \gg \mathcal{U}$ & $\mathcal{S} \simeq \mathcal{U}$  or DVP &  NDVP & $\mathcal{S} \ll \mathcal{U}$ \\
        \hline
$3$ & 0.50 & 0.80 & 0.86 & 1 \\
$4$ & 0.67 & 0.86 & 0.90 & 1 \\
$5$ & 0.75 & 0.89 & 0.92 & 1 \\
        \hline
      \end{tabular}
    \end{center}
    \tblcaption{Virial relations: $\om_a$ in terms of space dimension
$D$ and ratio $\mathcal{S}/\mathcal{U}$, see \eq{viriwa}}	
  \label{tbl:eqwa}
  \end{minipage}
   \hfill
	\\
	\\
  \begin{minipage}[t]{\textwidth}
    \begin{center}
      \begin{tabular}{|c||c|c|c|c|c||c|c|c|c|c|}
	\hline
		&  \multicolumn{5}{|c||}{DVP} & \multicolumn{5}{|c|}{NDVP} \\
	\hline
	$D$ & $\omega_a$ & $\omega_c$ & $\omega_s$ & $\om_{ch}$ & $\om_f$ & $\omega_a$ & $\omega_c$ & $\omega_s$ & $\om_{ch}$  & $\om_f$ \\
    	\hline
3 & 0.82 & 0.92 & 0.92 & 0.92 & 0.92 & 0.87  & 0.94  & 0.94 & 0.94 & 0.94\\
4 & 0.86 & 0.96 & 0.96 & 0.96 & 0.96 & 0.89  & 0.97  & 0.97 & 0.97 & 0.97 \\
5 & 0.882 & 0.983 & 0.993 & 0.983 & 0.983 & 0.910 & 0.985 & 0.996 & 0.991 & 0.985 \\
\hline
      \end{tabular}
    \end{center}
    \tblcaption{The critical values for classical stability, absolute stability and stability against fission in DVP and NDVP using \eqss{ABSCOND}{CLS}{chslope} and \eq{SAF}. The critical values are defined by $\left.\frac{dQ}{d\om}\right|_{\om_c}=\left.\frac{d^2S_\om}{d\om^2}\right|_{\om_s}=\left.\frac{d}{d\om}\bset{\frac{E_Q}{Q}}\right|_{\om_{ch}}=0$, $E_Q/Q|_{\om_a}=m$,
 and $\left.\frac{d\om}{dQ}\right|_{\om_f}=0$. The numerical values of $\om_a$ coincide with the analytic ones in \tbl{tbl:eqwa}. We have confirmed numerically that  $\om_c=\om_f \simeq \om_s\simeq \om_{ch}$.}
    \label{tbl:dwmc}
  \end{minipage}
\end{figure}

\paragraph*{Legendre relations}

\begin{figure}[htp]
    \begin{center}
    \subfigure{\label{fig:dlgrd1}\includegraphics[angle=-90,scale=0.21]{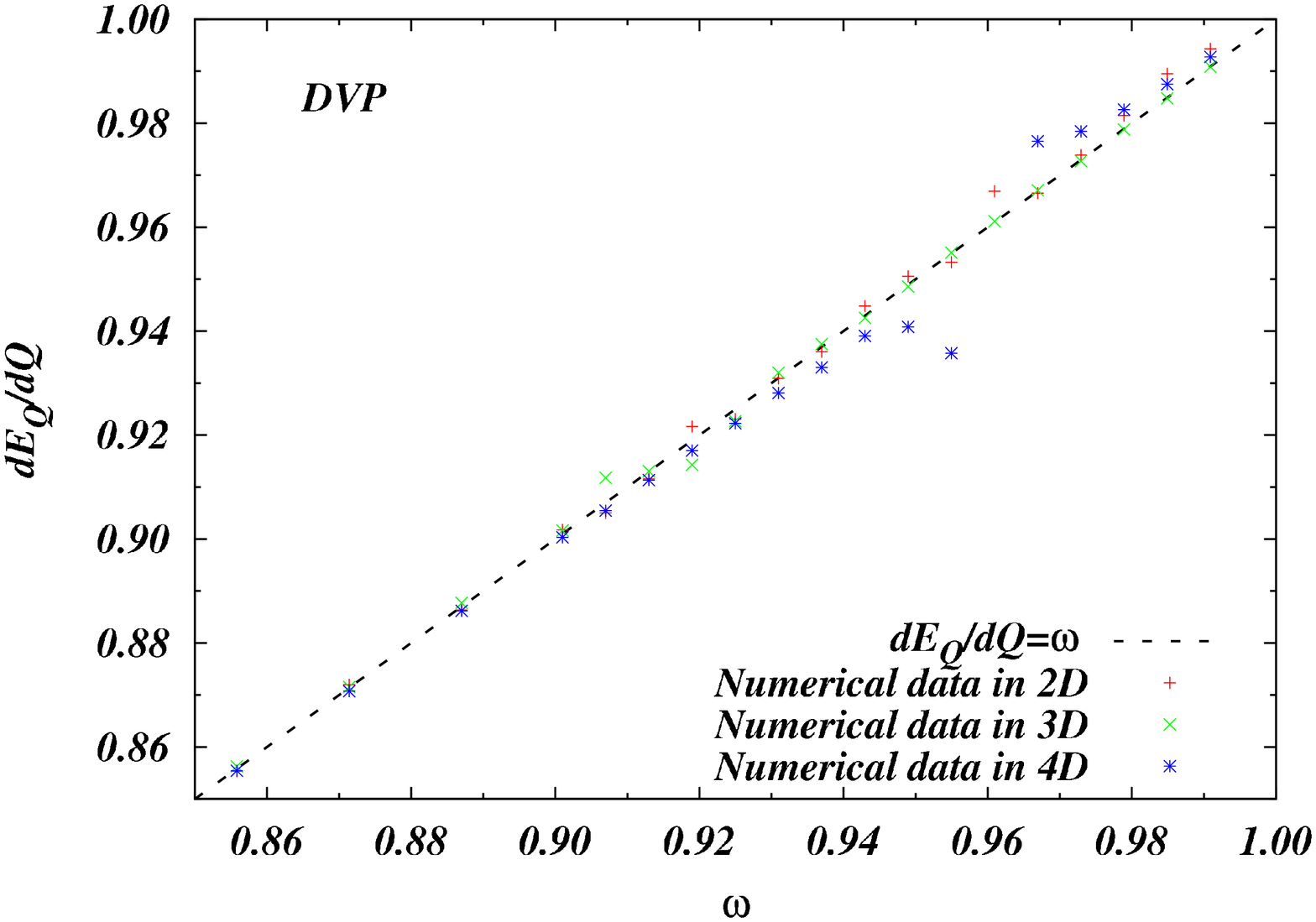}}
    \subfigure{\label{fig:dlgrd2}\includegraphics[angle=-90,scale=0.21]{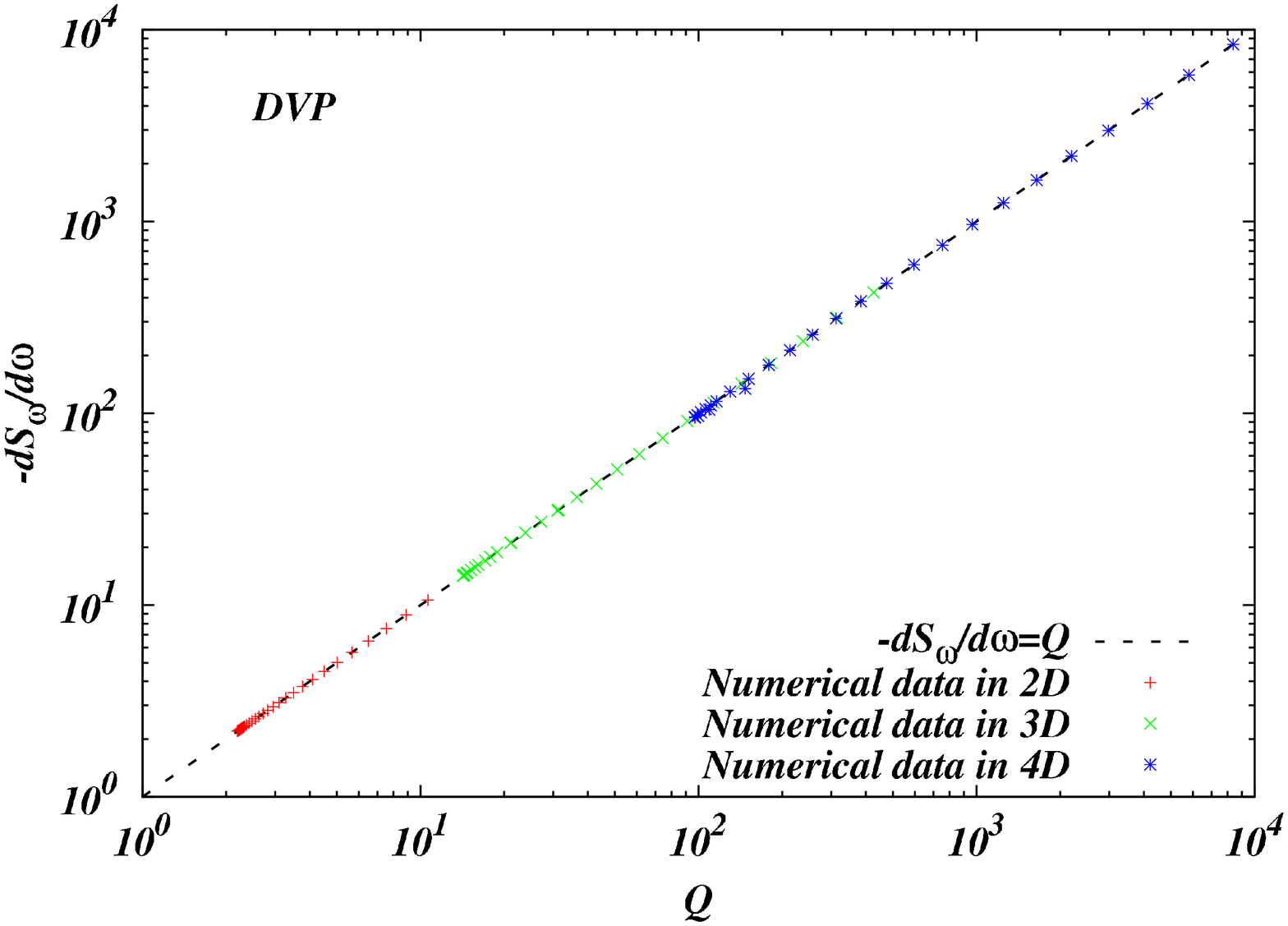}}
    \subfigure{\label{fig:dlgrd3}\includegraphics[angle=-90,scale=0.21]{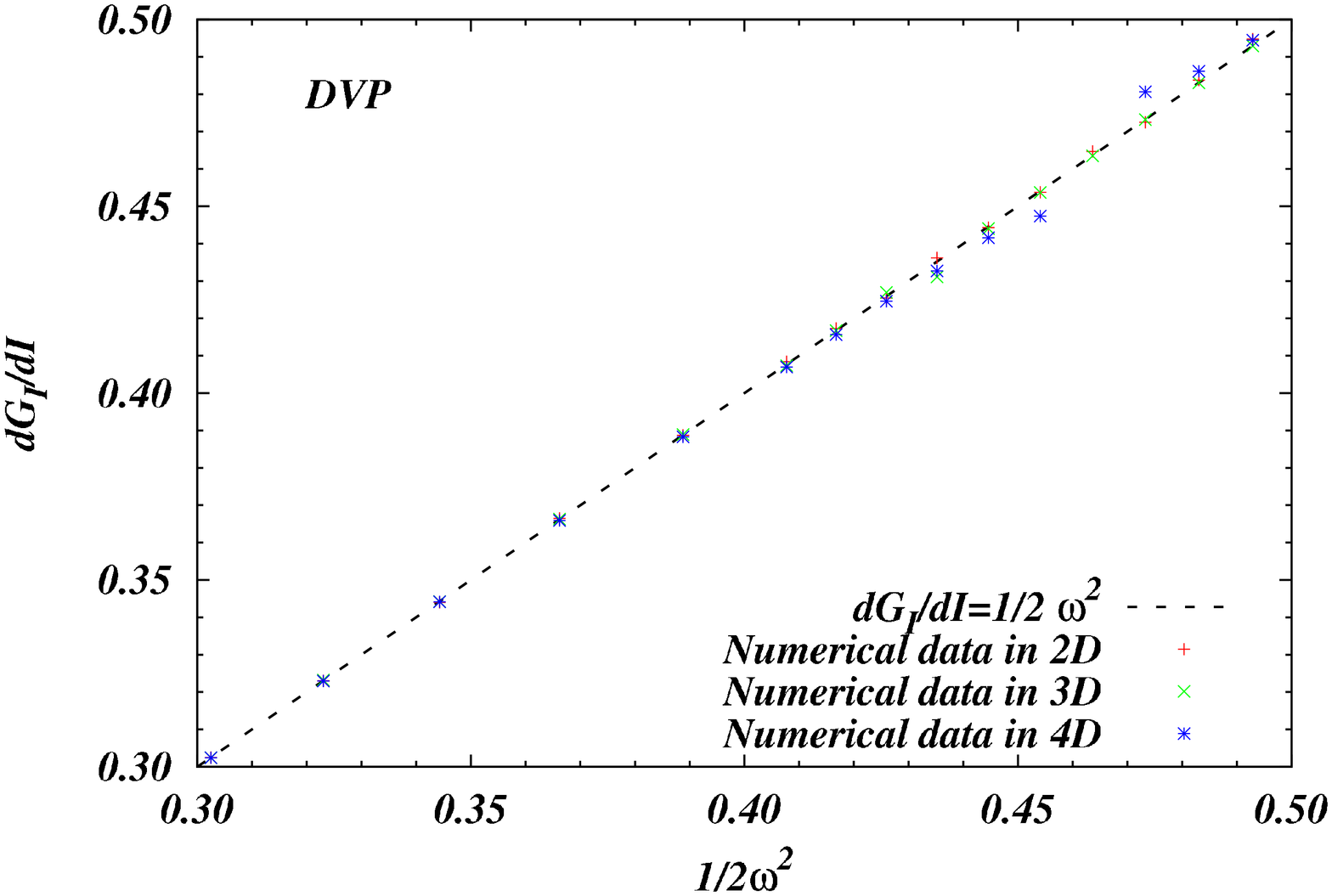}}\\
    \subfigure{\label{fig:ndlgrd1}\includegraphics[angle=-90,scale=0.21]{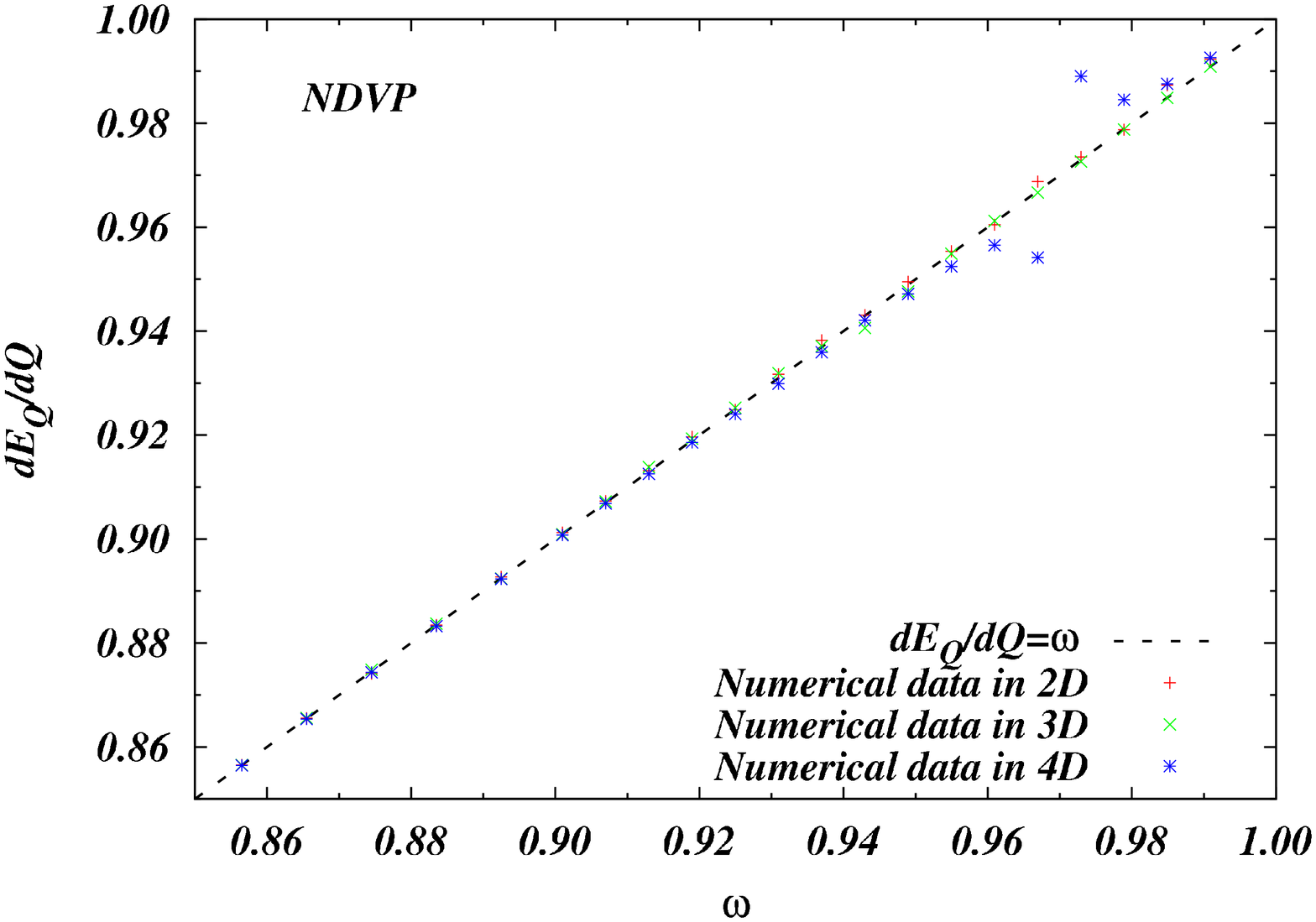}}
    \subfigure{\label{fig:ndlgrd2}\includegraphics[angle=-90,scale=0.21]{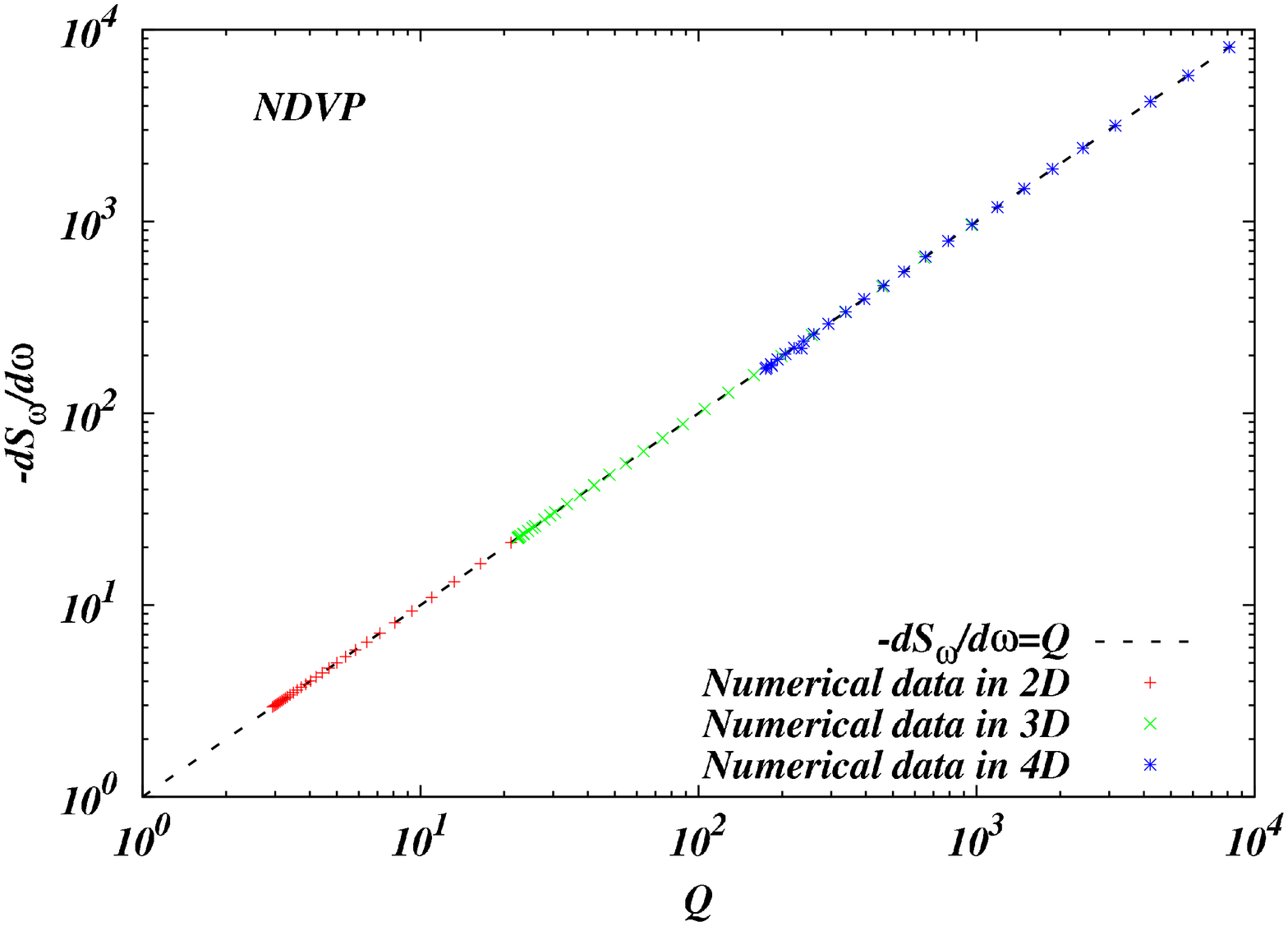}}
    \subfigure{\label{fig:ndlgrd3}\includegraphics[angle=-90,scale=0.21]{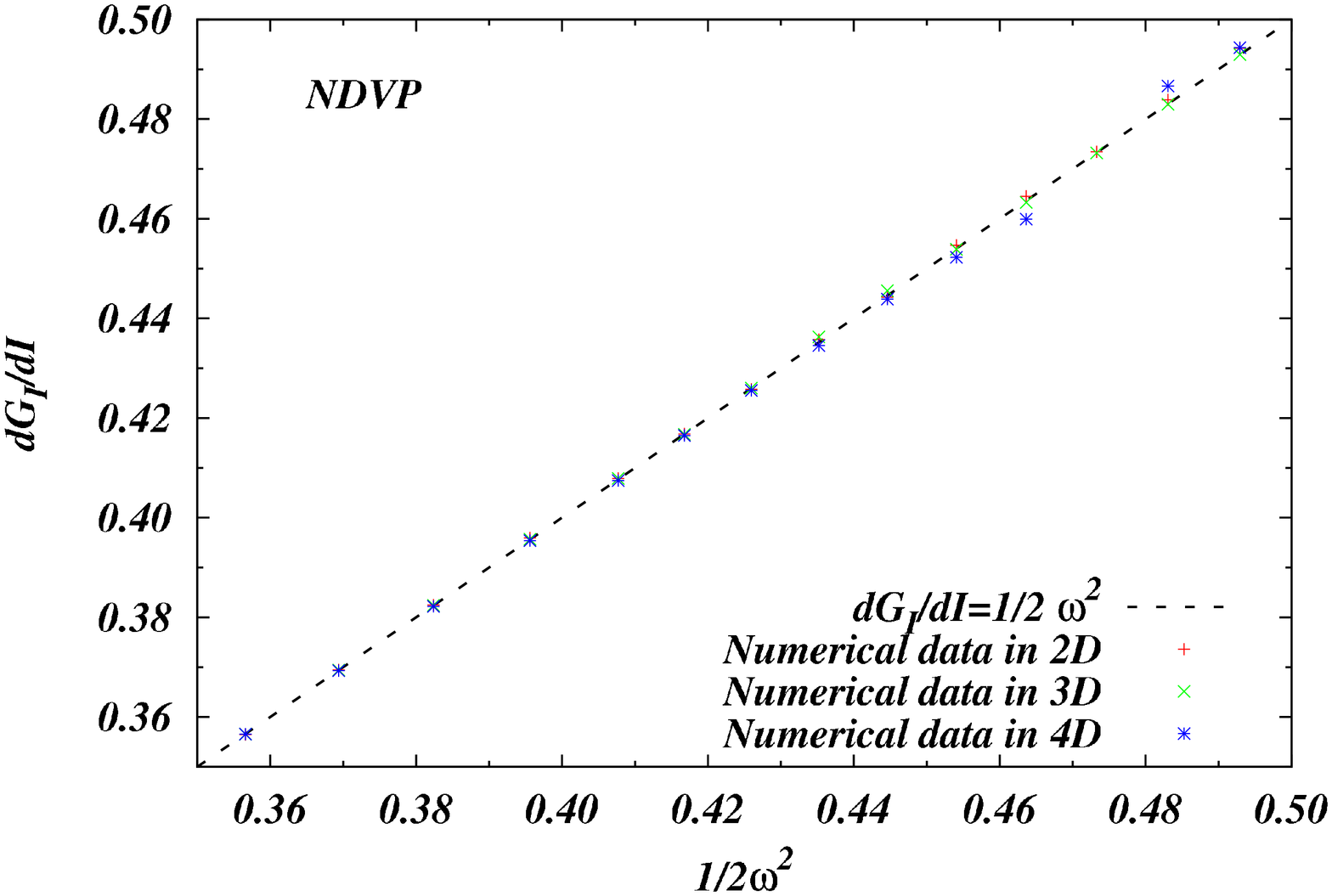}}
  \end{center}
  \figcaption{The Legendre relations of \eq{legendre}: $\frac{dE_Q}{dQ}=\om$ (left), $-\frac{d S_\omega}{d\omega}=Q$ (middle), and $\frac{dG_I}{dI}=\half \omega^2$ (right). Note the excellent  agreement between the analytical dotted lines and the numerical data (dots). }
  \label{fig:dlgrd}
\end{figure}

\fig{fig:dlgrd} shows the Legendre relations:$\frac{dE_Q}{dQ}$ v. $\om$, $-\frac{d S_\omega}{d\omega}$ v. $Q$, and $\frac{dG_I}{dI}$ v. $\half \om^2$ which can be used to check \eq{legendre}. We have also checked the validity of the Legendre transformations in \eqm{legtrns}{legtrns2}. Since the numerical results match our analytical ones, these results strengthen the validity of our analytic arguments.

\section{Conclusion}
\label{conc}
We have numerically and analytically explored the stationary properties of a single $Q$-ball for arbitrary spatial dimension $D$. With the time-dependent non-linear solutions in the system, the virial theorem induces the characteristic slopes \eq{virieq}, and gives the approximate critical values for $\om_a$ in \eq{viriwa} without requiring a knowledge of the detailed profiles and potential forms. By linearising the $Q$-ball \eq{QBeq} or rescaling in $\So$, we have been able to consider the two limiting cases called the thin and thick wall $Q$-balls. The step-like ansatz of \eq{equation} can describe thin wall $Q$-balls in the extreme limit $\om=\om_-$, whereas the modified ansatz \eq{thindanstz} is applicable to $\sigma_0\simeq \sigma_+$ which leads to wider range of parameter space $\om$ and of course  includes the previous limit. On the other hand, the limit $\om\simeq \om_+$ is used to describe thick wall $Q$-balls in both the Gaussian ansatz \eq{gaussansatz} and our modified ansatz for the thick wall case.

The thin wall approximation is valid for $D \ge 2$. Since the step-like ansatz in the thin wall approximation does not have surface effects, the characteristic slope is simply $E_Q/\om Q=1$, \eq{QEGYTHIN}. With the modified ansatz including
surface effects, the classical stability for thin
wall $Q$-balls does not depend on $D$ in \eq{q1class}, but the absolute stability condition \eq{ndvpwa} does. Throughout the analysis, we have
assumed \eqs{coreQ}{app2}, and imposed \eq{const} explicitly, which differ from the analysis in
\cite{PaccettiCorreia:2001uh}. Without these approximations, our calculations in particular \eqs{shellsw}{shllsw2} and \eq{swall} become inconsistent. The mechanical analogies and the numerical
results naturally explain and validate our underlying assumptions: the core sizes of the $Q$-balls are
much smaller than their corresponding  thickness as seen in the middle two panels of \fig{fig:sig}, and the
surface tension depends weakly on $\om$ as seen in \tbl{tbl:ndrq}. With these
assumptions, thin wall $Q$-balls for $\om < \om_a$ are absolutely stable. Moreover, the characteristic slopes coincide with those derived using the virial
theorem. This follows from our analysis of the relative contributions between the potential and surface energies. The slopes have two types in either non-degenerate vacua potentials (NDVPs) or degenerate vacua potentials (DVPs): NDVPs have a large energy from the charge, hence the surface energy is less effective than the potential energy. They support the existence of $Q$-matter in the extreme limit, $\om=\om_-$. DVPs, however, have negligible energy from the charge compared to surface and potential energies, thus the surface energy is well virialised with the potential energy. As seen in the left-bottom panel of \fig{fig:conf}, the configurations of energy density have peaks within the shells, which contribute to the surface energy. It would be worthwhile understanding these peaks in terms of our modified ansatz. Even in the extreme thin wall limit, the charge and energy of the $Q$-balls in NDVPs are not proportional to the volume, \ie no $Q$-matter.

Thick wall $Q$-ball solutions naturally tend to free charged and massive particle solutions \eq{freeengy}. With the simple Gaussian ansatz we have extremised $\So$ with respect to $\sigma_0$ and $R$ with fixed $\om$, while the approaches in \cite{Gleiser:2005iq} are that $E_Q$ is extremised with respect to only $R$. By extremising with respect to two degrees of freedom we are able to  recover the expected results of \eqs{apprxsigma0}{gausseq} unlike in \cite{Gleiser:2005iq}. The Gaussian ansatz, however, is valid only for $D=1$ because of \eq{gausscore}, and gives contradictory  results for the condition for classical stability. In order to remove these drawbacks in the Gaussian ansatz, we introduced another modified ansatz and used the Legendre relations to simplify the computations of $\So,\; Q,\; E_Q$. We obtained a consistent classical stability condition \eq{modcls} which depends on $D$ and a non-linear power $n$ of the polynomial potential \eq{thckpot}. Not surprisingly, our numerical results suggest the modified ansatz is much better than the Gaussian ansatz in the bottom two panels of \fig{fig:egy}. With the same panels, the validity condition \eq{validthck} in the modified ansatz has also been confirmed numerically.

In \eqs{viriwa}{ndvpwa} and \tbl{tbl:dwmc}, the analytical and numerical results found the critical value $\om_a$ with an assumption. The assumption says that the higher dimensional $Q$-balls could be applicable to the thin wall approximations over a wide range values of $\om$. Although this statement may not hold for extremely flat potentials \cite{Laine:1998rg, Asko:2002phd} because $\mu$ can be as small as $1/R_Q$ (see \eq{app2}) and the energy spectrum has the following proportionality $E_Q\propto Q^{D/D+1}$, we believe that it may apply to other large types of the $Q$-ball potentials. In summary, the higher dimensional $Q$-balls can be simplified into the thin and thick wall cases, while it is more challenging and interesting to understand stationary properties of one- and two- dimensional $Q$-balls. For example, those $Q$-balls embedded in $3D$ space (called $Q$-strings and $Q$-walls) may exist in the formation of three dimensional $Q$-balls \cite{mitmovie, MacKenzie:2001av}.

The properties of non-thermal $Q$-balls can lead to different consequences
compared to thermal ones, \ie in the evolution of the universe. The thermal
effects on $Q$-balls induce subsequent radiation and evaporation. The
Affleck-Dine condensate provides a natural homogeneous condensate with small
quantum fluctuation, these fluctuations are then amplified to non-linear objects namely $Q$-balls if the pressure of AD condensates is negative. The formation, dynamics, and thermalization might have phenomenological consequences in our present universe, \eg gravitational waves \cite{GarciaBellido:2007af} and baryon to photon ratio.
\paragraph*{Acknowledgement}

Our numerical simulations were carried out by UK National Cosmology Supercomputer, Cosmos, funded by STFC, HEFCE and silicon Graphics, and carried out by Nottingham HPC facility. M.I.T. developed the code based on LAT.field by Neil Bevis and Mark Hindmarsh. He would like to thank Kari Enqvist, Masahiro Kawasaki, and Osamu Seto for fruitful discussions and to the former two groups for their hospitalities. We are also grateful to Yoonbai Kim for useful correspondence. The work of M.I.T. is supported by Nottingham University studentships. EJC would like to acknowledge the Royal Society for financial support.


\end{document}